\pdfoutput=1
\documentclass[10pt,reqno,oneside]{amsart}

\usepackage{lmodern}
\usepackage[T1]{fontenc}
\usepackage[utf8]{inputenc}
\usepackage[british]{babel}
\usepackage{amssymb}
\usepackage{microtype}
\usepackage{booktabs}
\usepackage{graphicx}
\usepackage{xcolor}
\usepackage{geometry}
\usepackage{tikz}
\usepackage[colorlinks=true,allcolors=blue]{hyperref}
\definecolor{inkline}{gray}{0.10}
\definecolor{inkmid}{gray}{0.45}
\definecolor{inkpale}{gray}{0.72}
\definecolor{accent}{RGB}{178,24,43}
\definecolor{accentpale}{RGB}{223,150,158}
\definecolor{cool}{RGB}{33,80,140}
\definecolor{coolpale}{RGB}{146,177,211}

\tikzset{
  every picture/.style={line cap=round, line join=round},
  limb/.style        ={draw=inkmid,   line width=0.50pt},
  grat/.style        ={draw=inkpale,  line width=0.24pt},
  gratback/.style    ={draw=inkpale!40!white, line width=0.20pt,
                       dash pattern=on 0.4pt off 1.1pt},
  fieldback/.style   ={draw=inkpale!35!white, line width=0.22pt,
                       dash pattern=on 0.5pt off 1.2pt},
  equator/.style     ={draw=inkmid,   line width=0.45pt},
  equatorback/.style ={draw=inkpale,  line width=0.35pt,
                       dash pattern=on 1pt off 1pt},
  polemark/.style    ={fill=inkmid},
  orbit/.style       ={draw=inkline,  line width=0.75pt},
  orbitback/.style   ={draw=inkpale!85!black, line width=0.40pt,
                       dash pattern=on 1.1pt off 1.1pt},
  ghost/.style       ={draw=accent!55!white, line width=0.55pt},
  ghostback/.style   ={draw=accentpale!55!white, line width=0.35pt,
                       dash pattern=on 1pt off 1.2pt},
  turn/.style        ={draw=accent,   line width=0.50pt,
                       dash pattern=on 2pt off 1.4pt},
  turnback/.style    ={draw=accentpale, line width=0.35pt,
                       dash pattern=on 1.2pt off 1.4pt},
  curveA/.style      ={draw=inkline,  line width=0.70pt},
  cubicpart/.style   ={draw=inkmid,   line width=0.60pt,
                       dash pattern=on 2.4pt off 1.6pt},
  curveB/.style      ={draw=cool,     line width=0.70pt},
  curveBdash/.style  ={draw=cool,     line width=0.70pt,
                       dash pattern=on 3pt off 1.8pt},
  sep/.style         ={draw=accent,   line width=0.95pt},
  energy/.style      ={draw=accent,   line width=0.50pt,
                       dash pattern=on 2.2pt off 1.6pt},
  axis/.style        ={draw=inkmid,   line width=0.40pt},
  droph/.style       ={draw=inkpale,  line width=0.30pt,
                       dash pattern=on 1pt off 1.2pt},
  dropv/.style       ={draw=inkpale,  line width=0.30pt,
                       dash pattern=on 1pt off 1.2pt},
  dotmark/.style     ={fill=accent},
  ind1/.style        ={draw=inkmid,   line width=0.55pt,
                       dash pattern=on 2pt off 1.5pt},
  ind2/.style        ={draw=coolpale, line width=0.70pt},
  ind3/.style        ={draw=cool,     line width=0.70pt},
  ind4/.style        ={draw=accent,   line width=0.75pt},
  indcrit/.style     ={draw=accent,   line width=0.95pt,
                       dash pattern=on 3pt off 1.8pt},
}
\tikzset{
  fieldpos1/.style     ={draw=accent, line width=0.3pt},
  fieldpos1back/.style ={draw=accentpale!70!white, line width=0.3pt,
                          dash pattern=on 1.1pt off 1.3pt},
  fieldneg1/.style     ={draw=cool, line width=0.3pt,
                          dash pattern=on 2.6pt off 1.5pt},
  fieldneg1back/.style ={draw=coolpale!70!white, line width=0.3pt,
                          dash pattern=on 1.1pt off 1.3pt},
  fieldpos2/.style     ={draw=accent, line width=0.52pt},
  fieldpos2back/.style ={draw=accentpale!70!white, line width=0.52pt,
                          dash pattern=on 1.1pt off 1.3pt},
  fieldneg2/.style     ={draw=cool, line width=0.52pt,
                          dash pattern=on 2.6pt off 1.5pt},
  fieldneg2back/.style ={draw=coolpale!70!white, line width=0.52pt,
                          dash pattern=on 1.1pt off 1.3pt},
  fieldpos3/.style     ={draw=accent, line width=0.74pt},
  fieldpos3back/.style ={draw=accentpale!70!white, line width=0.74pt,
                          dash pattern=on 1.1pt off 1.3pt},
  fieldneg3/.style     ={draw=cool, line width=0.74pt,
                          dash pattern=on 2.6pt off 1.5pt},
  fieldneg3back/.style ={draw=coolpale!70!white, line width=0.74pt,
                          dash pattern=on 1.1pt off 1.3pt},
  fieldpos4/.style     ={draw=accent, line width=0.96pt},
  fieldpos4back/.style ={draw=accentpale!70!white, line width=0.96pt,
                          dash pattern=on 1.1pt off 1.3pt},
  fieldneg4/.style     ={draw=cool, line width=0.96pt,
                          dash pattern=on 2.6pt off 1.5pt},
  fieldneg4back/.style ={draw=coolpale!70!white, line width=0.96pt,
                          dash pattern=on 1.1pt off 1.3pt},
  fieldpos5/.style     ={draw=accent, line width=1.18pt},
  fieldpos5back/.style ={draw=accentpale!70!white, line width=1.18pt,
                          dash pattern=on 1.1pt off 1.3pt},
  fieldneg5/.style     ={draw=cool, line width=1.18pt,
                          dash pattern=on 2.6pt off 1.5pt},
  fieldneg5back/.style ={draw=coolpale!70!white, line width=1.18pt,
                          dash pattern=on 1.1pt off 1.3pt},
}

\newlength{\plotw}
\newlength{\ploth}
\newcommand{\plotunits}[2]{%
  \setlength{\plotw}{0.6\columnwidth}%
  \setlength{\ploth}{0.3708203932\columnwidth}
  \pgfmathsetlengthmacro{\plotux}{\plotw/#1}%
  \pgfmathsetlengthmacro{\plotuy}{\ploth/#2}%
}

\DeclareRobustCommand{\orcid}[1]{\,\href{https://orcid.org/#1}%
  {\raisebox{-0.5pt}{\includegraphics[width=8pt]{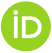}}}}

\theoremstyle{definition}
\newtheorem{convention}{Convention}
\theoremstyle{remark}
\newtheorem{remark}{Remark}

\newcommand{\Sph}{\mathbb{S}}
\newcommand{\Reals}{\mathbb{R}}

\def\d{{\rm d}}
\def\beq{\begin{equation}}
\def\eeq{\end{equation}}

\begin{document}

\title[Lissajous-type orbits and non-Riemannian geometry]%
{Lissajous-type orbits on a sphere and their non-Riemannian geometry}

\author[C. S. L\'opez-Monsalvo]{C\'esar S. L\'opez-Monsalvo\orcid{0000-0002-0378-0415}}
\address{Departamento de Ciencias B\'asicas, Universidad Aut\'onoma Metropolitana --
Azcapotzalco, Avenida San Pablo 420, Colonia Nueva El Rosario, Azcapotzalco 02128,
Ciudad de M\'exico, Mexico}
\email{cslm@azc.uam.mx}
\thanks{Corresponding author: C. S. L\'opez-Monsalvo.}

\author[S. Islas-Ram\'irez]{Sergio Islas-Ram\'irez}
\address{Departamento de Ciencias B\'asicas, Universidad Aut\'onoma Metropolitana --
Azcapotzalco, Avenida San Pablo 420, Colonia Nueva El Rosario, Azcapotzalco 02128,
Ciudad de M\'exico, Mexico}
\email{al2202803149@azc.uam.mx}

\author[A. Rubio Ponce]{Alberto Rubio Ponce\orcid{0000-0002-7562-2918}}
\address{Departamento de Ciencias B\'asicas, Universidad Aut\'onoma Metropolitana --
Azcapotzalco, Avenida San Pablo 420, Colonia Nueva El Rosario, Azcapotzalco 02128,
Ciudad de M\'exico, Mexico}
\email{arp@azc.uam.mx}


\begin{abstract}
A charged test particle is confined to the round sphere in a uniform ambient magnetic field. We reduce its motion to quadrature by elementary means and solve explicitly the family of trajectories which reaches the poles. On that family the azimuth advances at a constant rate and the latitude obeys a pendulum equation. The winding gained per oscillation is a complete elliptic integral of the first kind. That integral is a strictly increasing bijection onto the positive numbers. Every positive rational winding is therefore carried by exactly one value of the half-cyclotron frequency. Then we ask which geometry has these trajectories. No affine connection has them among its geodesics. A Finsler metric of Randers type has all of them. The threshold which separates the two dynamical regimes is precisely the condition for that metric to exist. The closed trajectories are its closed geodesics. The example places a non-Riemannian geometry within reach of an upper-level undergraduate or beginning graduate course in classical mechanics. We supply the derivations, the numerical recipes and the exercises.
\end{abstract}

\maketitle

\section{Introduction}
\label{sec.intro}

\emph{When is a trajectory a geodesic?}

When no force acts on a particle its motion is inertial. Whether the inertial motions of a theory are the straightest curves of a geometry is a question lying at the very core of the link between geometry and physics~\cite{lm2026newton}. On a surface with no forces they are, and the chain which establishes it --- a metric, its Levi-Civita connection, the curves that connection carries parallel to themselves --- is the one we learn to run in that order. Switch a force on and the chain returns nothing.

That is worth testing, and the simplest curved example available will serve. A charged test particle moves on a round sphere in a uniform ambient magnetic field, feeling the field induced on the surface. Nothing in the setting is unfamiliar --- the surface is the one every syllabus already carries, the field the one every laboratory already has. We solve it completely, and then ask what kind of object its orbits are.

No affine connection whatsoever has these orbits among its geodesics, as a criterion of Cartan settles in a single fourth derivative~\cite{cartan1924,ovsienko2005}. The chain from a metric to its connection to its geodesics thus breaks at a named link. What repairs it is forced by a homogeneity count. The Lorentz force is of degree one in the velocity, and only a length differing from a Riemannian one by a term of degree exactly one can absorb it. That length is \emph{Finslerian}, and the geometry is not imported into the example; it is produced by it.

The dynamics then confirms the geometry twice over. A single threshold in $\alpha$ separates orbits which cross the poles from orbits trapped in a cap, and it appears first as the separatrix of a pendulum. It is precisely the condition under which the Randers expression is a metric at all. The closed orbits, selected by an arithmetic condition on a rotation number, are its closed geodesics.

What should a reader take from the construction? Four things, and only the last is peculiar to this system. The first is that a conserved quantity need not be the one a symmetry naively suggests. The integral which reduces our problem is the momentum canonically conjugate to the longitude, not the angular momentum about the polar axis, and the two differ by the term the field contributes. The second is that recognising a normal form can be the whole of a calculation, since one substitution turns the equation for the latitude into a pendulum and everything qualitative then follows from a phase portrait already familiar. The third is that special functions arrive on schedule rather than by fiat, the complete elliptic integral appearing as the period of a nonlinear oscillator and the rotation number built from it deciding whether the motion repeats. The fourth is the one we care most about. Whether a family of curves is the geodesic family of some geometry is a question with a definite answer, reached by differentiation, and a negative answer for connections does not close the matter. A reader who follows the argument to its end has watched a non-Riemannian geometry be forced by a mechanical problem, which is worth seeing once in an upper-level undergraduate or beginning graduate course. Nothing is assumed beyond ordinary differential equations, elementary mechanics and the differential geometry of surfaces, with the calculus of variations wanted only in Section~\ref{sec.randers}; elliptic functions and Randers metrics are built where they are needed rather than presupposed. The six exercises test the four outcomes in order, and the development divides naturally into a guided reading or into problems set across several lectures.

We work throughout on one family of orbits, the one which reaches the poles. The remaining level sets are integrable as well, but their azimuthal advance is a complete elliptic integral of the third kind rather than of the first, a heavier computation with the same conclusions~\cite{lm2026closure}. We develop no general theory of Finsler spaces, using no result from it beyond the definition of a Randers metric.

The system has been studied from several directions. Saksida formulated the motion through the Neumann system and the spherical pendulum~\cite{saksida2002neumann}; Dragovi\'c, Gaji\'c and Jovanovi\'c proved the flow Liouville integrable and integrated it in elliptic functions~\cite{dragovic2023gyroscopic}. That magnetic flows of exact fields are Randers geodesic flows above a critical energy is classical~\cite{randers1941,baochernshen2000,baoroblesshen2004,gibbons2009}. We bring these strands together in a self-contained educational treatment of this one spherical problem, joining its explicit dynamical reduction to the identification of its trajectories as Randers geodesics.

The manuscript is structured as follows. In Section~\ref{sec.model} we set up the model and reduce it to a quadrature; in Section~\ref{sec.boundary} we specialise to the orbits which reach the poles, where the motion is a pendulum, and locate the threshold as its separatrix; in Section~\ref{sec.closure} we compute the winding per oscillation. Then, in Section~\ref{sec.noconnection} we show that no affine connection has these orbits, and in Section~\ref{sec.randers} we exhibit the Randers metric which does. Section~\ref{sec.closing} gathers what was shown.

\section{A charged particle on a round sphere}
\label{sec.model}

Let $\Sph^2$ be the unit sphere in $\Reals^3$, with colatitude $\theta \in (0,\pi)$ measured from the north pole and longitude $\varphi$, in which the round metric is
\beq \label{eq.metric}
g = \d\theta^2 + \sin^2\theta \, \d\varphi^2 .
\eeq
A uniform magnetic field of strength $B$ along the $z$ axis of the ambient space is the two-form $B \, \d x \wedge \d y$, and what a particle confined to the sphere feels is its restriction, which is not uniform. Let us pull it back along the inclusion
\beq \label{eq.inclusion}
\iota(\theta,\varphi)
 = \left( \sin\theta\cos\varphi, \ \sin\theta\sin\varphi, \ \cos\theta \right) ,
\eeq
which gives
\beq \label{eq.field}
F = \iota^{*}\!\left( B \, \d x \wedge \d y \right) = B \cos\theta \; {\rm Vol}_g ,
\qquad {\rm Vol}_g = \sin\theta \, \d\theta \wedge \d\varphi ,
\eeq
Thus the field on the surface is a \emph{function} multiple of the area form, positive on the northern hemisphere, negative on the southern one, and zero on the equator. It carries no flux through the sphere, so it is exact, and we shall use a primitive of it in Section~\ref{sec.randers}.

The equation of motion is the Lorentz force written covariantly, which is the formulation the problem asks for. The particle is confined to a surface, so the acceleration which appears must be the one intrinsic to it, and we formulate the force intrinsically rather than through the ambient cross product~\cite{lm2025monopole,lm2026closure}. Writing $\gamma$ for the trajectory, $\nabla$ for the Levi-Civita connection of the round metric~\eqref{eq.metric}, and $J$ for the rotation by a quarter turn in each tangent plane, a field $F = f \, {\rm Vol}_g$ acts through
\beq \label{eq.lorentz}
m \, \nabla_{\dot\gamma} \dot\gamma = q \, f \, J(\dot\gamma) ,
\eeq
which is the surface form of $m\ddot{\mathbf r} = q \, \dot{\mathbf r} \times \mathbf{B}$. In coordinates the covariant equation~\eqref{eq.lorentz} becomes the pair
\beq \label{eq.eom}
\ddot\theta = \sin\theta\cos\theta \, \dot\varphi \, (\dot\varphi - 2\alpha) ,
\qquad
\ddot\varphi = -2\cot\theta \, \dot\theta \, (\dot\varphi - \alpha) ,
\eeq
in which the entire dependence on the charge, the mass and the field has collected into the single constant
\beq \label{eq.alpha}
\alpha = \frac{qB}{2m} ,
\eeq
half the cyclotron frequency the same particle would have in the flat plane. Two conserved quantities follow, and we derive both.

The first is the speed. The magnetic force does no work, so
\beq \label{eq.speed}
v^2 = \dot\theta^2 + \sin^2\theta \, \dot\varphi^2
\eeq
is constant along every orbit, as we verify by differentiating it and substituting the equations of motion~\eqref{eq.eom}. The second is less obvious, and it is the quantity on which everything below turns. Differentiating
\beq \label{eq.ell}
\ell = \sin^2\theta \, \left( \dot\varphi - \alpha \right)
\eeq
along an orbit, the product rule gives $2\sin\theta\cos\theta \, \dot\theta (\dot\varphi - \alpha) + \sin^2\theta \, \ddot\varphi$, which the second equation of motion cancels exactly, so $\dot\ell = 0$. It is not the angular momentum about the polar axis, which is $\sin^2\theta\,\dot\varphi$ and is not conserved; the two differ by the term $\alpha \sin^2\theta$ that the field contributes. The distinction is the classical analogue of that between mechanical and canonical momentum in the presence of a gauge potential.

Solving the first integral~\eqref{eq.ell} for the azimuthal rate and substituting into the speed~\eqref{eq.speed} reduces the motion to a quadrature,
\beq \label{eq.quadrature}
\dot\varphi = \alpha + \frac{\ell}{\sin^2\theta} ,
\qquad
\dot\theta^2 = v^2 - \frac{\left( \ell + \alpha\sin^2\theta \right)^2}{\sin^2\theta} .
\eeq
Every orbit of the system is contained in that pair. Two consequences are immediate. Indeed, the right-hand side of the polar quadrature~\eqref{eq.quadrature} tends to $-\infty$ as $\theta \to 0$ or $\theta \to \pi$ unless $\ell$ vanishes, so \emph{a pole is reached if and only if $\ell = 0$}; every other orbit oscillates between two turning parallels and never leaves the band between them. Moreover, rescaling the time by the speed shows that the shape of an orbit depends on the parameters only through $\alpha/v$ and $\ell/v$.

\begin{convention}[Units and the parameters kept]
\label{conv.units}
We take the sphere to have unit radius and keep the speed $v$ and the half-cyclotron frequency $\alpha$ of~\eqref{eq.alpha} explicit. Numerical values are quoted at $v = 1$.
\end{convention}

In this sense the system has one control parameter, the ratio $\alpha/v$, and one label distinguishing its orbits, the constant $\ell$. Everything below works at $\ell = 0$.

\section{The orbits that reach the poles}
\label{sec.boundary}

\emph{Which orbits of this system move with two clean frequencies?} Only one family does. The azimuthal rate in the quadrature~\eqref{eq.quadrature} is $\alpha + \ell/\sin^2\theta$, which varies along an orbit as the latitude changes, unless $\ell$ vanishes. So we set
\beq \label{eq.ellzero}
\ell = 0
\eeq
and keep it so throughout. This is exactly the family of orbits that reaches the poles, and exactly the family on which
\beq \label{eq.phidot}
\dot\varphi = \alpha
\eeq
holds identically. The longitude advances at a strictly constant rate, fixed by the field alone, while the latitude does something else.

Substituting the constant azimuthal rate~\eqref{eq.phidot} into the equations of motion~\eqref{eq.eom} and into the speed~\eqref{eq.speed} leaves a single autonomous equation for the latitude, together with its first integral,
\beq \label{eq.polar}
\ddot\theta = -\alpha^2 \sin\theta\cos\theta ,
\qquad
\dot\theta^2 = v^2 - \alpha^2 \sin^2\theta .
\eeq
An effective potential $\alpha^2\sin^2\theta$ confines the latitude, with minima at the two poles and a maximum on the equator, and the particle carries the fixed energy $v^2$ against it; we shall meet that potential again in Section~\ref{sec.randers}, where it acquires a name. However, the polar equation~\eqref{eq.polar} is a pendulum in disguise. Put
\beq \label{eq.psi}
\psi = 2\theta ,
\eeq
so that the double angle turns $\sin\theta\cos\theta$ into $\tfrac12 \sin\psi$ and the polar equation~\eqref{eq.polar} becomes
\beq \label{eq.pendulum}
\ddot\psi = -\alpha^2 \sin\psi ,
\eeq
the equation of a plane pendulum of natural frequency $\alpha$. Its energy follows from the first integral in~\eqref{eq.polar} by the same substitution,
\beq \label{eq.energy}
\tfrac12 \dot\psi^2 - \alpha^2 \cos\psi = 2v^2 - \alpha^2 .
\eeq
The correspondence is exact and it inverts the intuition twice over. The stable equilibrium of the pendulum, $\psi = 0$, is a \emph{pole} of the sphere, and the unstable one, $\psi = \pi$, is the \emph{equator}.

\begin{figure}[tbp]
\centering
\plotunits{1.06}{1.85}%
\begin{tikzpicture}[x=\plotux,y=\plotuy]
  \draw[curveA] (0.0000,0.0000) -- (0.0134,0.0009) -- (0.0267,0.0034) -- (0.0534,0.0137) -- (0.0818,0.0317) -- (0.1102,0.0564) -- (0.1419,0.0911) -- (0.1753,0.1342) -- (0.2104,0.1846) -- (0.3105,0.3359) -- (0.3422,0.3792) -- (0.3706,0.4134) -- (0.4040,0.4468) -- (0.4207,0.4602) -- (0.4374,0.4713) -- (0.4541,0.4799) -- (0.4691,0.4854) -- (0.4841,0.4888) -- (0.4992,0.4900) -- (0.5125,0.4892) -- (0.5259,0.4868) -- (0.5526,0.4767) -- (0.5810,0.4590) -- (0.6093,0.4344) -- (0.6411,0.3999) -- (0.6745,0.3570) -- (0.7095,0.3067) -- (0.8114,0.1529) -- (0.8431,0.1098) -- (0.8715,0.0757) -- (0.9048,0.0425) -- (0.9215,0.0292) -- (0.9382,0.0182) -- (0.9549,0.0098) -- (0.9699,0.0044) -- (0.9850,0.0011) -- (1.0000,0.0000);
  \draw[curveBdash] (0.0000,0.0000) -- (0.0134,0.0030) -- (0.0267,0.0119) -- (0.0401,0.0266) -- (0.0534,0.0472) -- (0.0668,0.0733) -- (0.0818,0.1092) -- (0.0952,0.1466) -- (0.1102,0.1945) -- (0.1419,0.3142) -- (0.1753,0.4628) -- (0.2104,0.6367) -- (0.3105,1.1586) -- (0.3422,1.3078) -- (0.3706,1.4258) -- (0.3873,1.4869) -- (0.4040,1.5409) -- (0.4207,1.5873) -- (0.4374,1.6255) -- (0.4541,1.6551) -- (0.4691,1.6741) -- (0.4841,1.6858) -- (0.4925,1.6891) -- (0.4992,1.6900) -- (0.5125,1.6874) -- (0.5259,1.6789) -- (0.5392,1.6645) -- (0.5526,1.6443) -- (0.5659,1.6185) -- (0.5810,1.5830) -- (0.5943,1.5459) -- (0.6093,1.4983) -- (0.6411,1.3792) -- (0.6745,1.2312) -- (0.7095,1.0576) -- (0.8114,0.5273) -- (0.8431,0.3785) -- (0.8715,0.2610) -- (0.8881,0.2002) -- (0.9048,0.1466) -- (0.9215,0.1006) -- (0.9382,0.0628) -- (0.9549,0.0337) -- (0.9699,0.0150) -- (0.9850,0.0038) -- (0.9933,0.0007) -- (1.0000,0.0000);
  \draw[energy] (0,1) -- (1,1);
  \draw[axis] (0,0) -- (1.06,0);
  \draw[axis] (0,0) -- (0,1.85);
  \foreach \x/\lab in {0/{0}, 0.5/{\pi/2}, 1/{\pi}}
    {\draw[axis] (\x,0) -- (\x,-0.06); \node[below,font=\footnotesize] at (\x,-0.06) {$\lab$};}
  \draw[axis] (0,1) -- (-0.02,1);
  \node[left,font=\footnotesize] at (-0.02,1) {$v^2$};
  \node[below,font=\footnotesize] at (0.53,-0.19) {$\theta$};
\end{tikzpicture}\\[2pt]
{\footnotesize (a)}\\[8pt]
\plotunits{2.30}{2.10}%
\begin{tikzpicture}[x=\plotux,y=\plotuy]
  \input{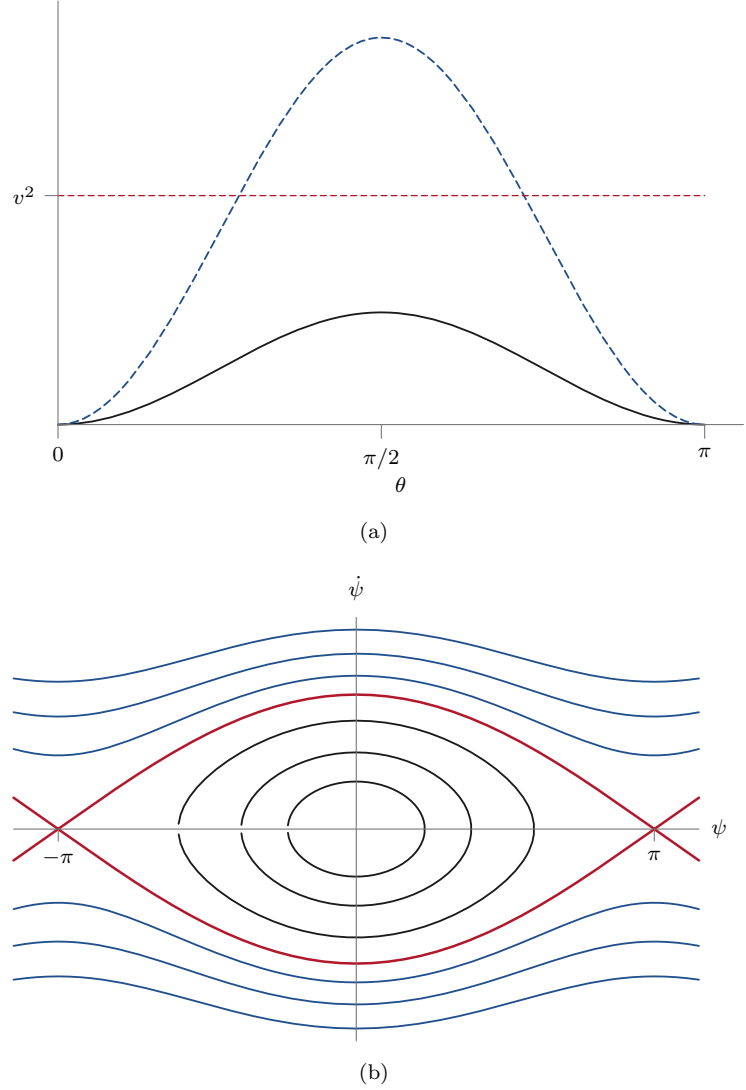}
  \foreach \x/\lab in {-1/{-\pi}, 1/{\pi}}
    {\draw[axis] (\x,0) -- (\x,-0.06); \node[below,font=\footnotesize] at (\x,-0.06) {$\lab$};}
  \node[right,font=\footnotesize] at (1.16,0) {$\psi$};
  \node[above,font=\footnotesize] at (0,1.09) {$\dot\psi$};
\end{tikzpicture}\\[2pt]
{\footnotesize (b)} \caption{\label{fig.pendulum}The threshold as a separatrix. \emph{(a)} The effective potential $\alpha^2\sin^2\theta$ of the polar equation~\eqref{eq.polar}, for $\alpha = 0.7$ solid and $\alpha = 1.3$ dashed at $v = 1$, against the energy $v^2$ the particle carries. Below the threshold the energy clears the barrier, above it the particle turns back. \emph{(b)} The same statement as the phase portrait of the pendulum~\eqref{eq.pendulum}, with the separatrix drawn heavily; the condition~\eqref{eq.threshold1} decides which side of it our particle is on.}
\end{figure}

\begin{figure}[tbp]
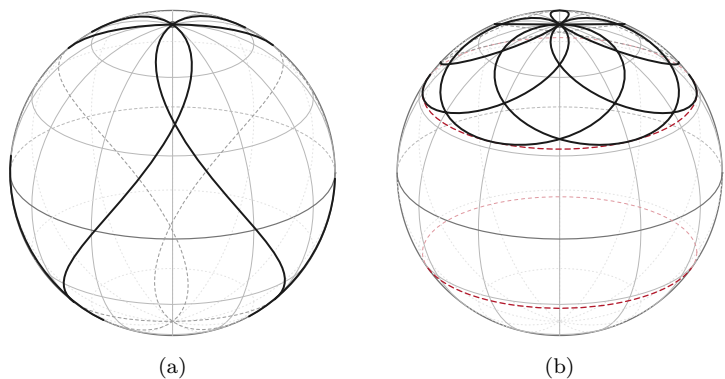

\centering
\begin{tabular}{@{}c@{\hspace{8mm}}c@{}}
\begin{tikzpicture}[x=2.15cm,y=2.15cm]
  \input{figures/tikz/frame.tikz}
  \draw[orbitback] (0.6179,0.7862) -- (0.6418,0.7645) -- (0.6601,0.7408) -- (0.6726,0.7156) -- (0.6793,0.6890) -- (0.6804,0.6601) -- (0.6758,0.6296) -- (0.6655,0.5968) -- (0.6495,0.5615) -- (0.6151,0.5037) -- (0.5662,0.4367) -- (0.5043,0.3613) -- (0.3233,0.1531) -- (0.2325,0.0440) -- (0.1435,-0.0719) -- (0.0686,-0.1816) -- (-0.0022,-0.3031) -- (-0.0309,-0.3613) -- (-0.0553,-0.4179) -- (-0.0757,-0.4739) -- (-0.0914,-0.5275) -- (-0.1024,-0.5785) -- (-0.1090,-0.6279) -- (-0.1108,-0.6817) -- (-0.1064,-0.7323) -- (-0.0958,-0.7792) -- (-0.0793,-0.8209) -- (-0.0569,-0.8580) -- (-0.0289,-0.8900) -- (0.0046,-0.9167) -- (0.0431,-0.9376) -- (0.0862,-0.9526) -- (0.1344,-0.9615) -- (0.1849,-0.9640) -- (0.2395,-0.9600) -- (0.2948,-0.9500) -- (0.3529,-0.9334) -- (0.4117,-0.9108) -- (0.4703,-0.8825);
  \draw[orbitback] (0.9994,-0.0351) -- (0.9993,0.0334) -- (0.9942,0.1032) -- (0.9839,0.1727) -- (0.9686,0.2414) -- (0.9482,0.3090) -- (0.9231,0.3754) -- (0.8932,0.4401) -- (0.8579,0.5044) -- (0.8192,0.5648) -- (0.7753,0.6238) -- (0.7277,0.6795) -- (0.6766,0.7315) -- (0.6225,0.7794) -- (0.5661,0.8227) -- (0.5064,0.8618) -- (0.4467,0.8947);
  \draw[orbitback] (0.6291,-0.7773) -- (0.5860,-0.8061) -- (0.5324,-0.8317) -- (0.4673,-0.8544) -- (0.3925,-0.8737) -- (0.3077,-0.8896) -- (0.2143,-0.9018) -- (0.1159,-0.9097) -- (0.0149,-0.9134) -- (-0.0880,-0.9126) -- (-0.1868,-0.9074) -- (-0.2806,-0.8979) -- (-0.3675,-0.8841) -- (-0.4429,-0.8670) -- (-0.5086,-0.8462) -- (-0.5637,-0.8219) -- (-0.6068,-0.7949);
  \draw[orbitback] (-0.4964,0.8681) -- (-0.5446,0.8385) -- (-0.5928,0.8045) -- (-0.6396,0.7672) -- (-0.6832,0.7279) -- (-0.7260,0.6846) -- (-0.7652,0.6400) -- (-0.8030,0.5918) -- (-0.8369,0.5429) -- (-0.8679,0.4922) -- (-0.8959,0.4399) -- (-0.9208,0.3863) -- (-0.9423,0.3315) -- (-0.9606,0.2758) -- (-0.9754,0.2192) -- (-0.9867,0.1621) -- (-0.9944,0.1058);
  \draw[orbitback] (-0.4250,-0.9052) -- (-0.3697,-0.9286) -- (-0.3136,-0.9470) -- (-0.2600,-0.9592) -- (-0.2083,-0.9657) -- (-0.1589,-0.9664) -- (-0.1125,-0.9612) -- (-0.0697,-0.9502) -- (-0.0310,-0.9334) -- (0.0023,-0.9118) -- (0.0316,-0.8843) -- (0.0551,-0.8527) -- (0.0738,-0.8157) -- (0.0868,-0.7745) -- (0.0941,-0.7295) -- (0.0958,-0.6813) -- (0.0917,-0.6290) -- (0.0835,-0.5812) -- (0.0712,-0.5320) -- (0.0545,-0.4801) -- (0.0333,-0.4260) -- (-0.0205,-0.3151) -- (-0.0911,-0.1974) -- (-0.1650,-0.0913) -- (-0.2524,0.0214) -- (-0.3405,0.1265) -- (-0.5189,0.3309) -- (-0.5817,0.4069) -- (-0.6323,0.4756) -- (-0.6682,0.5349) -- (-0.6852,0.5716) -- (-0.6960,0.6049) -- (-0.7012,0.6367) -- (-0.7007,0.6662) -- (-0.6943,0.6943) -- (-0.6823,0.7202) -- (-0.6644,0.7447) -- (-0.6402,0.7682);
  \draw[orbit] (0.0000,0.9135) -- (0.1029,0.9113) -- (0.2014,0.9047) -- (0.2950,0.8938) -- (0.3800,0.8790) -- (0.4551,0.8608) -- (0.5206,0.8390) -- (0.5745,0.8144) -- (0.6179,0.7862);
  \draw[orbit] (0.4703,-0.8825) -- (0.5242,-0.8513) -- (0.5782,-0.8147) -- (0.6305,-0.7739) -- (0.6794,-0.7303) -- (0.7269,-0.6820) -- (0.7704,-0.6319) -- (0.8108,-0.5790) -- (0.8477,-0.5238) -- (0.8810,-0.4663) -- (0.9098,-0.4087) -- (0.9353,-0.3479) -- (0.9563,-0.2875) -- (0.9736,-0.2245) -- (0.9863,-0.1623) -- (0.9949,-0.0996) -- (0.9994,-0.0351);
  \draw[orbit] (0.4467,0.8947) -- (0.3899,0.9203) -- (0.3332,0.9405) -- (0.2773,0.9549) -- (0.2231,0.9632) -- (0.1711,0.9654) -- (0.1221,0.9614) -- (0.0767,0.9512) -- (0.0364,0.9354) -- (0.0006,0.9140) -- (-0.0304,0.8872) -- (-0.0561,0.8553) -- (-0.0769,0.8178) -- (-0.0914,0.7769) -- (-0.1003,0.7309) -- (-0.1033,0.6815) -- (-0.1002,0.6277) -- (-0.0929,0.5798) -- (-0.0811,0.5290) -- (-0.0646,0.4755) -- (-0.0440,0.4212) -- (-0.0189,0.3648) -- (0.0096,0.3083) -- (0.0804,0.1887) -- (0.1549,0.0809) -- (0.2431,-0.0334) -- (0.3320,-0.1398) -- (0.5117,-0.3461) -- (0.5740,-0.4219) -- (0.6240,-0.4902) -- (0.6589,-0.5482) -- (0.6756,-0.5847) -- (0.6860,-0.6177) -- (0.6909,-0.6485) -- (0.6900,-0.6777) -- (0.6834,-0.7054) -- (0.6711,-0.7309) -- (0.6529,-0.7550) -- (0.6291,-0.7773);
  \draw[orbit] (-0.6068,-0.7949) -- (-0.6306,-0.7738) -- (-0.6491,-0.7506) -- (-0.6615,-0.7264) -- (-0.6684,-0.7001) -- (-0.6698,-0.6723) -- (-0.6656,-0.6422) -- (-0.6557,-0.6096) -- (-0.6400,-0.5746) -- (-0.6059,-0.5172) -- (-0.5574,-0.4505) -- (-0.4960,-0.3753) -- (-0.3135,-0.1650) -- (-0.2217,-0.0545) -- (-0.1321,0.0630) -- (-0.0567,0.1745) -- (-0.0192,0.2366) -- (0.0141,0.2979) -- (0.0423,0.3564) -- (0.0667,0.4146) -- (0.0864,0.4709) -- (0.1017,0.5261) -- (0.1121,0.5786) -- (0.1176,0.6281) -- (0.1182,0.6832) -- (0.1123,0.7350) -- (0.1001,0.7817) -- (0.0816,0.8240) -- (0.0569,0.8614) -- (0.0263,0.8935) -- (-0.0099,0.9198) -- (-0.0513,0.9400) -- (-0.0972,0.9539) -- (-0.1484,0.9614) -- (-0.2017,0.9621) -- (-0.2590,0.9560) -- (-0.3168,0.9435) -- (-0.3771,0.9242) -- (-0.4363,0.8993) -- (-0.4964,0.8681);
  \draw[orbit] (-0.9944,0.1058) -- (-0.9994,0.0295) -- (-0.9984,-0.0456) -- (-0.9914,-0.1205) -- (-0.9786,-0.1947) -- (-0.9600,-0.2681) -- (-0.9351,-0.3417) -- (-0.9052,-0.4121) -- (-0.8693,-0.4818) -- (-0.8281,-0.5488) -- (-0.7810,-0.6141) -- (-0.7305,-0.6743) -- (-0.6748,-0.7315) -- (-0.6156,-0.7838) -- (-0.5537,-0.8305) -- (-0.4898,-0.8712) -- (-0.4250,-0.9052);
  \draw[orbit] (-0.6402,0.7682) -- (-0.5966,0.7982) -- (-0.5410,0.8254) -- (-0.4731,0.8497) -- (-0.3949,0.8704) -- (-0.3063,0.8876) -- (-0.2089,0.9007) -- (-0.1065,0.9094) -- (-0.0000,0.9135);
\end{tikzpicture}
&
\begin{tikzpicture}[x=2.15cm,y=2.15cm]
  \input{figures/tikz/frame.tikz}
  \draw[turnback] (0.8097,0.5868) -- (0.7902,0.6105) -- (0.7636,0.6361) -- (0.7320,0.6607) -- (0.6956,0.6842) -- (0.6547,0.7065) -- (0.6095,0.7273) -- (0.5603,0.7465) -- (0.5075,0.7641) -- (0.4513,0.7799) -- (0.3922,0.7938) -- (0.3305,0.8057) -- (0.2666,0.8155) -- (0.1341,0.8287) -- (-0.0019,0.8331) -- (-0.1379,0.8285) -- (-0.2702,0.8150) -- (-0.3340,0.8051) -- (-0.3955,0.7931) -- (-0.5044,0.7650) -- (-0.5575,0.7476) -- (-0.6069,0.7284) -- (-0.6523,0.7077) -- (-0.6935,0.6855) -- (-0.7301,0.6621) -- (-0.7620,0.6375) -- (-0.7888,0.6120) -- (-0.8097,0.5868);
  \draw[turn] (0.8442,0.4897) -- (0.8420,0.5144) -- (0.8355,0.5389) -- (0.8246,0.5632) -- (0.8097,0.5868);
  \draw[turn] (-0.8097,0.5868) -- (-0.8299,0.5526) -- (-0.8417,0.5159) -- (-0.8440,0.4974) -- (-0.8437,0.4789) -- (-0.8411,0.4604) -- (-0.8360,0.4420) -- (-0.8185,0.4057) -- (-0.7915,0.3704) -- (-0.7553,0.3364) -- (-0.7103,0.3042) -- (-0.6571,0.2742) -- (-0.5962,0.2467) -- (-0.5343,0.2239) -- (-0.4609,0.2021) -- (-0.3821,0.1836) -- (-0.2988,0.1686) -- (-0.2121,0.1574) -- (-0.1229,0.1500) -- (-0.0322,0.1466) -- (0.0588,0.1472) -- (0.1491,0.1518) -- (0.2304,0.1594) -- (0.3165,0.1714) -- (0.3989,0.1871) -- (0.4767,0.2064) -- (0.5489,0.2289) -- (0.6147,0.2544) -- (0.6735,0.2827) -- (0.7243,0.3134) -- (0.7668,0.3461) -- (0.8004,0.3806) -- (0.8246,0.4162) -- (0.8393,0.4528) -- (0.8429,0.4712) -- (0.8442,0.4897);
  \draw[turnback] (0.8442,-0.4897) -- (0.8429,-0.4712) -- (0.8393,-0.4528) -- (0.8262,-0.4193) -- (0.8027,-0.3835) -- (0.7699,-0.3490) -- (0.7282,-0.3161) -- (0.6780,-0.2852) -- (0.6199,-0.2567) -- (0.5546,-0.2309) -- (0.4829,-0.2081) -- (0.4055,-0.1886) -- (0.3235,-0.1726) -- (0.2377,-0.1603) -- (0.1491,-0.1518) -- (0.0588,-0.1472) -- (-0.0322,-0.1466) -- (-0.1229,-0.1500) -- (-0.2121,-0.1574) -- (-0.2988,-0.1686) -- (-0.3821,-0.1836) -- (-0.4609,-0.2021) -- (-0.5343,-0.2239) -- (-0.6016,-0.2489) -- (-0.6571,-0.2742) -- (-0.7103,-0.3042) -- (-0.7553,-0.3364) -- (-0.7915,-0.3704) -- (-0.8185,-0.4057) -- (-0.8360,-0.4420) -- (-0.8411,-0.4604) -- (-0.8437,-0.4789) -- (-0.8440,-0.4974) -- (-0.8417,-0.5159) -- (-0.8299,-0.5526) -- (-0.8097,-0.5868);
  \draw[turnback] (0.8097,-0.5868) -- (0.8246,-0.5632) -- (0.8355,-0.5389) -- (0.8420,-0.5144) -- (0.8442,-0.4897);
  \draw[turn] (-0.8097,-0.5868) -- (-0.7888,-0.6120) -- (-0.7620,-0.6375) -- (-0.7301,-0.6621) -- (-0.6935,-0.6855) -- (-0.6523,-0.7077) -- (-0.6069,-0.7284) -- (-0.5575,-0.7476) -- (-0.5044,-0.7650) -- (-0.4481,-0.7807) -- (-0.3888,-0.7945) -- (-0.3270,-0.8063) -- (-0.2630,-0.8160) -- (-0.1304,-0.8290) -- (-0.0019,-0.8331) -- (0.1341,-0.8287) -- (0.2666,-0.8155) -- (0.3305,-0.8057) -- (0.3922,-0.7938) -- (0.4513,-0.7799) -- (0.5075,-0.7641) -- (0.5603,-0.7465) -- (0.6095,-0.7273) -- (0.6547,-0.7065) -- (0.6956,-0.6842) -- (0.7320,-0.6607) -- (0.7636,-0.6361) -- (0.7902,-0.6105) -- (0.8097,-0.5868);%
  \draw[orbitback] (0.4006,0.9162) -- (0.1714,0.8895) -- (0.0104,0.8676) -- (-0.1579,0.8405) -- (-0.3141,0.8108) -- (-0.4634,0.7771) -- (-0.5897,0.7424) -- (-0.6835,0.7092) -- (-0.7138,0.6947) -- (-0.7308,0.6826);
  \draw[orbitback] (0.0022,1.0000) -- (-0.0365,0.9976) -- (-0.0846,0.9899) -- (-0.1398,0.9772) -- (-0.1985,0.9605) -- (-0.3228,0.9160) -- (-0.4474,0.8604) -- (-0.5595,0.7995) -- (-0.6570,0.7353) -- (-0.6995,0.7024) -- (-0.7372,0.6695) -- (-0.7696,0.6367) -- (-0.7958,0.6056);
  \draw[orbitback] (0.7958,0.6056) -- (0.7696,0.6367) -- (0.7372,0.6695) -- (0.6995,0.7024) -- (0.6570,0.7353) -- (0.5595,0.7995) -- (0.4474,0.8604) -- (0.3228,0.9160) -- (0.1985,0.9605) -- (0.1398,0.9772) -- (0.0846,0.9899) -- (0.0365,0.9976) -- (-0.0022,1.0000);
  \draw[orbitback] (0.7308,0.6826) -- (0.7138,0.6947) -- (0.6835,0.7092) -- (0.5897,0.7424) -- (0.4634,0.7771) -- (0.3141,0.8108) -- (0.1579,0.8405) -- (-0.0104,0.8676) -- (-0.1714,0.8895) -- (-0.4006,0.9162);
  \draw[orbit] (0.0000,0.9135) -- (0.4006,0.9162);
  \draw[orbit] (-0.7308,0.6826) -- (-0.7341,0.6752) -- (-0.7284,0.6702) -- (-0.7143,0.6678) -- (-0.6919,0.6681) -- (-0.6284,0.6762) -- (-0.5450,0.6939) -- (-0.4575,0.7173) -- (-0.3653,0.7463) -- (-0.2743,0.7789) -- (-0.1875,0.8142) -- (-0.1130,0.8486) -- (-0.0501,0.8820) -- (0.0002,0.9137) -- (0.0353,0.9422) -- (0.0535,0.9661) -- (0.0561,0.9762) -- (0.0540,0.9846) -- (0.0474,0.9912) -- (0.0364,0.9961) -- (0.0022,1.0000);
  \draw[orbit] (-0.7958,0.6056) -- (-0.8200,0.5691) -- (-0.8360,0.5342) -- (-0.8435,0.5023) -- (-0.8426,0.4720) -- (-0.8332,0.4448) -- (-0.8154,0.4210) -- (-0.7892,0.4011) -- (-0.7563,0.3860) -- (-0.7164,0.3752) -- (-0.6700,0.3692) -- (-0.6177,0.3682) -- (-0.5622,0.3723) -- (-0.5028,0.3816) -- (-0.4425,0.3955) -- (-0.3824,0.4137) -- (-0.3218,0.4365) -- (-0.2656,0.4617) -- (-0.2128,0.4893) -- (-0.1622,0.5201) -- (-0.1165,0.5522) -- (-0.0745,0.5865) -- (-0.0371,0.6224) -- (-0.0063,0.6580) -- (0.0190,0.6941) -- (0.0380,0.7301) -- (0.0504,0.7653) -- (0.0559,0.7992) -- (0.0541,0.8311) -- (0.0451,0.8605) -- (0.0298,0.8856) -- (0.0071,0.9082) -- (-0.0206,0.9260) -- (-0.0644,0.9428) -- (-0.1157,0.9522) -- (-0.1729,0.9539) -- (-0.2344,0.9479) -- (-0.2981,0.9345) -- (-0.3623,0.9140) -- (-0.4251,0.8872) -- (-0.4851,0.8546) -- (-0.5390,0.8185) -- (-0.5893,0.7774) -- (-0.6317,0.7346) -- (-0.6684,0.6882) -- (-0.6973,0.6405) -- (-0.7180,0.5922) -- (-0.7303,0.5441) -- (-0.7342,0.4955) -- (-0.7301,0.4528) -- (-0.7185,0.4106) -- (-0.7003,0.3720) -- (-0.6747,0.3348) -- (-0.6425,0.3008) -- (-0.6044,0.2703) -- (-0.5606,0.2435) -- (-0.5119,0.2207) -- (-0.4588,0.2021) -- (-0.4000,0.1876) -- (-0.3379,0.1779) -- (-0.2753,0.1733) -- (-0.2089,0.1736) -- (-0.1437,0.1789) -- (-0.0785,0.1890) -- (-0.0139,0.2040) -- (0.0593,0.2277) -- (0.1271,0.2568) -- (0.1903,0.2917) -- (0.2476,0.3320) -- (0.2962,0.3758) -- (0.3367,0.4236) -- (0.3674,0.4731) -- (0.3885,0.5247) -- (0.3994,0.5777) -- (0.3996,0.6291) -- (0.3893,0.6796) -- (0.3686,0.7280) -- (0.3377,0.7731) -- (0.2971,0.8137) -- (0.2479,0.8487) -- (0.1933,0.8761) -- (0.1435,0.8935) -- (0.0930,0.9053) -- (0.0386,0.9121) -- (-0.0144,0.9134) -- (-0.0671,0.9093) -- (-0.1206,0.8995) -- (-0.1698,0.8851) -- (-0.2177,0.8650) -- (-0.2597,0.8413) -- (-0.2986,0.8125) -- (-0.3305,0.7814) -- (-0.3578,0.7460) -- (-0.3786,0.7081) -- (-0.3925,0.6684) -- (-0.3997,0.6275) -- (-0.4001,0.5860) -- (-0.3942,0.5461) -- (-0.3823,0.5068) -- (-0.3649,0.4683) -- (-0.3410,0.4296) -- (-0.3132,0.3942) -- (-0.2794,0.3594) -- (-0.2412,0.3270) -- (-0.1974,0.2962) -- (-0.1500,0.2685) -- (-0.0997,0.2441) -- (-0.0470,0.2232) -- (0.0097,0.2052) -- (0.0677,0.1912) -- (0.1263,0.1811) -- (0.1851,0.1750) -- (0.2455,0.1728) -- (0.3027,0.1747) -- (0.3602,0.1808) -- (0.4155,0.1909) -- (0.4680,0.2049) -- (0.5170,0.2228) -- (0.5622,0.2443) -- (0.6030,0.2693) -- (0.6378,0.2965) -- (0.6688,0.3278) -- (0.6933,0.3605) -- (0.7124,0.3956) -- (0.7256,0.4328) -- (0.7328,0.4717) -- (0.7339,0.5106) -- (0.7289,0.5519) -- (0.7180,0.5922) -- (0.6988,0.6374) -- (0.6726,0.6821) -- (0.6394,0.7257) -- (0.6012,0.7662) -- (0.5559,0.8057) -- (0.5072,0.8407) -- (0.4546,0.8721) -- (0.3990,0.8992) -- (0.3395,0.9220) -- (0.2815,0.9386) -- (0.2222,0.9496) -- (0.1672,0.9541) -- (0.1139,0.9520) -- (0.0675,0.9436) -- (0.0271,0.9292) -- (-0.0061,0.9090) -- (-0.0282,0.8876) -- (-0.0441,0.8627) -- (-0.0534,0.8348) -- (-0.0561,0.8045) -- (-0.0521,0.7722) -- (-0.0416,0.7386) -- (-0.0250,0.7042) -- (-0.0014,0.6681) -- (0.0267,0.6337) -- (0.0610,0.5988) -- (0.0998,0.5652) -- (0.1442,0.5322) -- (0.1919,0.5015) -- (0.2419,0.4736) -- (0.2956,0.4477) -- (0.3501,0.4253) -- (0.4086,0.4053) -- (0.4680,0.3891) -- (0.5255,0.3775) -- (0.5817,0.3703) -- (0.6341,0.3680) -- (0.6832,0.3703) -- (0.7267,0.3774) -- (0.7650,0.3892) -- (0.7953,0.4049) -- (0.8191,0.4249) -- (0.8352,0.4486) -- (0.8432,0.4754) -- (0.8431,0.5051) -- (0.8353,0.5362) -- (0.8194,0.5702) -- (0.7958,0.6056);
  \draw[orbit] (-0.0022,1.0000) -- (-0.0364,0.9961) -- (-0.0474,0.9912) -- (-0.0540,0.9846) -- (-0.0561,0.9762) -- (-0.0535,0.9661) -- (-0.0353,0.9422) -- (-0.0002,0.9137) -- (0.0501,0.8820) -- (0.1130,0.8486) -- (0.1875,0.8142) -- (0.2743,0.7789) -- (0.3653,0.7463) -- (0.4575,0.7173) -- (0.5450,0.6939) -- (0.6284,0.6762) -- (0.6919,0.6681) -- (0.7143,0.6678) -- (0.7284,0.6702) -- (0.7341,0.6752) -- (0.7308,0.6826);
  \draw[orbit] (-0.4006,0.9162) -- (-0.0000,0.9135);
\end{tikzpicture}
\\[2pt]
{\footnotesize (a)} & {\footnotesize (b)}
\end{tabular}
\caption{\label{fig.regimes}The two regimes, three polar periods of each, with the far side of each orbit drawn faintly. \emph{(a)} Rotation, $\alpha = 0.6 < v$: the trajectory runs through both poles. \emph{(b)} Libration, $\alpha = 1.1846 > v$: it is confined to a polar cap bounded by the turning parallels $\sin\theta = v/\alpha$, drawn dashed, and never crosses the equator. Because $\ell$ vanishes it nevertheless reaches the pole at the centre of the cap, which Convention~\ref{conv.theta} carries it through smoothly. This value of $\alpha$ is chosen so that the trapped orbit closes after exactly the three periods drawn, which is what makes the figure symmetric.}
\end{figure}

Recall that a pendulum either swings or goes over the top, and that the two are separated by the energy $\alpha^2$ at which it arrives at $\psi = \pi$ with no speed left. Comparing that against the energy~\eqref{eq.energy} our particle carries,
\beq \label{eq.threshold1}
2v^2 - \alpha^2 = \alpha^2
\qquad \Longleftrightarrow \qquad
\alpha = v ,
\eeq
so the separatrix of the pendulum~\eqref{eq.pendulum} sits precisely where the half-cyclotron frequency equals the speed (Figure~\ref{fig.pendulum}). Above it, $\alpha < v$, the pendulum circulates, the latitude sweeps every value, and the orbit runs through both poles. Below it, $\alpha > v$, the first integral in~\eqref{eq.polar} forces $\sin\theta \le v/\alpha$ and the particle is trapped in a cap about whichever pole it started near. Figure~\ref{fig.regimes} shows one orbit of each kind. Everything below concerns the circulating case, and we shall meet the same inequality~\eqref{eq.threshold1} again in Section~\ref{sec.randers} in an entirely different dress.

\begin{convention}[The latitude is run continuously]
\label{conv.theta}
An orbit of this family passes through the poles, where the longitude of a point on the sphere is undefined. We keep the equations~\eqref{eq.polar} smooth by letting $\theta$ run over the whole real line and locating the particle by
\beq \label{eq.embedding}
(x,y,z) = ( \sin\theta\cos\varphi, \ \sin\theta\sin\varphi, \ \cos\theta ) ,
\eeq
which is defined for every real $\theta$. Continuing $\theta$ past $\pi$ carries the point to the antipodal meridian, which is what a trajectory through a pole actually does, while $\varphi$ stays smooth and obeys the constant rate~\eqref{eq.phidot} throughout. A full oscillation in latitude is then an advance of $2\pi$ in $\theta$, containing one passage through each pole.
\end{convention}

With the latitude run continuously, the polar equation~\eqref{eq.polar} integrates in Jacobi elliptic functions. Writing
\beq \label{eq.modulus}
k = \frac{\alpha}{v} \in (0,1)
\eeq
for the modulus, which by the threshold~\eqref{eq.threshold1} measures how close the field is to the separatrix, the solution starting from the north pole at $\tau = 0$ is
\beq \label{eq.jacobi}
\sin\theta = {\rm sn}(v\tau, k) ,
\qquad
\cos\theta = {\rm cn}(v\tau, k) ,
\qquad
\varphi = \alpha \tau .
\eeq
Here ${\rm sn}$ and ${\rm cn}$ are the Jacobi elliptic sine and cosine, the functions which invert the incomplete elliptic integral of the first kind and reduce to $\sin$ and $\cos$ at $k = 0$. Verifying the solution takes one line, and we leave it as the first exercise. Since ${\rm sn}$ has period $4K(k)$ in its argument, where $K$ is the complete elliptic integral of the first kind, the latitude completes one oscillation in the time
\beq \label{eq.period}
T_\theta = \frac{4K(k)}{v} ,
\eeq
and over that time the longitude, advancing at the constant rate~\eqref{eq.phidot}, gains
\beq \label{eq.advance}
\Delta\varphi = \alpha \, T_\theta = 4 k K(k) ,
\eeq
where the modulus~\eqref{eq.modulus} has absorbed both parameters. The elliptic integral in the period~\eqref{eq.period} is the period of a pendulum, arriving exactly where one expects it rather than as a special function imposed from outside.

Two motions now compose the orbit, and their rates are no longer ours to set. The shape of the trajectory is governed by the ratio of the two, which we compute in the next section.

\section{Closure, and one winding for each value of $\alpha$}
\label{sec.closure}

Define the \emph{rotation number} of an orbit of the family as the azimuth it gains during one oscillation in latitude, measured in turns,
\beq \label{eq.rho}
\rho = \frac{\Delta\varphi}{2\pi} = \frac{2}{\pi} \, k \, K(k) ,
\eeq
using the advance~\eqref{eq.advance}. The trajectory returns to its starting position with its starting velocity exactly when some whole number of latitude oscillations carries the longitude through a whole number of turns.

\begin{convention}[The winding of a closed orbit]
\label{conv.pq}
When the rotation number~\eqref{eq.rho} is rational we write $\rho = p/q$ with $p$ and $q$ coprime positive integers. The orbit then closes after exactly $q$ oscillations in latitude and $p$ turns in longitude, and after no smaller number of either.
\end{convention}

Thus closure is decided by an arithmetic property of the number~\eqref{eq.rho}. A rational rotation number gives a closed orbit, an irrational one a trajectory returning arbitrarily close to its start without ever repeating. By itself that settles little, since it says nothing about which values of $\rho$ the system can produce. The content is in the following observation.

\medskip \noindent\textbf{The rotation number is a bijection.} \emph{The map $k \mapsto \rho(k) = (2/\pi)kK(k)$ is a strictly increasing bijection from $(0,1)$ onto $(0,\infty)$.}

\medskip \noindent The proof is available directly from the properties of $K$. On $(0,1)$ the factor $k$ is positive and strictly increasing, and so is $K(k)$, whose integrand increases pointwise with $k$; a product of two such functions is strictly increasing. At the lower end $K(0) = \pi/2$, so $\rho \to 0$; at the upper end $K$ diverges logarithmically, so $\rho \to \infty$. Continuity supplies everything between. \qed

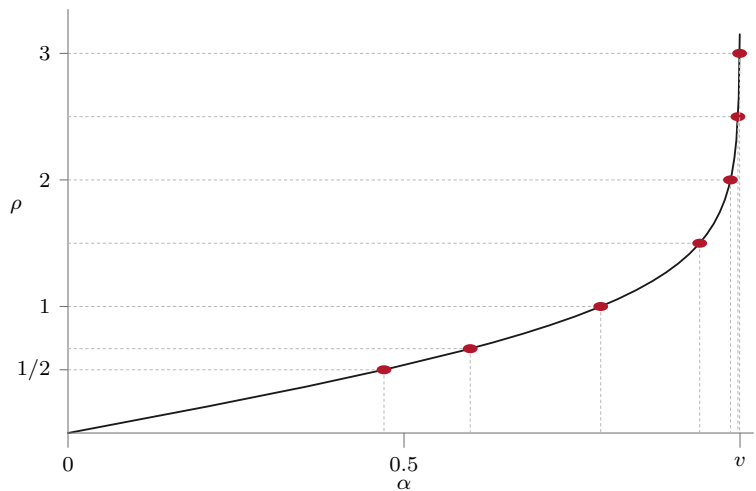
\begin{figure}[tbp]
\centering
\plotunits{1.02}{1.03}%
\begin{tikzpicture}[x=\plotux,y=\plotuy]
  \draw[curveA] (0.0001,0.0000) -- (0.2086,0.0649) -- (0.3551,0.1130) -- (0.4796,0.1574) -- (0.5848,0.1993) -- (0.6329,0.2204) -- (0.6768,0.2413) -- (0.7171,0.2622) -- (0.7539,0.2830) -- (0.7873,0.3041) -- (0.8176,0.3253) -- (0.8451,0.3471) -- (0.8696,0.3693) -- (0.8899,0.3903) -- (0.9081,0.4122) -- (0.9241,0.4347) -- (0.9384,0.4585) -- (0.9506,0.4833) -- (0.9611,0.5095) -- (0.9698,0.5368) -- (0.9774,0.5670) -- (0.9831,0.5974) -- (0.9879,0.6312) -- (0.9916,0.6685) -- (0.9946,0.7129) -- (0.9981,0.8165) -- (0.9996,0.9697);
  \draw[axis] (0,0) -- (1.02,0);
  \draw[axis] (0,0) -- (0,1.03);
  \draw[droph] (0,0.1538) -- (0.4702,0.1538);
  \draw[dropv] (0.4702,0.1538) -- (0.4702,0);
  \fill[dotmark] (0.4702,0.1538) circle (0.011);
  \draw[droph] (0,0.2051) -- (0.5985,0.2051);
  \draw[dropv] (0.5985,0.2051) -- (0.5985,0);
  \fill[dotmark] (0.5985,0.2051) circle (0.011);
  \draw[droph] (0,0.3077) -- (0.7927,0.3077);
  \draw[dropv] (0.7927,0.3077) -- (0.7927,0);
  \fill[dotmark] (0.7927,0.3077) circle (0.011);
  \draw[droph] (0,0.4615) -- (0.9400,0.4615);
  \draw[dropv] (0.9400,0.4615) -- (0.9400,0);
  \fill[dotmark] (0.9400,0.4615) circle (0.011);
  \draw[droph] (0,0.6154) -- (0.9858,0.6154);
  \draw[dropv] (0.9858,0.6154) -- (0.9858,0);
  \fill[dotmark] (0.9858,0.6154) circle (0.011);
  \draw[droph] (0,0.7692) -- (0.9969,0.7692);
  \draw[dropv] (0.9969,0.7692) -- (0.9969,0);
  \fill[dotmark] (0.9969,0.7692) circle (0.011);
  \draw[droph] (0,0.9231) -- (0.9994,0.9231);
  \draw[dropv] (0.9994,0.9231) -- (0.9994,0);
  \fill[dotmark] (0.9994,0.9231) circle (0.011);
  \foreach \x/\lab in {0/{0}, 0.5/{0.5}, 1/{v}}
    {\draw[axis] (\x,0) -- (\x,-0.035); \node[below,font=\footnotesize] at (\x,-0.035) {$\lab$};}
  \foreach \y/\lab in {0.153846/{1/2}, 0.307692/{1}, 0.615385/{2}, 0.923077/{3}}
    {\draw[axis] (0,\y) -- (-0.012,\y); \node[left,font=\footnotesize] at (-0.012,\y) {$\lab$};}
  \node[below,font=\footnotesize] at (0.5,-0.09) {$\alpha$};
  \node[left,font=\footnotesize] at (-0.055,0.55) {$\rho$};
\end{tikzpicture}
\caption{\label{fig.rho}The rotation number~\eqref{eq.rho} against $\alpha$, at $v = 1$. It rises from zero without a single flat place and runs off to infinity as $\alpha$ approaches the threshold~\eqref{eq.threshold1}, so every horizontal line meets it exactly once. The marked points are the entries of Table~\ref{tab.resonances}, each found by reading across from a rational and down to the axis.}
\end{figure}

\medskip \noindent Figure~\ref{fig.rho} is the graph, and the bijection is legible in it, since every horizontal line meets the curve once. The consequence deserves stating in words. \emph{For every rational number $p/q$ there is exactly one value of $\alpha$ which makes the orbit close with that winding --- one, and never several, and never none.} The rotation number~\eqref{eq.rho} is not a quantity we tune; it is one the system produces, and the map producing it happens to be invertible. Two periodic motions compose such an orbit, one in latitude and one in longitude, and whether it closes is settled by the rationality of the ratio of their rates. Curves on the sphere with both angles advancing uniformly, $\theta = m_2\tau$ and $\varphi = m_1\tau$, are known as spherical Lissajous curves~\cite{erb2020}. Ours belong to that family with the uniform latitude replaced by the Jacobi amplitude, since the solution~\eqref{eq.jacobi} is $\theta = {\rm am}(v\tau,k)$, and they reduce to it as $\alpha \to 0$. They are Lissajous-type, then, rather than spherical Lissajous curves in the strict sense. Whereas in the planar figures the two frequencies are dials the experimenter sets, here the single ratio $\alpha/v$ must supply both, and the arithmetic still comes out.

\begin{figure*}[tbp]
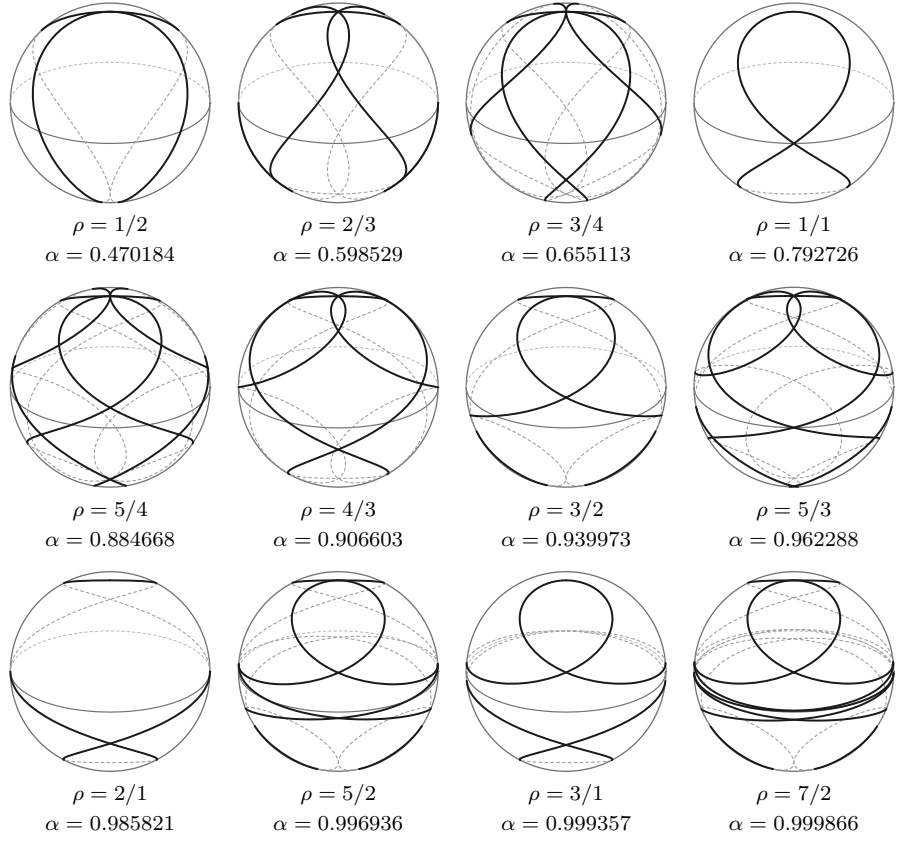

\centering
\begin{tabular}{@{}cccc@{}}
\begin{tikzpicture}[x=1.32cm,y=1.32cm]
  \input{figures/tikz/frame_min.tikz}
  \draw[orbitback] (0.6866,0.7270) -- (0.7182,0.6927) -- (0.7424,0.6566) -- (0.7599,0.6183) -- (0.7710,0.5765) -- (0.7753,0.5324) -- (0.7729,0.4862) -- (0.7638,0.4363) -- (0.7477,0.3826) -- (0.7269,0.3299) -- (0.7000,0.2735) -- (0.6671,0.2133) -- (0.6273,0.1472) -- (0.5349,0.0086) -- (0.3152,-0.2996) -- (0.2195,-0.4394) -- (0.1720,-0.5129) -- (0.1299,-0.5823) -- (0.0944,-0.6452) -- (0.0638,-0.7050) -- (0.0341,-0.7722) -- (0.0137,-0.8310) -- (0.0024,-0.8823) -- (0.0003,-0.9239) -- (0.0027,-0.9415) -- (0.0075,-0.9575) -- (0.0144,-0.9702) -- (0.0239,-0.9810) -- (0.0351,-0.9887) -- (0.0490,-0.9941) -- (0.0651,-0.9967) -- (0.0830,-0.9966);
  \draw[orbitback] (-0.0830,-0.9966) -- (-0.0651,-0.9967) -- (-0.0490,-0.9941) -- (-0.0351,-0.9887) -- (-0.0239,-0.9810) -- (-0.0144,-0.9702) -- (-0.0075,-0.9575) -- (-0.0027,-0.9415) -- (-0.0003,-0.9239) -- (-0.0024,-0.8823) -- (-0.0137,-0.8310) -- (-0.0341,-0.7722) -- (-0.0638,-0.7050) -- (-0.0944,-0.6452) -- (-0.1299,-0.5823) -- (-0.1720,-0.5129) -- (-0.2195,-0.4394) -- (-0.3152,-0.2996) -- (-0.5349,0.0086) -- (-0.6273,0.1472) -- (-0.6671,0.2133) -- (-0.7000,0.2735) -- (-0.7269,0.3299) -- (-0.7477,0.3826) -- (-0.7638,0.4363) -- (-0.7729,0.4862) -- (-0.7753,0.5324) -- (-0.7710,0.5765) -- (-0.7599,0.6183) -- (-0.7424,0.6566) -- (-0.7182,0.6927) -- (-0.6866,0.7270);
  \draw[orbit] (0.0000,0.9135) -- (0.1101,0.9103) -- (0.2177,0.9006) -- (0.3204,0.8847) -- (0.4138,0.8632) -- (0.4987,0.8361) -- (0.5719,0.8046) -- (0.6054,0.7866) -- (0.6346,0.7683) -- (0.6624,0.7480) -- (0.6866,0.7270);
  \draw[orbit] (0.0830,-0.9966) -- (0.1073,-0.9926) -- (0.1350,-0.9842) -- (0.1651,-0.9716) -- (0.1960,-0.9555) -- (0.2284,-0.9356) -- (0.2621,-0.9121) -- (0.3307,-0.8561) -- (0.3994,-0.7893) -- (0.4645,-0.7154) -- (0.5260,-0.6349) -- (0.5810,-0.5513) -- (0.6279,-0.4687) -- (0.6697,-0.3819) -- (0.7040,-0.2964) -- (0.7322,-0.2083) -- (0.7534,-0.1209) -- (0.7674,-0.0349) -- (0.7745,0.0516) -- (0.7744,0.1355) -- (0.7675,0.2161) -- (0.7542,0.2930) -- (0.7342,0.3682) -- (0.7075,0.4410) -- (0.6752,0.5086) -- (0.6367,0.5729) -- (0.5922,0.6333) -- (0.5419,0.6891) -- (0.4862,0.7398) -- (0.4254,0.7850) -- (0.3602,0.8240) -- (0.2935,0.8554) -- (0.2214,0.8811) -- (0.1493,0.8990) -- (0.0757,0.9098) -- (0.0013,0.9135) -- (-0.0731,0.9101) -- (-0.1493,0.8990) -- (-0.2214,0.8811) -- (-0.2935,0.8554) -- (-0.3602,0.8240) -- (-0.4254,0.7850) -- (-0.4862,0.7398) -- (-0.5419,0.6891) -- (-0.5922,0.6333) -- (-0.6367,0.5729) -- (-0.6752,0.5086) -- (-0.7075,0.4410) -- (-0.7342,0.3682) -- (-0.7542,0.2930) -- (-0.7675,0.2161) -- (-0.7744,0.1355) -- (-0.7745,0.0516) -- (-0.7674,-0.0349) -- (-0.7534,-0.1209) -- (-0.7322,-0.2083) -- (-0.7040,-0.2964) -- (-0.6697,-0.3819) -- (-0.6279,-0.4687) -- (-0.5810,-0.5513) -- (-0.5260,-0.6349) -- (-0.4645,-0.7154) -- (-0.3994,-0.7893) -- (-0.3307,-0.8561) -- (-0.2621,-0.9121) -- (-0.2284,-0.9356) -- (-0.1960,-0.9555) -- (-0.1651,-0.9716) -- (-0.1350,-0.9842) -- (-0.1073,-0.9926) -- (-0.0830,-0.9966);
  \draw[orbit] (-0.6866,0.7270) -- (-0.6624,0.7480) -- (-0.6346,0.7683) -- (-0.6054,0.7866) -- (-0.5719,0.8046) -- (-0.4987,0.8361) -- (-0.4138,0.8632) -- (-0.3204,0.8847) -- (-0.2177,0.9006) -- (-0.1101,0.9103) -- (-0.0000,0.9135);
\end{tikzpicture} &
\begin{tikzpicture}[x=1.32cm,y=1.32cm]
  \input{figures/tikz/frame_min.tikz}
  \draw[orbitback] (0.6186,0.7857) -- (0.6429,0.7634) -- (0.6612,0.7397) -- (0.6736,0.7144) -- (0.6803,0.6877) -- (0.6814,0.6594) -- (0.6769,0.6285) -- (0.6664,0.5947) -- (0.6503,0.5592) -- (0.6158,0.5011) -- (0.5659,0.4326) -- (0.5041,0.3570) -- (0.3248,0.1500) -- (0.2351,0.0420) -- (0.1457,-0.0747) -- (0.0715,-0.1835) -- (0.0340,-0.2447) -- (0.0011,-0.3042) -- (-0.0274,-0.3615) -- (-0.0522,-0.4187) -- (-0.0722,-0.4731) -- (-0.0881,-0.5266) -- (-0.0996,-0.5788) -- (-0.1063,-0.6273) -- (-0.1085,-0.6818) -- (-0.1047,-0.7312) -- (-0.0948,-0.7771) -- (-0.0790,-0.8189) -- (-0.0573,-0.8563) -- (-0.0299,-0.8886) -- (0.0029,-0.9156) -- (0.0408,-0.9369) -- (0.0832,-0.9522) -- (0.1297,-0.9615) -- (0.1796,-0.9645) -- (0.2323,-0.9614) -- (0.2872,-0.9521) -- (0.3457,-0.9361) -- (0.4027,-0.9148) -- (0.4610,-0.8874);
  \draw[orbitback] (1.0000,-0.0000) -- (0.9977,0.0663) -- (0.9909,0.1316) -- (0.9796,0.1965) -- (0.9631,0.2632) -- (0.9428,0.3264) -- (0.9172,0.3908) -- (0.8884,0.4512) -- (0.8543,0.5120) -- (0.8161,0.5706) -- (0.7741,0.6266) -- (0.7286,0.6795) -- (0.6800,0.7291) -- (0.6287,0.7750) -- (0.5751,0.8167) -- (0.5197,0.8539) -- (0.4610,0.8874);
  \draw[orbitback] (0.6186,-0.7857) -- (0.5761,-0.8134) -- (0.5220,-0.8383) -- (0.4575,-0.8600) -- (0.3814,-0.8787) -- (0.2966,-0.8935) -- (0.2024,-0.9045) -- (0.1032,-0.9113) -- (0.0013,-0.9135) -- (-0.1005,-0.9114) -- (-0.1999,-0.9048) -- (-0.2942,-0.8938) -- (-0.3792,-0.8791) -- (-0.4556,-0.8606) -- (-0.5204,-0.8390) -- (-0.5748,-0.8141) -- (-0.5979,-0.8004) -- (-0.6186,-0.7857);
  \draw[orbitback] (-0.4610,0.8874) -- (-0.5197,0.8539) -- (-0.5751,0.8167) -- (-0.6287,0.7750) -- (-0.6800,0.7291) -- (-0.7286,0.6795) -- (-0.7741,0.6266) -- (-0.8161,0.5706) -- (-0.8543,0.5120) -- (-0.8884,0.4512) -- (-0.9172,0.3908) -- (-0.9428,0.3264) -- (-0.9631,0.2632) -- (-0.9796,0.1965) -- (-0.9909,0.1316) -- (-0.9977,0.0663) -- (-1.0000,-0.0000);
  \draw[orbitback] (-0.4610,-0.8874) -- (-0.4027,-0.9148) -- (-0.3457,-0.9361) -- (-0.2872,-0.9521) -- (-0.2323,-0.9614) -- (-0.1796,-0.9645) -- (-0.1297,-0.9615) -- (-0.0832,-0.9522) -- (-0.0408,-0.9369) -- (-0.0029,-0.9156) -- (0.0299,-0.8886) -- (0.0573,-0.8563) -- (0.0790,-0.8189) -- (0.0948,-0.7771) -- (0.1047,-0.7312) -- (0.1085,-0.6818) -- (0.1063,-0.6273) -- (0.0996,-0.5788) -- (0.0881,-0.5266) -- (0.0722,-0.4731) -- (0.0522,-0.4187) -- (0.0274,-0.3615) -- (-0.0011,-0.3042) -- (-0.0340,-0.2447) -- (-0.0715,-0.1835) -- (-0.1457,-0.0747) -- (-0.2351,0.0420) -- (-0.3248,0.1500) -- (-0.5041,0.3570) -- (-0.5659,0.4326) -- (-0.6158,0.5011) -- (-0.6503,0.5592) -- (-0.6664,0.5947) -- (-0.6769,0.6285) -- (-0.6814,0.6594) -- (-0.6803,0.6877) -- (-0.6736,0.7144) -- (-0.6612,0.7397) -- (-0.6429,0.7634) -- (-0.6186,0.7857);
  \draw[orbit] (0.0000,0.9135) -- (0.1019,0.9113) -- (0.2012,0.9047) -- (0.2954,0.8937) -- (0.3803,0.8789) -- (0.4566,0.8603) -- (0.5212,0.8387) -- (0.5755,0.8137) -- (0.6186,0.7857);
  \draw[orbit] (0.4610,-0.8874) -- (0.5187,-0.8546) -- (0.5740,-0.8175) -- (0.6277,-0.7758) -- (0.6790,-0.7301) -- (0.7277,-0.6805) -- (0.7732,-0.6276) -- (0.8153,-0.5717) -- (0.8536,-0.5132) -- (0.8878,-0.4524) -- (0.9167,-0.3920) -- (0.9424,-0.3277) -- (0.9627,-0.2645) -- (0.9793,-0.1978) -- (0.9907,-0.1329) -- (0.9976,-0.0676) -- (1.0000,-0.0000);
  \draw[orbit] (0.4610,0.8874) -- (0.4016,0.9153) -- (0.3446,0.9364) -- (0.2883,0.9518) -- (0.2334,0.9613) -- (0.1786,0.9645) -- (0.1288,0.9614) -- (0.0824,0.9520) -- (0.0400,0.9365) -- (0.0023,0.9151) -- (-0.0293,0.8892) -- (-0.0568,0.8569) -- (-0.0787,0.8197) -- (-0.0946,0.7779) -- (-0.1045,0.7321) -- (-0.1085,0.6828) -- (-0.1064,0.6284) -- (-0.0998,0.5799) -- (-0.0884,0.5277) -- (-0.0725,0.4742) -- (-0.0526,0.4199) -- (-0.0279,0.3627) -- (0.0005,0.3053) -- (0.0334,0.2458) -- (0.0707,0.1846) -- (0.1449,0.0758) -- (0.2343,-0.0410) -- (0.3240,-0.1491) -- (0.5047,-0.3578) -- (0.5664,-0.4334) -- (0.6162,-0.5018) -- (0.6507,-0.5598) -- (0.6666,-0.5953) -- (0.6770,-0.6291) -- (0.6815,-0.6600) -- (0.6803,-0.6882) -- (0.6735,-0.7149) -- (0.6609,-0.7401) -- (0.6425,-0.7639) -- (0.6186,-0.7857);
  \draw[orbit] (-0.6186,-0.7857) -- (-0.6425,-0.7639) -- (-0.6609,-0.7401) -- (-0.6735,-0.7149) -- (-0.6803,-0.6882) -- (-0.6815,-0.6600) -- (-0.6770,-0.6291) -- (-0.6666,-0.5953) -- (-0.6507,-0.5598) -- (-0.6162,-0.5018) -- (-0.5664,-0.4334) -- (-0.5047,-0.3578) -- (-0.3240,-0.1491) -- (-0.2343,-0.0410) -- (-0.1449,0.0758) -- (-0.0707,0.1846) -- (-0.0334,0.2458) -- (-0.0005,0.3053) -- (0.0279,0.3627) -- (0.0526,0.4199) -- (0.0725,0.4742) -- (0.0884,0.5277) -- (0.0998,0.5799) -- (0.1064,0.6284) -- (0.1085,0.6828) -- (0.1045,0.7321) -- (0.0946,0.7779) -- (0.0787,0.8197) -- (0.0568,0.8569) -- (0.0293,0.8892) -- (-0.0023,0.9151) -- (-0.0400,0.9365) -- (-0.0824,0.9520) -- (-0.1288,0.9614) -- (-0.1786,0.9645) -- (-0.2334,0.9613) -- (-0.2883,0.9518) -- (-0.3446,0.9364) -- (-0.4016,0.9153) -- (-0.4610,0.8874);
  \draw[orbit] (-1.0000,-0.0000) -- (-0.9976,-0.0676) -- (-0.9907,-0.1329) -- (-0.9793,-0.1978) -- (-0.9627,-0.2645) -- (-0.9424,-0.3277) -- (-0.9167,-0.3920) -- (-0.8878,-0.4524) -- (-0.8536,-0.5132) -- (-0.8153,-0.5717) -- (-0.7732,-0.6276) -- (-0.7277,-0.6805) -- (-0.6790,-0.7301) -- (-0.6277,-0.7758) -- (-0.5740,-0.8175) -- (-0.5187,-0.8546) -- (-0.4610,-0.8874);
  \draw[orbit] (-0.6186,0.7857) -- (-0.5755,0.8137) -- (-0.5212,0.8387) -- (-0.4566,0.8603) -- (-0.3803,0.8789) -- (-0.2954,0.8937) -- (-0.2012,0.9047) -- (-0.1019,0.9113) -- (-0.0000,0.9135);
\end{tikzpicture} &
\begin{tikzpicture}[x=1.32cm,y=1.32cm]
  \input{figures/tikz/frame_min.tikz}
  \draw[orbitback] (0.5919,0.8060) -- (0.6131,0.7878) -- (0.6286,0.7682) -- (0.6386,0.7473) -- (0.6431,0.7251) -- (0.6423,0.7018) -- (0.6360,0.6760) -- (0.6239,0.6478) -- (0.6064,0.6181) -- (0.5721,0.5717) -- (0.5246,0.5178) -- (0.4666,0.4580) -- (0.2975,0.2924) -- (0.2124,0.2059) -- (0.1270,0.1128) -- (0.0511,0.0222) -- (-0.0230,-0.0778) -- (-0.0563,-0.1282) -- (-0.0860,-0.1772) -- (-0.1120,-0.2245) -- (-0.1355,-0.2721) -- (-0.1563,-0.3200) -- (-0.1734,-0.3656) -- (-0.1881,-0.4134) -- (-0.1991,-0.4585) -- (-0.2067,-0.5030) -- (-0.2110,-0.5467) -- (-0.2118,-0.5893) -- (-0.2089,-0.6307) -- (-0.2024,-0.6705) -- (-0.1928,-0.7066) -- (-0.1790,-0.7428) -- (-0.1626,-0.7749) -- (-0.1417,-0.8063) -- (-0.1187,-0.8335) -- (-0.0911,-0.8594) -- (-0.0620,-0.8810) -- (-0.0301,-0.8997) -- (0.0042,-0.9152) -- (0.0366,-0.9263) -- (0.0728,-0.9353) -- (0.1083,-0.9410) -- (0.1474,-0.9442) -- (0.1853,-0.9444) -- (0.2264,-0.9417) -- (0.2681,-0.9360) -- (0.3103,-0.9274) -- (0.3501,-0.9166) -- (0.3923,-0.9025) -- (0.4342,-0.8856) -- (0.4756,-0.8661) -- (0.5162,-0.8442) -- (0.5558,-0.8198) -- (0.5943,-0.7932) -- (0.6314,-0.7645) -- (0.6650,-0.7357) -- (0.6991,-0.7033) -- (0.7295,-0.6713) -- (0.7601,-0.6357) -- (0.7870,-0.6009) -- (0.8136,-0.5627) -- (0.8367,-0.5256) -- (0.8591,-0.4853) -- (0.8781,-0.4464) -- (0.8960,-0.4044) -- (0.9117,-0.3616) -- (0.9250,-0.3183) -- (0.9360,-0.2744) -- (0.9446,-0.2302) -- (0.9508,-0.1857) -- (0.9548,-0.1409) -- (0.9558,-0.0538) -- (0.9484,0.0333) -- (0.9328,0.1198) -- (0.9085,0.2077) -- (0.8762,0.2939) -- (0.8354,0.3804) -- (0.7874,0.4640) -- (0.7314,0.5463) -- (0.6642,0.6304) -- (0.5916,0.7084) -- (0.5133,0.7809) -- (0.4310,0.8462) -- (0.3469,0.9022) -- (0.3058,0.9257) -- (0.2640,0.9468) -- (0.2257,0.9636) -- (0.1876,0.9773) -- (0.1519,0.9871) -- (0.1199,0.9928);
  \draw[orbitback] (0.9470,-0.3213) -- (0.9314,-0.3562) -- (0.9087,-0.3910) -- (0.8786,-0.4262) -- (0.8401,-0.4631) -- (0.7926,-0.5018) -- (0.7378,-0.5411) -- (0.6742,-0.5822) -- (0.5998,-0.6263) -- (0.4984,-0.6819) -- (0.3812,-0.7416) -- (0.2546,-0.8021) -- (0.1231,-0.8614) -- (-0.0027,-0.9146) -- (-0.1024,-0.9530) -- (-0.1698,-0.9743) -- (-0.1918,-0.9787) -- (-0.2054,-0.9787);
  \draw[orbitback] (0.2054,-0.9787) -- (0.1918,-0.9787) -- (0.1698,-0.9743) -- (0.1024,-0.9530) -- (0.0027,-0.9146) -- (-0.1231,-0.8614) -- (-0.2546,-0.8021) -- (-0.3812,-0.7416) -- (-0.4984,-0.6819) -- (-0.5998,-0.6263) -- (-0.6742,-0.5822) -- (-0.7378,-0.5411) -- (-0.7926,-0.5018) -- (-0.8401,-0.4631) -- (-0.8786,-0.4262) -- (-0.9087,-0.3910) -- (-0.9314,-0.3562) -- (-0.9470,-0.3213);
  \draw[orbitback] (-0.1199,0.9928) -- (-0.1519,0.9871) -- (-0.1876,0.9773) -- (-0.2257,0.9636) -- (-0.2640,0.9468) -- (-0.3058,0.9257) -- (-0.3469,0.9022) -- (-0.4310,0.8462) -- (-0.5133,0.7809) -- (-0.5916,0.7084) -- (-0.6642,0.6304) -- (-0.7314,0.5463) -- (-0.7874,0.4640) -- (-0.8354,0.3804) -- (-0.8762,0.2939) -- (-0.9085,0.2077) -- (-0.9328,0.1198) -- (-0.9484,0.0333) -- (-0.9558,-0.0538) -- (-0.9548,-0.1409) -- (-0.9508,-0.1857) -- (-0.9446,-0.2302) -- (-0.9360,-0.2744) -- (-0.9250,-0.3183) -- (-0.9117,-0.3616) -- (-0.8960,-0.4044) -- (-0.8781,-0.4464) -- (-0.8591,-0.4853) -- (-0.8367,-0.5256) -- (-0.8136,-0.5627) -- (-0.7870,-0.6009) -- (-0.7601,-0.6357) -- (-0.7295,-0.6713) -- (-0.6991,-0.7033) -- (-0.6650,-0.7357) -- (-0.6314,-0.7645) -- (-0.5943,-0.7932) -- (-0.5558,-0.8198) -- (-0.5162,-0.8442) -- (-0.4756,-0.8661) -- (-0.4342,-0.8856) -- (-0.3923,-0.9025) -- (-0.3501,-0.9166) -- (-0.3103,-0.9274) -- (-0.2681,-0.9360) -- (-0.2264,-0.9417) -- (-0.1853,-0.9444) -- (-0.1474,-0.9442) -- (-0.1083,-0.9410) -- (-0.0728,-0.9353) -- (-0.0366,-0.9263) -- (-0.0042,-0.9152) -- (0.0301,-0.8997) -- (0.0620,-0.8810) -- (0.0911,-0.8594) -- (0.1187,-0.8335) -- (0.1417,-0.8063) -- (0.1626,-0.7749) -- (0.1790,-0.7428) -- (0.1928,-0.7066) -- (0.2024,-0.6705) -- (0.2089,-0.6307) -- (0.2118,-0.5893) -- (0.2110,-0.5467) -- (0.2067,-0.5030) -- (0.1991,-0.4585) -- (0.1881,-0.4134) -- (0.1734,-0.3656) -- (0.1563,-0.3200) -- (0.1355,-0.2721) -- (0.1120,-0.2245) -- (0.0860,-0.1772) -- (0.0563,-0.1282) -- (0.0230,-0.0778) -- (-0.0511,0.0222) -- (-0.1270,0.1128) -- (-0.2124,0.2059) -- (-0.2975,0.2924) -- (-0.4666,0.4580) -- (-0.5246,0.5178) -- (-0.5721,0.5717) -- (-0.6064,0.6181) -- (-0.6239,0.6478) -- (-0.6360,0.6760) -- (-0.6423,0.7018) -- (-0.6431,0.7251) -- (-0.6386,0.7473) -- (-0.6286,0.7682) -- (-0.6131,0.7878) -- (-0.5919,0.8060);
  \draw[orbit] (0.0000,0.9135) -- (0.0992,0.9117) -- (0.1957,0.9060) -- (0.2870,0.8967) -- (0.3687,0.8842) -- (0.4416,0.8685) -- (0.5028,0.8502) -- (0.5533,0.8291) -- (0.5919,0.8060);
  \draw[orbit] (0.1199,0.9928) -- (0.0962,0.9945) -- (0.0753,0.9936) -- (0.0556,0.9900) -- (0.0396,0.9840) -- (0.0255,0.9750) -- (0.0149,0.9641) -- (0.0071,0.9508) -- (0.0022,0.9354) -- (0.0001,0.9177) -- (0.0009,0.8980) -- (0.0046,0.8764) -- (0.0116,0.8516) -- (0.0334,0.7997) -- (0.0668,0.7408) -- (0.1088,0.6793) -- (0.1606,0.6130) -- (0.2230,0.5411) -- (0.2933,0.4667) -- (0.3638,0.3968) -- (0.4420,0.3229) -- (0.7563,0.0388) -- (0.8217,-0.0258) -- (0.8714,-0.0812) -- (0.8984,-0.1161) -- (0.9198,-0.1486) -- (0.9359,-0.1792) -- (0.9475,-0.2091) -- (0.9544,-0.2385) -- (0.9564,-0.2664) -- (0.9540,-0.2940) -- (0.9470,-0.3213);
  \draw[orbit] (-0.2054,-0.9787) -- (-0.2116,-0.9743) -- (-0.2097,-0.9652) -- (-0.1998,-0.9513) -- (-0.1808,-0.9314) -- (0.0728,-0.7075) -- (0.1495,-0.6351) -- (0.2206,-0.5636) -- (0.2751,-0.5050) -- (0.3250,-0.4477) -- (0.3716,-0.3901) -- (0.4146,-0.3326) -- (0.4613,-0.2640) -- (0.5015,-0.1980) -- (0.5377,-0.1301) -- (0.5684,-0.0633) -- (0.5936,0.0023) -- (0.6141,0.0687) -- (0.6289,0.1331) -- (0.6387,0.1978) -- (0.6433,0.2702) -- (0.6410,0.3394) -- (0.6319,0.4075) -- (0.6164,0.4716) -- (0.5943,0.5337) -- (0.5655,0.5931) -- (0.5301,0.6492) -- (0.4884,0.7015) -- (0.4407,0.7493) -- (0.3895,0.7904) -- (0.3312,0.8276) -- (0.2710,0.8576) -- (0.2048,0.8823) -- (0.1385,0.8995) -- (0.0704,0.9099) -- (0.0014,0.9135) -- (-0.0676,0.9102) -- (-0.1358,0.9000) -- (-0.2022,0.8831) -- (-0.2685,0.8587) -- (-0.3289,0.8290) -- (-0.3873,0.7919) -- (-0.4387,0.7510) -- (-0.4867,0.7034) -- (-0.5287,0.6513) -- (-0.5643,0.5953) -- (-0.5933,0.5360) -- (-0.6157,0.4741) -- (-0.6314,0.4100) -- (-0.6408,0.3419) -- (-0.6434,0.2727) -- (-0.6390,0.2004) -- (-0.6294,0.1357) -- (-0.6148,0.0713) -- (-0.5945,0.0048) -- (-0.5695,-0.0608) -- (-0.5389,-0.1277) -- (-0.5029,-0.1956) -- (-0.4628,-0.2617) -- (-0.4163,-0.3304) -- (-0.3733,-0.3879) -- (-0.3268,-0.4456) -- (-0.2751,-0.5050) -- (-0.2206,-0.5636) -- (-0.1495,-0.6351) -- (-0.0728,-0.7075) -- (0.1808,-0.9314) -- (0.1998,-0.9513) -- (0.2097,-0.9652) -- (0.2116,-0.9743) -- (0.2054,-0.9787);
  \draw[orbit] (-0.9470,-0.3213) -- (-0.9540,-0.2940) -- (-0.9564,-0.2664) -- (-0.9544,-0.2385) -- (-0.9475,-0.2091) -- (-0.9359,-0.1792) -- (-0.9198,-0.1486) -- (-0.8984,-0.1161) -- (-0.8714,-0.0812) -- (-0.8217,-0.0258) -- (-0.7563,0.0388) -- (-0.4420,0.3229) -- (-0.3638,0.3968) -- (-0.2933,0.4667) -- (-0.2230,0.5411) -- (-0.1606,0.6130) -- (-0.1088,0.6793) -- (-0.0668,0.7408) -- (-0.0334,0.7997) -- (-0.0116,0.8516) -- (-0.0046,0.8764) -- (-0.0009,0.8980) -- (-0.0001,0.9177) -- (-0.0022,0.9354) -- (-0.0071,0.9508) -- (-0.0149,0.9641) -- (-0.0255,0.9750) -- (-0.0396,0.9840) -- (-0.0556,0.9900) -- (-0.0753,0.9936) -- (-0.0962,0.9945) -- (-0.1199,0.9928);
  \draw[orbit] (-0.5919,0.8060) -- (-0.5533,0.8291) -- (-0.5028,0.8502) -- (-0.4416,0.8685) -- (-0.3687,0.8842) -- (-0.2870,0.8967) -- (-0.1957,0.9060) -- (-0.0992,0.9117) -- (-0.0000,0.9135);
\end{tikzpicture} &
\begin{tikzpicture}[x=1.32cm,y=1.32cm]
  \input{figures/tikz/frame_min.tikz}
  \draw[orbitback] (0.5333,-0.8459) -- (0.5025,-0.8609) -- (0.4614,-0.8738) -- (0.4079,-0.8854) -- (0.3444,-0.8950) -- (0.2693,-0.9029) -- (0.1866,-0.9087) -- (0.0955,-0.9123) -- (0.0015,-0.9135) -- (-0.0925,-0.9124) -- (-0.1837,-0.9088) -- (-0.2693,-0.9029) -- (-0.3444,-0.8950) -- (-0.4079,-0.8854) -- (-0.4614,-0.8738) -- (-0.5025,-0.8609) -- (-0.5333,-0.8459);
  \draw[orbit] (-0.0000,0.9135) -- (-0.0669,0.9100) -- (-0.1328,0.8996) -- (-0.1966,0.8824) -- (-0.2600,0.8575) -- (-0.3166,0.8275) -- (-0.3707,0.7902) -- (-0.4169,0.7494) -- (-0.4589,0.7020) -- (-0.4925,0.6529) -- (-0.5206,0.5982) -- (-0.5414,0.5409) -- (-0.5547,0.4815) -- (-0.5607,0.4208) -- (-0.5594,0.3566) -- (-0.5507,0.2922) -- (-0.5342,0.2255) -- (-0.5169,0.1753) -- (-0.4945,0.1232) -- (-0.4682,0.0721) -- (-0.4367,0.0196) -- (-0.4016,-0.0315) -- (-0.3632,-0.0811) -- (-0.3199,-0.1314) -- (-0.2718,-0.1820) -- (-0.2145,-0.2367) -- (-0.1501,-0.2926) -- (-0.0813,-0.3472) -- (-0.0066,-0.4021) -- (0.0657,-0.4516) -- (0.1442,-0.5024) -- (0.4262,-0.6747) -- (0.4731,-0.7061) -- (0.5090,-0.7333) -- (0.5296,-0.7517) -- (0.5447,-0.7683) -- (0.5550,-0.7839) -- (0.5602,-0.7979) -- (0.5607,-0.8112) -- (0.5564,-0.8236) -- (0.5471,-0.8353) -- (0.5333,-0.8459);
  \draw[orbit] (-0.5333,-0.8459) -- (-0.5471,-0.8353) -- (-0.5564,-0.8236) -- (-0.5607,-0.8112) -- (-0.5602,-0.7979) -- (-0.5550,-0.7839) -- (-0.5447,-0.7683) -- (-0.5296,-0.7517) -- (-0.5090,-0.7333) -- (-0.4731,-0.7061) -- (-0.4262,-0.6747) -- (-0.1442,-0.5024) -- (-0.0657,-0.4516) -- (0.0066,-0.4021) -- (0.0813,-0.3472) -- (0.1501,-0.2926) -- (0.2145,-0.2367) -- (0.2718,-0.1820) -- (0.3199,-0.1314) -- (0.3632,-0.0811) -- (0.4016,-0.0315) -- (0.4367,0.0196) -- (0.4682,0.0721) -- (0.4945,0.1232) -- (0.5169,0.1753) -- (0.5342,0.2255) -- (0.5507,0.2922) -- (0.5594,0.3566) -- (0.5607,0.4208) -- (0.5547,0.4815) -- (0.5414,0.5409) -- (0.5206,0.5982) -- (0.4925,0.6529) -- (0.4589,0.7020) -- (0.4169,0.7494) -- (0.3707,0.7902) -- (0.3166,0.8275) -- (0.2600,0.8575) -- (0.1966,0.8824) -- (0.1328,0.8996) -- (0.0669,0.9100) -- (-0.0000,0.9135);
\end{tikzpicture} \\[1pt]
\shortstack{{\footnotesize $\rho=1/2$}\\[1pt]{\footnotesize $\alpha=0.470184$}} &
\shortstack{{\footnotesize $\rho=2/3$}\\[1pt]{\footnotesize $\alpha=0.598529$}} &
\shortstack{{\footnotesize $\rho=3/4$}\\[1pt]{\footnotesize $\alpha=0.655113$}} &
\shortstack{{\footnotesize $\rho=1/1$}\\[1pt]{\footnotesize $\alpha=0.792726$}} \\[6pt]
\begin{tikzpicture}[x=1.32cm,y=1.32cm]
  \input{figures/tikz/frame_min.tikz}
  \draw[orbitback] (0.4983,0.8670) -- (0.5076,0.8603) -- (0.5130,0.8533) -- (0.5146,0.8457) -- (0.5123,0.8372) -- (0.5059,0.8281) -- (0.4956,0.8184) -- (0.4627,0.7967) -- (0.4241,0.7764) -- (0.3706,0.7516) -- (0.0410,0.6090) -- (-0.0447,0.5688) -- (-0.1254,0.5284) -- (-0.2210,0.4767) -- (-0.3089,0.4242) -- (-0.3884,0.3718) -- (-0.4617,0.3180) -- (-0.5283,0.2631) -- (-0.5879,0.2076) -- (-0.6425,0.1494) -- (-0.6895,0.0910) -- (-0.7291,0.0329) -- (-0.7629,-0.0272) -- (-0.7903,-0.0891) -- (-0.8103,-0.1497) -- (-0.8196,-0.1890) -- (-0.8261,-0.2286) -- (-0.8298,-0.2686) -- (-0.8306,-0.3088) -- (-0.8285,-0.3492) -- (-0.8233,-0.3896) -- (-0.8152,-0.4298) -- (-0.8048,-0.4670) -- (-0.7907,-0.5067) -- (-0.7748,-0.5430) -- (-0.7548,-0.5814) -- (-0.7336,-0.6164) -- (-0.7079,-0.6529) -- (-0.6815,-0.6858) -- (-0.6528,-0.7174) -- (-0.6219,-0.7476) -- (-0.5863,-0.7783) -- (-0.5485,-0.8070) -- (-0.5088,-0.8334) -- (-0.4704,-0.8558) -- (-0.4278,-0.8773) -- (-0.3873,-0.8947) -- (-0.3430,-0.9106) -- (-0.3015,-0.9225) -- (-0.2569,-0.9321) -- (-0.2158,-0.9381) -- (-0.1723,-0.9412) -- (-0.1330,-0.9410) -- (-0.0922,-0.9374) -- (-0.0561,-0.9310) -- (-0.0195,-0.9208) -- (0.0122,-0.9083) -- (0.0455,-0.8905) -- (0.0769,-0.8677) -- (0.1036,-0.8412) -- (0.1254,-0.8114) -- (0.1412,-0.7807) -- (0.1529,-0.7454) -- (0.1593,-0.7080) -- (0.1605,-0.6687) -- (0.1564,-0.6281) -- (0.1473,-0.5865) -- (0.1334,-0.5443) -- (0.1136,-0.4992) -- (0.0906,-0.4570) -- (0.0619,-0.4126) -- (0.0292,-0.3691) -- (-0.0092,-0.3242) -- (-0.0462,-0.2857) -- (-0.0881,-0.2463) -- (-0.1321,-0.2087) -- (-0.1806,-0.1709) -- (-0.2305,-0.1354) -- (-0.2841,-0.1004) -- (-0.3412,-0.0664) -- (-0.3982,-0.0352) -- (-0.5045,0.0162) -- (-0.6158,0.0619) -- (-0.6846,0.0864) -- (-0.7543,0.1088) -- (-0.9530,0.1643) -- (-0.9768,0.1747) -- (-0.9817,0.1795) -- (-0.9828,0.1847);
  \draw[orbitback] (-0.1745,0.9847) -- (-0.2291,0.9721) -- (-0.2899,0.9522) -- (-0.3510,0.9270) -- (-0.4135,0.8963) -- (-0.4764,0.8606) -- (-0.5387,0.8203) -- (-0.5992,0.7761) -- (-0.6545,0.7306) -- (-0.7069,0.6825) -- (-0.7535,0.6345) -- (-0.7986,0.5823) -- (-0.8374,0.5310) -- (-0.8736,0.4761) -- (-0.9035,0.4228) -- (-0.9287,0.3689) -- (-0.9498,0.3129);
  \draw[orbitback] (0.1601,-0.9871) -- (0.1592,-0.9849) -- (0.1502,-0.9794) -- (0.1117,-0.9609) -- (-0.0258,-0.9031) -- (-0.1893,-0.8392) -- (-0.5415,-0.7053) -- (-0.6761,-0.6510) -- (-0.7344,-0.6248) -- (-0.7781,-0.6021) -- (-0.8078,-0.5824) -- (-0.8251,-0.5650);
  \draw[orbitback] (0.8251,-0.5650) -- (0.8078,-0.5824) -- (0.7781,-0.6021) -- (0.7344,-0.6248) -- (0.6761,-0.6510) -- (0.5415,-0.7053) -- (0.1893,-0.8392) -- (0.0258,-0.9031) -- (-0.1117,-0.9609) -- (-0.1502,-0.9794) -- (-0.1592,-0.9849) -- (-0.1601,-0.9871);
  \draw[orbitback] (0.9498,0.3129) -- (0.9287,0.3689) -- (0.9035,0.4228) -- (0.8736,0.4761) -- (0.8374,0.5310) -- (0.7986,0.5823) -- (0.7535,0.6345) -- (0.7069,0.6825) -- (0.6545,0.7306) -- (0.5992,0.7761) -- (0.5387,0.8203) -- (0.4764,0.8606) -- (0.4135,0.8963) -- (0.3510,0.9270) -- (0.2899,0.9522) -- (0.2291,0.9721) -- (0.1745,0.9847);
  \draw[orbitback] (0.9828,0.1847) -- (0.9817,0.1795) -- (0.9768,0.1747) -- (0.9530,0.1643) -- (0.7543,0.1088) -- (0.6846,0.0864) -- (0.6158,0.0619) -- (0.5045,0.0162) -- (0.3982,-0.0352) -- (0.3412,-0.0664) -- (0.2841,-0.1004) -- (0.2305,-0.1354) -- (0.1806,-0.1709) -- (0.1321,-0.2087) -- (0.0881,-0.2463) -- (0.0462,-0.2857) -- (0.0092,-0.3242) -- (-0.0292,-0.3691) -- (-0.0619,-0.4126) -- (-0.0906,-0.4570) -- (-0.1136,-0.4992) -- (-0.1334,-0.5443) -- (-0.1473,-0.5865) -- (-0.1564,-0.6281) -- (-0.1605,-0.6687) -- (-0.1593,-0.7080) -- (-0.1529,-0.7454) -- (-0.1412,-0.7807) -- (-0.1254,-0.8114) -- (-0.1036,-0.8412) -- (-0.0769,-0.8677) -- (-0.0455,-0.8905) -- (-0.0122,-0.9083) -- (0.0195,-0.9208) -- (0.0561,-0.9310) -- (0.0922,-0.9374) -- (0.1330,-0.9410) -- (0.1723,-0.9412) -- (0.2158,-0.9381) -- (0.2569,-0.9321) -- (0.3015,-0.9225) -- (0.3430,-0.9106) -- (0.3873,-0.8947) -- (0.4278,-0.8773) -- (0.4704,-0.8558) -- (0.5088,-0.8334) -- (0.5485,-0.8070) -- (0.5863,-0.7783) -- (0.6219,-0.7476) -- (0.6528,-0.7174) -- (0.6815,-0.6858) -- (0.7079,-0.6529) -- (0.7336,-0.6164) -- (0.7548,-0.5814) -- (0.7748,-0.5430) -- (0.7907,-0.5067) -- (0.8048,-0.4670) -- (0.8152,-0.4298) -- (0.8233,-0.3896) -- (0.8285,-0.3492) -- (0.8306,-0.3088) -- (0.8298,-0.2686) -- (0.8261,-0.2286) -- (0.8196,-0.1890) -- (0.8103,-0.1497) -- (0.7903,-0.0891) -- (0.7629,-0.0272) -- (0.7291,0.0329) -- (0.6895,0.0910) -- (0.6425,0.1494) -- (0.5879,0.2076) -- (0.5283,0.2631) -- (0.4617,0.3180) -- (0.3884,0.3718) -- (0.3089,0.4242) -- (0.2210,0.4767) -- (0.1254,0.5284) -- (0.0447,0.5688) -- (-0.0410,0.6090) -- (-0.3706,0.7516) -- (-0.4241,0.7764) -- (-0.4627,0.7967) -- (-0.4956,0.8184) -- (-0.5059,0.8281) -- (-0.5123,0.8372) -- (-0.5146,0.8457) -- (-0.5130,0.8533) -- (-0.5076,0.8603) -- (-0.4983,0.8670);
  \draw[orbit] (0.0000,0.9135) -- (0.1801,0.9103) -- (0.3308,0.9009) -- (0.3898,0.8943) -- (0.4383,0.8862) -- (0.4739,0.8772) -- (0.4983,0.8670);
  \draw[orbit] (-0.9828,0.1847) -- (-0.9803,0.1896) -- (-0.9746,0.1949) -- (-0.9532,0.2077) -- (-0.7723,0.2881) -- (-0.6736,0.3349) -- (-0.5650,0.3914) -- (-0.4636,0.4497) -- (-0.3693,0.5096) -- (-0.2804,0.5725) -- (-0.2027,0.6343) -- (-0.1337,0.6972) -- (-0.1015,0.7305) -- (-0.0732,0.7632) -- (-0.0490,0.7950) -- (-0.0294,0.8254) -- (-0.0146,0.8542) -- (-0.0049,0.8809) -- (-0.0003,0.9054) -- (-0.0009,0.9259) -- (-0.0060,0.9439) -- (-0.0166,0.9602) -- (-0.0321,0.9731) -- (-0.0525,0.9825) -- (-0.0773,0.9885) -- (-0.1063,0.9908) -- (-0.1390,0.9895) -- (-0.1745,0.9847);
  \draw[orbit] (-0.9498,0.3129) -- (-0.9629,0.2682) -- (-0.9725,0.2244) -- (-0.9790,0.1807) -- (-0.9824,0.1371) -- (-0.9827,0.0936) -- (-0.9799,0.0502) -- (-0.9740,0.0072) -- (-0.9645,-0.0383) -- (-0.9525,-0.0807) -- (-0.9367,-0.1253) -- (-0.9188,-0.1668) -- (-0.8968,-0.2104) -- (-0.8719,-0.2535) -- (-0.8440,-0.2959) -- (-0.8135,-0.3376) -- (-0.7783,-0.3810) -- (-0.7403,-0.4236) -- (-0.6975,-0.4675) -- (-0.6521,-0.5104) -- (-0.6043,-0.5521) -- (-0.5519,-0.5946) -- (-0.4977,-0.6357) -- (-0.3797,-0.7168) -- (-0.3140,-0.7579) -- (-0.2431,-0.7996) -- (-0.0970,-0.8776) -- (-0.0171,-0.9163) -- (0.0592,-0.9503) -- (0.1264,-0.9772) -- (0.1601,-0.9871);
  \draw[orbit] (-0.8251,-0.5650) -- (-0.8293,-0.5570) -- (-0.8307,-0.5496) -- (-0.8295,-0.5424) -- (-0.8255,-0.5350) -- (-0.8091,-0.5205) -- (-0.7810,-0.5057) -- (-0.7083,-0.4791) -- (-0.4310,-0.3924) -- (-0.2784,-0.3369) -- (-0.1927,-0.3012) -- (-0.1089,-0.2625) -- (-0.0313,-0.2225) -- (0.0424,-0.1803) -- (0.1086,-0.1380) -- (0.1697,-0.0945) -- (0.2280,-0.0480) -- (0.2805,-0.0008) -- (0.3292,0.0490) -- (0.3716,0.0987) -- (0.4095,0.1505) -- (0.4408,0.2015) -- (0.4721,0.2650) -- (0.4944,0.3273) -- (0.5087,0.3905) -- (0.5144,0.4510) -- (0.5120,0.5111) -- (0.5011,0.5701) -- (0.4920,0.6002) -- (0.4816,0.6270) -- (0.4680,0.6558) -- (0.4537,0.6812) -- (0.4357,0.7082) -- (0.4174,0.7317) -- (0.3952,0.7563) -- (0.3732,0.7776) -- (0.3471,0.7995) -- (0.3217,0.8180) -- (0.2668,0.8506) -- (0.2037,0.8781) -- (0.1398,0.8973) -- (0.0698,0.9096) -- (0.0017,0.9135) -- (-0.0664,0.9099) -- (-0.1365,0.8980) -- (-0.2006,0.8792) -- (-0.2639,0.8521) -- (-0.3191,0.8197) -- (-0.3446,0.8014) -- (-0.3709,0.7796) -- (-0.3931,0.7585) -- (-0.4155,0.7340) -- (-0.4339,0.7106) -- (-0.4521,0.6837) -- (-0.4667,0.6584) -- (-0.4805,0.6297) -- (-0.4910,0.6029) -- (-0.5004,0.5728) -- (-0.5117,0.5139) -- (-0.5145,0.4539) -- (-0.5087,0.3905) -- (-0.4944,0.3273) -- (-0.4721,0.2650) -- (-0.4408,0.2015) -- (-0.4095,0.1505) -- (-0.3716,0.0987) -- (-0.3292,0.0490) -- (-0.2805,-0.0008) -- (-0.2280,-0.0480) -- (-0.1697,-0.0945) -- (-0.1086,-0.1380) -- (-0.0424,-0.1803) -- (0.0313,-0.2225) -- (0.1089,-0.2625) -- (0.1927,-0.3012) -- (0.2784,-0.3369) -- (0.4310,-0.3924) -- (0.7083,-0.4791) -- (0.7810,-0.5057) -- (0.8091,-0.5205) -- (0.8255,-0.5350) -- (0.8295,-0.5424) -- (0.8307,-0.5496) -- (0.8293,-0.5570) -- (0.8251,-0.5650);
  \draw[orbit] (-0.1601,-0.9871) -- (-0.1264,-0.9772) -- (-0.0592,-0.9503) -- (0.0171,-0.9163) -- (0.0970,-0.8776) -- (0.2431,-0.7996) -- (0.3140,-0.7579) -- (0.3797,-0.7168) -- (0.4977,-0.6357) -- (0.5519,-0.5946) -- (0.6043,-0.5521) -- (0.6521,-0.5104) -- (0.6975,-0.4675) -- (0.7403,-0.4236) -- (0.7783,-0.3810) -- (0.8135,-0.3376) -- (0.8440,-0.2959) -- (0.8719,-0.2535) -- (0.8968,-0.2104) -- (0.9188,-0.1668) -- (0.9367,-0.1253) -- (0.9525,-0.0807) -- (0.9645,-0.0383) -- (0.9740,0.0072) -- (0.9799,0.0502) -- (0.9827,0.0936) -- (0.9824,0.1371) -- (0.9790,0.1807) -- (0.9725,0.2244) -- (0.9629,0.2682) -- (0.9498,0.3129);
  \draw[orbit] (0.1745,0.9847) -- (0.1390,0.9895) -- (0.1063,0.9908) -- (0.0773,0.9885) -- (0.0525,0.9825) -- (0.0321,0.9731) -- (0.0166,0.9602) -- (0.0060,0.9439) -- (0.0009,0.9259) -- (0.0003,0.9054) -- (0.0049,0.8809) -- (0.0146,0.8542) -- (0.0294,0.8254) -- (0.0490,0.7950) -- (0.0732,0.7632) -- (0.1015,0.7305) -- (0.1337,0.6972) -- (0.2027,0.6343) -- (0.2804,0.5725) -- (0.3693,0.5096) -- (0.4636,0.4497) -- (0.5650,0.3914) -- (0.6736,0.3349) -- (0.7723,0.2881) -- (0.9532,0.2077) -- (0.9746,0.1949) -- (0.9803,0.1896) -- (0.9828,0.1847);
  \draw[orbit] (-0.4983,0.8670) -- (-0.4739,0.8772) -- (-0.4383,0.8862) -- (-0.3898,0.8943) -- (-0.3308,0.9009) -- (-0.1801,0.9103) -- (-0.0000,0.9135);
\end{tikzpicture} &
\begin{tikzpicture}[x=1.32cm,y=1.32cm]
  \input{figures/tikz/frame_min.tikz}
  \draw[orbitback] (0.4904,0.8715) -- (0.4986,0.8657) -- (0.5033,0.8595) -- (0.5044,0.8528) -- (0.5017,0.8451) -- (0.4855,0.8289) -- (0.4532,0.8094) -- (0.4140,0.7907) -- (0.3615,0.7684) -- (0.0293,0.6363) -- (-0.0646,0.5957) -- (-0.1503,0.5560) -- (-0.2484,0.5067) -- (-0.3386,0.4568) -- (-0.4227,0.4052) -- (-0.5002,0.3521) -- (-0.5705,0.2979) -- (-0.6333,0.2430) -- (-0.6906,0.1853) -- (-0.7400,0.1275) -- (-0.7832,0.0673) -- (-0.8183,0.0076) -- (-0.8467,-0.0541) -- (-0.8673,-0.1147) -- (-0.8775,-0.1567) -- (-0.8841,-0.1965) -- (-0.8877,-0.2366) -- (-0.8884,-0.2771) -- (-0.8860,-0.3178) -- (-0.8806,-0.3587) -- (-0.8721,-0.3995) -- (-0.8605,-0.4403) -- (-0.8457,-0.4808) -- (-0.8279,-0.5209) -- (-0.8069,-0.5604) -- (-0.7848,-0.5965) -- (-0.7582,-0.6345) -- (-0.7309,-0.6689) -- (-0.6990,-0.7046) -- (-0.6670,-0.7365) -- (-0.6277,-0.7715) -- (-0.5888,-0.8022) -- (-0.5481,-0.8307) -- (-0.5058,-0.8569) -- (-0.4592,-0.8821) -- (-0.4148,-0.9027) -- (-0.3700,-0.9203) -- (-0.3252,-0.9348) -- (-0.2777,-0.9466) -- (-0.2342,-0.9541) -- (-0.1920,-0.9580) -- (-0.1517,-0.9583) -- (-0.1109,-0.9547) -- (-0.0756,-0.9476) -- (-0.0434,-0.9369) -- (-0.0145,-0.9229) -- (0.0106,-0.9055) -- (0.0317,-0.8851) -- (0.0486,-0.8618) -- (0.0618,-0.8340) -- (0.0696,-0.8055) -- (0.0728,-0.7751) -- (0.0715,-0.7429) -- (0.0653,-0.7069) -- (0.0550,-0.6721) -- (0.0395,-0.6340) -- (0.0210,-0.5980) -- (-0.0029,-0.5593) -- (-0.0306,-0.5207) -- (-0.0617,-0.4826) -- (-0.0958,-0.4452) -- (-0.1352,-0.4064) -- (-0.1771,-0.3690) -- (-0.2210,-0.3332) -- (-0.2666,-0.2991) -- (-0.3161,-0.2650) -- (-0.3664,-0.2331) -- (-0.4199,-0.2019) -- (-0.4761,-0.1718) -- (-0.5313,-0.1447) -- (-0.5879,-0.1193) -- (-0.6477,-0.0949) -- (-0.7638,-0.0546) -- (-0.8741,-0.0247) -- (-1.0000,-0.0000);
  \draw[orbitback] (0.4904,-0.8715) -- (0.4672,-0.8808) -- (0.4325,-0.8890) -- (0.3870,-0.8960) -- (0.3288,-0.9021) -- (0.1787,-0.9106) -- (-0.0018,-0.9135) -- (-0.1787,-0.9106) -- (-0.3288,-0.9021) -- (-0.3870,-0.8960) -- (-0.4325,-0.8890) -- (-0.4672,-0.8808) -- (-0.4904,-0.8715);
  \draw[orbitback] (1.0000,-0.0000) -- (0.8741,-0.0247) -- (0.7638,-0.0546) -- (0.6477,-0.0949) -- (0.5879,-0.1193) -- (0.5313,-0.1447) -- (0.4761,-0.1718) -- (0.4199,-0.2019) -- (0.3664,-0.2331) -- (0.3161,-0.2650) -- (0.2666,-0.2991) -- (0.2210,-0.3332) -- (0.1771,-0.3690) -- (0.1352,-0.4064) -- (0.0958,-0.4452) -- (0.0617,-0.4826) -- (0.0306,-0.5207) -- (0.0029,-0.5593) -- (-0.0210,-0.5980) -- (-0.0395,-0.6340) -- (-0.0550,-0.6721) -- (-0.0653,-0.7069) -- (-0.0715,-0.7429) -- (-0.0728,-0.7751) -- (-0.0696,-0.8055) -- (-0.0618,-0.8340) -- (-0.0486,-0.8618) -- (-0.0317,-0.8851) -- (-0.0106,-0.9055) -- (0.0145,-0.9229) -- (0.0434,-0.9369) -- (0.0756,-0.9476) -- (0.1109,-0.9547) -- (0.1517,-0.9583) -- (0.1920,-0.9580) -- (0.2342,-0.9541) -- (0.2777,-0.9466) -- (0.3252,-0.9348) -- (0.3700,-0.9203) -- (0.4148,-0.9027) -- (0.4592,-0.8821) -- (0.5058,-0.8569) -- (0.5481,-0.8307) -- (0.5888,-0.8022) -- (0.6277,-0.7715) -- (0.6670,-0.7365) -- (0.6990,-0.7046) -- (0.7309,-0.6689) -- (0.7582,-0.6345) -- (0.7848,-0.5965) -- (0.8069,-0.5604) -- (0.8279,-0.5209) -- (0.8457,-0.4808) -- (0.8605,-0.4403) -- (0.8721,-0.3995) -- (0.8806,-0.3587) -- (0.8860,-0.3178) -- (0.8884,-0.2771) -- (0.8877,-0.2366) -- (0.8841,-0.1965) -- (0.8775,-0.1567) -- (0.8673,-0.1147) -- (0.8467,-0.0541) -- (0.8183,0.0076) -- (0.7832,0.0673) -- (0.7400,0.1275) -- (0.6906,0.1853) -- (0.6333,0.2430) -- (0.5705,0.2979) -- (0.5002,0.3521) -- (0.4227,0.4052) -- (0.3386,0.4568) -- (0.2484,0.5067) -- (0.1503,0.5560) -- (0.0646,0.5957) -- (-0.0293,0.6363) -- (-0.3615,0.7684) -- (-0.4140,0.7907) -- (-0.4532,0.8094) -- (-0.4855,0.8289) -- (-0.5017,0.8451) -- (-0.5044,0.8528) -- (-0.5033,0.8595) -- (-0.4986,0.8657) -- (-0.4904,0.8715);
  \draw[orbit] (0.0000,0.9135) -- (0.1804,0.9105) -- (0.3301,0.9020) -- (0.3881,0.8959) -- (0.4334,0.8889) -- (0.4679,0.8806) -- (0.4904,0.8715);
  \draw[orbit] (-1.0000,-0.0000) -- (-0.8750,0.0245) -- (-0.7649,0.0542) -- (-0.6490,0.0944) -- (-0.5893,0.1187) -- (-0.5328,0.1440) -- (-0.4775,0.1711) -- (-0.4214,0.2011) -- (-0.3679,0.2322) -- (-0.3176,0.2640) -- (-0.2680,0.2981) -- (-0.2225,0.3321) -- (-0.1785,0.3679) -- (-0.1365,0.4053) -- (-0.0970,0.4440) -- (-0.0628,0.4814) -- (-0.0315,0.5194) -- (-0.0037,0.5580) -- (0.0202,0.5968) -- (0.0401,0.6353) -- (0.0555,0.6734) -- (0.0656,0.7081) -- (0.0717,0.7441) -- (0.0728,0.7762) -- (0.0694,0.8066) -- (0.0615,0.8349) -- (0.0491,0.8609) -- (0.0324,0.8843) -- (0.0114,0.9048) -- (-0.0136,0.9223) -- (-0.0423,0.9365) -- (-0.0744,0.9472) -- (-0.1096,0.9545) -- (-0.1503,0.9583) -- (-0.1906,0.9581) -- (-0.2327,0.9543) -- (-0.2761,0.9469) -- (-0.3236,0.9352) -- (-0.3684,0.9209) -- (-0.4133,0.9033) -- (-0.4576,0.8828) -- (-0.5043,0.8578) -- (-0.5466,0.8317) -- (-0.5874,0.8032) -- (-0.6264,0.7726) -- (-0.6658,0.7377) -- (-0.6978,0.7059) -- (-0.7298,0.6702) -- (-0.7572,0.6359) -- (-0.7839,0.5979) -- (-0.8061,0.5618) -- (-0.8272,0.5223) -- (-0.8451,0.4822) -- (-0.8600,0.4417) -- (-0.8717,0.4010) -- (-0.8804,0.3601) -- (-0.8859,0.3193) -- (-0.8884,0.2786) -- (-0.8878,0.2381) -- (-0.8842,0.1979) -- (-0.8778,0.1581) -- (-0.8677,0.1160) -- (-0.8472,0.0555) -- (-0.8189,-0.0062) -- (-0.7840,-0.0660) -- (-0.7410,-0.1263) -- (-0.6917,-0.1842) -- (-0.6345,-0.2419) -- (-0.5718,-0.2968) -- (-0.5016,-0.3511) -- (-0.4242,-0.4042) -- (-0.3401,-0.4559) -- (-0.2500,-0.5058) -- (-0.1519,-0.5552) -- (-0.0662,-0.5950) -- (0.0278,-0.6357) -- (0.3626,-0.7688) -- (0.4149,-0.7911) -- (0.4538,-0.8098) -- (0.4851,-0.8286) -- (0.5015,-0.8448) -- (0.5044,-0.8525) -- (0.5034,-0.8592) -- (0.4988,-0.8655) -- (0.4904,-0.8715);
  \draw[orbit] (-0.4904,-0.8715) -- (-0.4988,-0.8655) -- (-0.5034,-0.8592) -- (-0.5044,-0.8525) -- (-0.5015,-0.8448) -- (-0.4851,-0.8286) -- (-0.4538,-0.8098) -- (-0.4149,-0.7911) -- (-0.3626,-0.7688) -- (-0.0278,-0.6357) -- (0.0662,-0.5950) -- (0.1519,-0.5552) -- (0.2500,-0.5058) -- (0.3401,-0.4559) -- (0.4242,-0.4042) -- (0.5016,-0.3511) -- (0.5718,-0.2968) -- (0.6345,-0.2419) -- (0.6917,-0.1842) -- (0.7410,-0.1263) -- (0.7840,-0.0660) -- (0.8189,-0.0062) -- (0.8472,0.0555) -- (0.8677,0.1160) -- (0.8778,0.1581) -- (0.8842,0.1979) -- (0.8878,0.2381) -- (0.8884,0.2786) -- (0.8859,0.3193) -- (0.8804,0.3601) -- (0.8717,0.4010) -- (0.8600,0.4417) -- (0.8451,0.4822) -- (0.8272,0.5223) -- (0.8061,0.5618) -- (0.7839,0.5979) -- (0.7572,0.6359) -- (0.7298,0.6702) -- (0.6978,0.7059) -- (0.6658,0.7377) -- (0.6264,0.7726) -- (0.5874,0.8032) -- (0.5466,0.8317) -- (0.5043,0.8578) -- (0.4576,0.8828) -- (0.4133,0.9033) -- (0.3684,0.9209) -- (0.3236,0.9352) -- (0.2761,0.9469) -- (0.2327,0.9543) -- (0.1906,0.9581) -- (0.1503,0.9583) -- (0.1096,0.9545) -- (0.0744,0.9472) -- (0.0423,0.9365) -- (0.0136,0.9223) -- (-0.0114,0.9048) -- (-0.0324,0.8843) -- (-0.0491,0.8609) -- (-0.0615,0.8349) -- (-0.0694,0.8066) -- (-0.0728,0.7762) -- (-0.0717,0.7441) -- (-0.0656,0.7081) -- (-0.0555,0.6734) -- (-0.0401,0.6353) -- (-0.0202,0.5968) -- (0.0037,0.5580) -- (0.0315,0.5194) -- (0.0628,0.4814) -- (0.0970,0.4440) -- (0.1365,0.4053) -- (0.1785,0.3679) -- (0.2225,0.3321) -- (0.2680,0.2981) -- (0.3176,0.2640) -- (0.3679,0.2322) -- (0.4214,0.2011) -- (0.4775,0.1711) -- (0.5328,0.1440) -- (0.5893,0.1187) -- (0.6490,0.0944) -- (0.7649,0.0542) -- (0.8750,0.0245) -- (1.0000,-0.0000);
  \draw[orbit] (-0.4904,0.8715) -- (-0.4679,0.8806) -- (-0.4334,0.8889) -- (-0.3881,0.8959) -- (-0.3301,0.9020) -- (-0.1804,0.9105) -- (-0.0000,0.9135);
\end{tikzpicture} &
\begin{tikzpicture}[x=1.32cm,y=1.32cm]
  \input{figures/tikz/frame_min.tikz}
  \draw[orbitback] (0.4786,0.8780) -- (0.4859,0.8728) -- (0.4893,0.8673) -- (0.4890,0.8613) -- (0.4850,0.8550) -- (0.4658,0.8405) -- (0.4310,0.8234) -- (0.3564,0.7945) -- (0.0856,0.7007) -- (-0.0722,0.6420) -- (-0.2275,0.5774) -- (-0.3631,0.5129) -- (-0.4763,0.4510) -- (-0.5762,0.3881) -- (-0.6651,0.3227) -- (-0.7049,0.2896) -- (-0.7422,0.2555) -- (-0.7816,0.2157) -- (-0.8175,0.1747) -- (-0.8497,0.1326) -- (-0.8766,0.0920) -- (-0.8999,0.0504) -- (-0.9195,0.0080) -- (-0.9354,-0.0352) -- (-0.9473,-0.0792) -- (-0.9553,-0.1239) -- (-0.9592,-0.1691) -- (-0.9589,-0.2149) -- (-0.9548,-0.2582) -- (-0.9464,-0.3047) -- (-0.9337,-0.3513) -- (-0.9168,-0.3980) -- (-0.8972,-0.4417);
  \draw[orbitback] (-0.1902,-0.9817) -- (-0.1502,-0.9878) -- (-0.1158,-0.9899) -- (-0.0852,-0.9883) -- (-0.0587,-0.9831) -- (-0.0354,-0.9736) -- (-0.0187,-0.9612) -- (-0.0072,-0.9455) -- (-0.0011,-0.9267) -- (-0.0004,-0.9051) -- (-0.0052,-0.8810) -- (-0.0153,-0.8546) -- (-0.0307,-0.8263) -- (-0.0511,-0.7965) -- (-0.0782,-0.7632) -- (-0.1078,-0.7314) -- (-0.1438,-0.6969) -- (-0.2207,-0.6327) -- (-0.3132,-0.5665) -- (-0.4145,-0.5034) -- (-0.5259,-0.4427) -- (-0.6409,-0.3882) -- (-0.7583,-0.3405) -- (-0.8697,-0.3033) -- (-0.9220,-0.2892) -- (-0.9596,-0.2815);
  \draw[orbitback] (0.9596,-0.2815) -- (0.9220,-0.2892) -- (0.8697,-0.3033) -- (0.7583,-0.3405) -- (0.6409,-0.3882) -- (0.5259,-0.4427) -- (0.4145,-0.5034) -- (0.3132,-0.5665) -- (0.2207,-0.6327) -- (0.1438,-0.6969) -- (0.1078,-0.7314) -- (0.0782,-0.7632) -- (0.0511,-0.7965) -- (0.0307,-0.8263) -- (0.0153,-0.8546) -- (0.0052,-0.8810) -- (0.0004,-0.9051) -- (0.0011,-0.9267) -- (0.0072,-0.9455) -- (0.0187,-0.9612) -- (0.0354,-0.9736) -- (0.0587,-0.9831) -- (0.0852,-0.9883) -- (0.1158,-0.9899) -- (0.1502,-0.9878) -- (0.1902,-0.9817);
  \draw[orbitback] (0.8972,-0.4417) -- (0.9168,-0.3980) -- (0.9337,-0.3513) -- (0.9464,-0.3047) -- (0.9548,-0.2582) -- (0.9589,-0.2149) -- (0.9592,-0.1691) -- (0.9553,-0.1239) -- (0.9473,-0.0792) -- (0.9354,-0.0352) -- (0.9195,0.0080) -- (0.8999,0.0504) -- (0.8766,0.0920) -- (0.8497,0.1326) -- (0.8175,0.1747) -- (0.7816,0.2157) -- (0.7422,0.2555) -- (0.7049,0.2896) -- (0.6651,0.3227) -- (0.5762,0.3881) -- (0.4763,0.4510) -- (0.3631,0.5129) -- (0.2275,0.5774) -- (0.0722,0.6420) -- (-0.0856,0.7007) -- (-0.3564,0.7945) -- (-0.4310,0.8234) -- (-0.4658,0.8405) -- (-0.4850,0.8550) -- (-0.4890,0.8613) -- (-0.4893,0.8673) -- (-0.4859,0.8728) -- (-0.4786,0.8780);
  \draw[orbit] (0.0000,0.9135) -- (0.1804,0.9110) -- (0.3255,0.9039) -- (0.3822,0.8987) -- (0.4259,0.8928) -- (0.4572,0.8861) -- (0.4786,0.8780);
  \draw[orbit] (-0.8972,-0.4417) -- (-0.8739,-0.4851) -- (-0.8453,-0.5310) -- (-0.8149,-0.5734) -- (-0.7789,-0.6178) -- (-0.7420,-0.6584) -- (-0.6995,-0.7005) -- (-0.6541,-0.7409) -- (-0.6063,-0.7793) -- (-0.5563,-0.8156) -- (-0.5049,-0.8493) -- (-0.4524,-0.8801) -- (-0.3963,-0.9093) -- (-0.3439,-0.9331) -- (-0.2894,-0.9542) -- (-0.2371,-0.9707) -- (-0.1902,-0.9817);
  \draw[orbit] (-0.9596,-0.2815) -- (-0.8865,-0.2890) -- (-0.7914,-0.2908) -- (-0.6932,-0.2867) -- (-0.5896,-0.2766) -- (-0.4876,-0.2612) -- (-0.3885,-0.2410) -- (-0.2923,-0.2162) -- (-0.1975,-0.1863) -- (-0.1092,-0.1532) -- (-0.0251,-0.1161) -- (0.0538,-0.0756) -- (0.1266,-0.0323) -- (0.1958,0.0155) -- (0.2575,0.0653) -- (0.3140,0.1188) -- (0.3621,0.1732) -- (0.4056,0.2334) -- (0.4410,0.2964) -- (0.4663,0.3583) -- (0.4756,0.3897) -- (0.4827,0.4214) -- (0.4873,0.4532) -- (0.4894,0.4850) -- (0.4891,0.5167) -- (0.4865,0.5450) -- (0.4812,0.5761) -- (0.4742,0.6037) -- (0.4639,0.6338) -- (0.4524,0.6602) -- (0.4372,0.6886) -- (0.4213,0.7134) -- (0.4012,0.7397) -- (0.3811,0.7622) -- (0.3565,0.7859) -- (0.3325,0.8058) -- (0.3038,0.8263) -- (0.2763,0.8430) -- (0.2441,0.8598) -- (0.2137,0.8731) -- (0.1786,0.8858) -- (0.1460,0.8952) -- (0.1089,0.9035) -- (0.0749,0.9088) -- (0.0366,0.9124) -- (0.0019,0.9135) -- (-0.0328,0.9126) -- (-0.0711,0.9093) -- (-0.1052,0.9041) -- (-0.1424,0.8961) -- (-0.1751,0.8869) -- (-0.2102,0.8745) -- (-0.2407,0.8614) -- (-0.2731,0.8448) -- (-0.3008,0.8282) -- (-0.3297,0.8079) -- (-0.3539,0.7882) -- (-0.3787,0.7647) -- (-0.3991,0.7423) -- (-0.4194,0.7160) -- (-0.4355,0.6914) -- (-0.4510,0.6631) -- (-0.4628,0.6367) -- (-0.4733,0.6068) -- (-0.4812,0.5761) -- (-0.4865,0.5450) -- (-0.4891,0.5167) -- (-0.4894,0.4850) -- (-0.4873,0.4532) -- (-0.4827,0.4214) -- (-0.4756,0.3897) -- (-0.4663,0.3583) -- (-0.4410,0.2964) -- (-0.4056,0.2334) -- (-0.3621,0.1732) -- (-0.3140,0.1188) -- (-0.2575,0.0653) -- (-0.1958,0.0155) -- (-0.1266,-0.0323) -- (-0.0538,-0.0756) -- (0.0251,-0.1161) -- (0.1092,-0.1532) -- (0.1975,-0.1863) -- (0.2923,-0.2162) -- (0.3885,-0.2410) -- (0.4876,-0.2612) -- (0.5896,-0.2766) -- (0.6932,-0.2867) -- (0.7914,-0.2908) -- (0.8865,-0.2890) -- (0.9596,-0.2815);
  \draw[orbit] (0.1902,-0.9817) -- (0.2371,-0.9707) -- (0.2894,-0.9542) -- (0.3439,-0.9331) -- (0.3963,-0.9093) -- (0.4524,-0.8801) -- (0.5049,-0.8493) -- (0.5563,-0.8156) -- (0.6063,-0.7793) -- (0.6541,-0.7409) -- (0.6995,-0.7005) -- (0.7420,-0.6584) -- (0.7789,-0.6178) -- (0.8149,-0.5734) -- (0.8453,-0.5310) -- (0.8739,-0.4851) -- (0.8972,-0.4417);
  \draw[orbit] (-0.4786,0.8780) -- (-0.4572,0.8861) -- (-0.4259,0.8928) -- (-0.3822,0.8987) -- (-0.3255,0.9039) -- (-0.1804,0.9110) -- (0.0000,0.9135);
\end{tikzpicture} &
\begin{tikzpicture}[x=1.32cm,y=1.32cm]
  \input{figures/tikz/frame_min.tikz}
  \draw[orbitback] (0.4709,0.8822) -- (0.4771,0.8778) -- (0.4800,0.8728) -- (0.4791,0.8675) -- (0.4746,0.8617) -- (0.4550,0.8490) -- (0.4207,0.8340) -- (0.3472,0.8081) -- (0.0925,0.7274) -- (-0.0536,0.6782) -- (-0.1970,0.6248) -- (-0.3281,0.5699) -- (-0.4329,0.5206) -- (-0.5293,0.4695) -- (-0.6164,0.4170) -- (-0.6966,0.3615) -- (-0.7740,0.2986) -- (-0.8089,0.2659) -- (-0.8410,0.2324) -- (-0.8702,0.1982) -- (-0.8964,0.1634) -- (-0.9196,0.1278) -- (-0.9382,0.0943) -- (-0.9551,0.0576) -- (-0.9679,0.0231) -- (-0.9783,-0.0146) -- (-0.9849,-0.0500) -- (-0.9886,-0.0884) -- (-0.9889,-0.1245) -- (-0.9862,-0.1607) -- (-0.9802,-0.1979);
  \draw[orbitback] (0.0436,-0.9990) -- (0.0624,-0.9949) -- (0.0673,-0.9904) -- (0.0688,-0.9843) -- (0.0670,-0.9768) -- (0.0617,-0.9678) -- (0.0426,-0.9470) -- (0.0117,-0.9220) -- (-0.0300,-0.8934) -- (-0.0814,-0.8621) -- (-0.1440,-0.8272) -- (-0.2649,-0.7662) -- (-0.4053,-0.7023) -- (-0.5686,-0.6343) -- (-0.8129,-0.5400) -- (-0.8438,-0.5260) -- (-0.8559,-0.5172);
  \draw[orbitback] (0.9890,0.1479) -- (0.9819,0.1601) -- (0.9656,0.1720) -- (0.9402,0.1836) -- (0.9076,0.1940) -- (0.8692,0.2028) -- (0.8264,0.2097) -- (0.7804,0.2147) -- (0.7296,0.2178) -- (0.6276,0.2178) -- (0.5241,0.2105) -- (0.4164,0.1958) -- (0.3110,0.1745) -- (0.2145,0.1488) -- (0.1212,0.1177) -- (0.0325,0.0818) -- (-0.0541,0.0398) -- (-0.1335,-0.0062) -- (-0.2051,-0.0553) -- (-0.2681,-0.1066) -- (-0.3251,-0.1622) -- (-0.3748,-0.2217) -- (-0.4143,-0.2813) -- (-0.4317,-0.3138) -- (-0.4453,-0.3435) -- (-0.4579,-0.3769) -- (-0.4669,-0.4074) -- (-0.4743,-0.4415) -- (-0.4784,-0.4722) -- (-0.4801,-0.5030) -- (-0.4792,-0.5336) -- (-0.4759,-0.5640) -- (-0.4700,-0.5940) -- (-0.4616,-0.6235) -- (-0.4506,-0.6524) -- (-0.4370,-0.6804) -- (-0.4209,-0.7076) -- (-0.4023,-0.7336) -- (-0.3814,-0.7584) -- (-0.3581,-0.7818) -- (-0.3326,-0.8037) -- (-0.3051,-0.8239) -- (-0.2790,-0.8404) -- (-0.2480,-0.8572) -- (-0.2153,-0.8719) -- (-0.1813,-0.8845) -- (-0.1461,-0.8950) -- (-0.1099,-0.9032) -- (-0.0729,-0.9090) -- (-0.0021,-0.9135) -- (0.0688,-0.9095) -- (0.1058,-0.9039) -- (0.1421,-0.8960) -- (0.1774,-0.8858) -- (0.2116,-0.8734) -- (0.2444,-0.8589) -- (0.2756,-0.8424) -- (0.3051,-0.8239) -- (0.3326,-0.8037) -- (0.3581,-0.7818) -- (0.3814,-0.7584) -- (0.4023,-0.7336) -- (0.4209,-0.7076) -- (0.4370,-0.6804) -- (0.4506,-0.6524) -- (0.4616,-0.6235) -- (0.4700,-0.5940) -- (0.4759,-0.5640) -- (0.4792,-0.5336) -- (0.4801,-0.5030) -- (0.4784,-0.4722) -- (0.4743,-0.4415) -- (0.4669,-0.4074) -- (0.4579,-0.3769) -- (0.4453,-0.3435) -- (0.4317,-0.3138) -- (0.4143,-0.2813) -- (0.3748,-0.2217) -- (0.3251,-0.1622) -- (0.2681,-0.1066) -- (0.2051,-0.0553) -- (0.1335,-0.0062) -- (0.0541,0.0398) -- (-0.0325,0.0818) -- (-0.1212,0.1177) -- (-0.2145,0.1488) -- (-0.3110,0.1745) -- (-0.4164,0.1958) -- (-0.5241,0.2105) -- (-0.6276,0.2178) -- (-0.7296,0.2178) -- (-0.7804,0.2147) -- (-0.8264,0.2097) -- (-0.8692,0.2028) -- (-0.9076,0.1940) -- (-0.9402,0.1836) -- (-0.9656,0.1720) -- (-0.9819,0.1601) -- (-0.9890,0.1479);
  \draw[orbitback] (0.8559,-0.5172) -- (0.8438,-0.5260) -- (0.8129,-0.5400) -- (0.5686,-0.6343) -- (0.4053,-0.7023) -- (0.2649,-0.7662) -- (0.1440,-0.8272) -- (0.0814,-0.8621) -- (0.0300,-0.8934) -- (-0.0117,-0.9220) -- (-0.0426,-0.9470) -- (-0.0617,-0.9678) -- (-0.0670,-0.9768) -- (-0.0688,-0.9843) -- (-0.0673,-0.9904) -- (-0.0624,-0.9949) -- (-0.0436,-0.9990);
  \draw[orbitback] (0.9802,-0.1979) -- (0.9862,-0.1607) -- (0.9889,-0.1245) -- (0.9886,-0.0884) -- (0.9849,-0.0500) -- (0.9783,-0.0146) -- (0.9679,0.0231) -- (0.9551,0.0576) -- (0.9382,0.0943) -- (0.9196,0.1278) -- (0.8964,0.1634) -- (0.8702,0.1982) -- (0.8410,0.2324) -- (0.8089,0.2659) -- (0.7740,0.2986) -- (0.6966,0.3615) -- (0.6164,0.4170) -- (0.5293,0.4695) -- (0.4329,0.5206) -- (0.3281,0.5699) -- (0.1970,0.6248) -- (0.0536,0.6782) -- (-0.0925,0.7274) -- (-0.3472,0.8081) -- (-0.4207,0.8340) -- (-0.4550,0.8490) -- (-0.4746,0.8617) -- (-0.4791,0.8675) -- (-0.4800,0.8728) -- (-0.4771,0.8778) -- (-0.4709,0.8822);
  \draw[orbit] (0.0000,0.9135) -- (0.1794,0.9113) -- (0.3252,0.9050) -- (0.4199,0.8954) -- (0.4513,0.8892) -- (0.4709,0.8822);
  \draw[orbit] (-0.9802,-0.1979) -- (-0.9646,-0.2564) -- (-0.9409,-0.3158) -- (-0.9095,-0.3751) -- (-0.8686,-0.4369) -- (-0.8223,-0.4951) -- (-0.7666,-0.5551) -- (-0.7043,-0.6137) -- (-0.6330,-0.6728) -- (-0.5497,-0.7339) -- (-0.4622,-0.7910) -- (-0.3690,-0.8452) -- (-0.2729,-0.8946) -- (-0.1748,-0.9384) -- (-0.0841,-0.9722) -- (-0.0100,-0.9928) -- (0.0207,-0.9980) -- (0.0436,-0.9990);
  \draw[orbit] (-0.8559,-0.5172) -- (-0.8561,-0.5138) -- (-0.8519,-0.5110) -- (-0.8297,-0.5071) -- (-0.6266,-0.4981) -- (-0.5099,-0.4903) -- (-0.3812,-0.4773) -- (-0.2572,-0.4599) -- (-0.1394,-0.4387) -- (-0.0272,-0.4136) -- (0.0815,-0.3842) -- (0.1850,-0.3508) -- (0.2859,-0.3121) -- (0.3793,-0.2700) -- (0.4678,-0.2228) -- (0.5470,-0.1727) -- (0.6166,-0.1205) -- (0.6791,-0.0641) -- (0.7085,-0.0331) -- (0.7335,-0.0037) -- (0.7564,0.0265) -- (0.7772,0.0575) -- (0.7989,0.0952) -- (0.8160,0.1309) -- (0.8313,0.1705) -- (0.8423,0.2078) -- (0.8507,0.2490) -- (0.8552,0.2877) -- (0.8565,0.3268) -- (0.8544,0.3663) -- (0.8489,0.4060) -- (0.8401,0.4458) -- (0.8278,0.4856) -- (0.8122,0.5251) -- (0.7931,0.5643) -- (0.7708,0.6029) -- (0.7452,0.6407) -- (0.7189,0.6744) -- (0.6847,0.7130) -- (0.6502,0.7471) -- (0.6099,0.7822) -- (0.5703,0.8124) -- (0.5252,0.8427) -- (0.4819,0.8680) -- (0.4373,0.8906) -- (0.3920,0.9103) -- (0.3424,0.9280) -- (0.2968,0.9408) -- (0.2519,0.9501) -- (0.2081,0.9558) -- (0.1626,0.9577) -- (0.1228,0.9555) -- (0.0858,0.9496) -- (0.0519,0.9399) -- (0.0192,0.9255) -- (-0.0068,0.9085) -- (-0.0302,0.8865) -- (-0.0482,0.8610) -- (-0.0607,0.8324) -- (-0.0674,0.8011) -- (-0.0685,0.7674) -- (-0.0639,0.7319) -- (-0.0538,0.6950) -- (-0.0385,0.6571) -- (-0.0181,0.6186) -- (0.0069,0.5800) -- (0.0361,0.5417) -- (0.0718,0.5011) -- (0.1113,0.4616) -- (0.1540,0.4236) -- (0.2027,0.3848) -- (0.2538,0.3483) -- (0.3099,0.3122) -- (0.3672,0.2791) -- (0.4282,0.2475) -- (0.4887,0.2195) -- (0.5514,0.1939) -- (0.6119,0.1724) -- (0.6786,0.1521) -- (0.7433,0.1364) -- (0.8065,0.1249) -- (0.8634,0.1188) -- (0.9152,0.1180) -- (0.9559,0.1230) -- (0.9711,0.1277) -- (0.9817,0.1335) -- (0.9877,0.1402) -- (0.9890,0.1479);
  \draw[orbit] (-0.9890,0.1479) -- (-0.9877,0.1402) -- (-0.9817,0.1335) -- (-0.9711,0.1277) -- (-0.9559,0.1230) -- (-0.9152,0.1180) -- (-0.8634,0.1188) -- (-0.8065,0.1249) -- (-0.7433,0.1364) -- (-0.6786,0.1521) -- (-0.6119,0.1724) -- (-0.5514,0.1939) -- (-0.4887,0.2195) -- (-0.4282,0.2475) -- (-0.3672,0.2791) -- (-0.3099,0.3122) -- (-0.2538,0.3483) -- (-0.2027,0.3848) -- (-0.1540,0.4236) -- (-0.1113,0.4616) -- (-0.0718,0.5011) -- (-0.0361,0.5417) -- (-0.0069,0.5800) -- (0.0181,0.6186) -- (0.0385,0.6571) -- (0.0538,0.6950) -- (0.0639,0.7319) -- (0.0685,0.7674) -- (0.0674,0.8011) -- (0.0607,0.8324) -- (0.0482,0.8610) -- (0.0302,0.8865) -- (0.0068,0.9085) -- (-0.0192,0.9255) -- (-0.0519,0.9399) -- (-0.0858,0.9496) -- (-0.1228,0.9555) -- (-0.1626,0.9577) -- (-0.2081,0.9558) -- (-0.2519,0.9501) -- (-0.2968,0.9408) -- (-0.3424,0.9280) -- (-0.3920,0.9103) -- (-0.4373,0.8906) -- (-0.4819,0.8680) -- (-0.5252,0.8427) -- (-0.5703,0.8124) -- (-0.6099,0.7822) -- (-0.6502,0.7471) -- (-0.6847,0.7130) -- (-0.7189,0.6744) -- (-0.7452,0.6407) -- (-0.7708,0.6029) -- (-0.7931,0.5643) -- (-0.8122,0.5251) -- (-0.8278,0.4856) -- (-0.8401,0.4458) -- (-0.8489,0.4060) -- (-0.8544,0.3663) -- (-0.8565,0.3268) -- (-0.8552,0.2877) -- (-0.8507,0.2490) -- (-0.8423,0.2078) -- (-0.8313,0.1705) -- (-0.8160,0.1309) -- (-0.7989,0.0952) -- (-0.7772,0.0575) -- (-0.7564,0.0265) -- (-0.7335,-0.0037) -- (-0.7085,-0.0331) -- (-0.6791,-0.0641) -- (-0.6166,-0.1205) -- (-0.5470,-0.1727) -- (-0.4678,-0.2228) -- (-0.3793,-0.2700) -- (-0.2859,-0.3121) -- (-0.1850,-0.3508) -- (-0.0815,-0.3842) -- (0.0272,-0.4136) -- (0.1394,-0.4387) -- (0.2572,-0.4599) -- (0.3812,-0.4773) -- (0.5099,-0.4903) -- (0.6266,-0.4981) -- (0.8297,-0.5071) -- (0.8519,-0.5110) -- (0.8561,-0.5138) -- (0.8559,-0.5172);
  \draw[orbit] (-0.0436,-0.9990) -- (-0.0207,-0.9980) -- (0.0100,-0.9928) -- (0.0841,-0.9722) -- (0.1748,-0.9384) -- (0.2729,-0.8946) -- (0.3690,-0.8452) -- (0.4622,-0.7910) -- (0.5497,-0.7339) -- (0.6330,-0.6728) -- (0.7043,-0.6137) -- (0.7666,-0.5551) -- (0.8223,-0.4951) -- (0.8686,-0.4369) -- (0.9095,-0.3751) -- (0.9409,-0.3158) -- (0.9646,-0.2564) -- (0.9802,-0.1979);
  \draw[orbit] (-0.4709,0.8822) -- (-0.4513,0.8892) -- (-0.4199,0.8954) -- (-0.3252,0.9050) -- (-0.1794,0.9113) -- (-0.0000,0.9135);
\end{tikzpicture} \\[1pt]
\shortstack{{\footnotesize $\rho=5/4$}\\[1pt]{\footnotesize $\alpha=0.884668$}} &
\shortstack{{\footnotesize $\rho=4/3$}\\[1pt]{\footnotesize $\alpha=0.906603$}} &
\shortstack{{\footnotesize $\rho=3/2$}\\[1pt]{\footnotesize $\alpha=0.939973$}} &
\shortstack{{\footnotesize $\rho=5/3$}\\[1pt]{\footnotesize $\alpha=0.962288$}} \\[6pt]
\begin{tikzpicture}[x=1.32cm,y=1.32cm]
  \input{figures/tikz/frame_min.tikz}
  \draw[orbitback] (0.4629,0.8864) -- (0.4681,0.8828) -- (0.4704,0.8785) -- (0.4691,0.8739) -- (0.4644,0.8690) -- (0.4452,0.8580) -- (0.4115,0.8448) -- (0.3401,0.8221) -- (0.1000,0.7529) -- (-0.0366,0.7112) -- (-0.1741,0.6653) -- (-0.2982,0.6194) -- (-0.3971,0.5787) -- (-0.4881,0.5372) -- (-0.5745,0.4933) -- (-0.6515,0.4494) -- (-0.7325,0.3964) -- (-0.8014,0.3435) -- (-0.8611,0.2887) -- (-0.9108,0.2323) -- (-0.9318,0.2036) -- (-0.9500,0.1745) -- (-0.9655,0.1453) -- (-0.9781,0.1158) -- (-0.9879,0.0861) -- (-0.9943,0.0591) -- (-0.9986,0.0292) -- (-1.0000,0.0000);
  \draw[orbitback] (0.4629,-0.8864) -- (0.4452,-0.8925) -- (0.4156,-0.8979) -- (0.3207,-0.9063) -- (0.1789,-0.9116) -- (-0.0025,-0.9135) -- (-0.1834,-0.9115) -- (-0.3242,-0.9061) -- (-0.4156,-0.8979) -- (-0.4452,-0.8925) -- (-0.4629,-0.8864);
  \draw[orbitback] (1.0000,0.0000) -- (0.9986,0.0292) -- (0.9943,0.0591) -- (0.9879,0.0861) -- (0.9781,0.1158) -- (0.9655,0.1453) -- (0.9500,0.1745) -- (0.9318,0.2036) -- (0.9108,0.2323) -- (0.8611,0.2887) -- (0.8014,0.3435) -- (0.7325,0.3964) -- (0.6515,0.4494) -- (0.5745,0.4933) -- (0.4881,0.5372) -- (0.3971,0.5787) -- (0.2982,0.6194) -- (0.1741,0.6653) -- (0.0366,0.7112) -- (-0.1000,0.7529) -- (-0.3401,0.8221) -- (-0.4115,0.8448) -- (-0.4452,0.8580) -- (-0.4644,0.8690) -- (-0.4691,0.8739) -- (-0.4704,0.8785) -- (-0.4681,0.8828) -- (-0.4629,0.8864);
  \draw[orbit] (0.0000,0.9135) -- (0.1812,0.9116) -- (0.3225,0.9062) -- (0.4145,0.8980) -- (0.4445,0.8926) -- (0.4629,0.8864);
  \draw[orbit] (-1.0000,0.0000) -- (-0.9987,-0.0279) -- (-0.9946,-0.0577) -- (-0.9875,-0.0875) -- (-0.9776,-0.1171) -- (-0.9661,-0.1439) -- (-0.9508,-0.1732) -- (-0.9327,-0.2022) -- (-0.9119,-0.2310) -- (-0.8624,-0.2874) -- (-0.8029,-0.3423) -- (-0.7308,-0.3976) -- (-0.6496,-0.4505) -- (-0.5725,-0.4944) -- (-0.4902,-0.5362) -- (-0.3993,-0.5777) -- (-0.3004,-0.6185) -- (-0.1764,-0.6645) -- (-0.0388,-0.7105) -- (0.0979,-0.7522) -- (0.3415,-0.8225) -- (0.4105,-0.8444) -- (0.4445,-0.8577) -- (0.4641,-0.8687) -- (0.4690,-0.8737) -- (0.4704,-0.8783) -- (0.4683,-0.8826) -- (0.4629,-0.8864);
  \draw[orbit] (-0.4629,-0.8864) -- (-0.4683,-0.8826) -- (-0.4704,-0.8783) -- (-0.4690,-0.8737) -- (-0.4641,-0.8687) -- (-0.4445,-0.8577) -- (-0.4105,-0.8444) -- (-0.3415,-0.8225) -- (-0.0979,-0.7522) -- (0.0388,-0.7105) -- (0.1764,-0.6645) -- (0.3004,-0.6185) -- (0.3993,-0.5777) -- (0.4902,-0.5362) -- (0.5725,-0.4944) -- (0.6496,-0.4505) -- (0.7308,-0.3976) -- (0.8029,-0.3423) -- (0.8624,-0.2874) -- (0.9119,-0.2310) -- (0.9327,-0.2022) -- (0.9508,-0.1732) -- (0.9661,-0.1439) -- (0.9776,-0.1171) -- (0.9875,-0.0875) -- (0.9946,-0.0577) -- (0.9987,-0.0279) -- (1.0000,0.0000);
  \draw[orbit] (-0.4629,0.8864) -- (-0.4445,0.8926) -- (-0.4145,0.8980) -- (-0.3225,0.9062) -- (-0.1812,0.9116) -- (-0.0000,0.9135);
\end{tikzpicture} &
\begin{tikzpicture}[x=1.32cm,y=1.32cm]
  \input{figures/tikz/frame_min.tikz}
  \draw[orbitback] (0.4592,0.8883) -- (0.4641,0.8849) -- (0.4659,0.8810) -- (0.4645,0.8768) -- (0.4598,0.8723) -- (0.4395,0.8615) -- (0.4060,0.8492) -- (0.3396,0.8291) -- (0.1027,0.7637) -- (-0.0282,0.7256) -- (-0.1646,0.6823) -- (-0.2846,0.6402) -- (-0.3800,0.6033) -- (-0.4722,0.5640) -- (-0.5549,0.5249) -- (-0.6279,0.4865) -- (-0.7089,0.4384) -- (-0.7815,0.3883) -- (-0.8447,0.3365) -- (-0.8951,0.2864) -- (-0.9359,0.2353) -- (-0.9535,0.2081) -- (-0.9667,0.1837) -- (-0.9787,0.1563) -- (-0.9870,0.1318) -- (-0.9935,0.1044) -- (-0.9969,0.0782);
  \draw[orbitback] (0.9288,-0.3705) -- (0.9286,-0.3654) -- (0.9230,-0.3615) -- (0.8976,-0.3587) -- (0.8580,-0.3623) -- (0.8064,-0.3721) -- (0.7536,-0.3857) -- (0.6927,-0.4045) -- (0.6296,-0.4271) -- (0.5665,-0.4525) -- (0.5055,-0.4797) -- (0.4480,-0.5077) -- (0.3374,-0.5686) -- (0.2860,-0.6005) -- (0.2366,-0.6338) -- (0.1898,-0.6683) -- (0.1507,-0.6999) -- (0.1113,-0.7352) -- (0.0801,-0.7668) -- (0.0535,-0.7977) -- (0.0319,-0.8275) -- (0.0156,-0.8559) -- (0.0050,-0.8824) -- (0.0003,-0.9066) -- (0.0014,-0.9283) -- (0.0085,-0.9470) -- (0.0213,-0.9625) -- (0.0397,-0.9746) -- (0.0662,-0.9837) -- (0.0952,-0.9881) -- (0.1285,-0.9887) -- (0.1655,-0.9855) -- (0.2080,-0.9781);
  \draw[orbitback] (0.8374,-0.5465) -- (0.8711,-0.4891) -- (0.8984,-0.4282) -- (0.9172,-0.3676) -- (0.9274,-0.3078) -- (0.9289,-0.2492) -- (0.9220,-0.1921) -- (0.9067,-0.1369) -- (0.8961,-0.1101) -- (0.8816,-0.0801) -- (0.8502,-0.0296) -- (0.8085,0.0217) -- (0.7591,0.0697) -- (0.7024,0.1144) -- (0.6391,0.1555) -- (0.5699,0.1929) -- (0.4955,0.2265) -- (0.4111,0.2581) -- (0.3111,0.2882) -- (0.2070,0.3127) -- (0.0941,0.3322) -- (-0.0202,0.3455) -- (-0.1344,0.3525) -- (-0.2470,0.3535) -- (-0.3621,0.3483) -- (-0.4719,0.3371) -- (-0.5797,0.3193) -- (-0.6782,0.2960) -- (-0.7658,0.2678) -- (-0.8068,0.2513) -- (-0.8441,0.2338) -- (-0.8778,0.2154) -- (-0.9075,0.1963) -- (-0.9332,0.1765) -- (-0.9547,0.1562) -- (-0.9719,0.1355) -- (-0.9848,0.1146) -- (-0.9933,0.0937) -- (-0.9975,0.0713);
  \draw[orbitback] (0.9975,0.0713) -- (0.9933,0.0937) -- (0.9848,0.1146) -- (0.9719,0.1355) -- (0.9547,0.1562) -- (0.9332,0.1765) -- (0.9075,0.1963) -- (0.8778,0.2154) -- (0.8441,0.2338) -- (0.8068,0.2513) -- (0.7658,0.2678) -- (0.6782,0.2960) -- (0.5797,0.3193) -- (0.4719,0.3371) -- (0.3621,0.3483) -- (0.2470,0.3535) -- (0.1344,0.3525) -- (0.0202,0.3455) -- (-0.0941,0.3322) -- (-0.2070,0.3127) -- (-0.3111,0.2882) -- (-0.4111,0.2581) -- (-0.4955,0.2265) -- (-0.5699,0.1929) -- (-0.6391,0.1555) -- (-0.7024,0.1144) -- (-0.7591,0.0697) -- (-0.8085,0.0217) -- (-0.8502,-0.0296) -- (-0.8816,-0.0801) -- (-0.8961,-0.1101) -- (-0.9067,-0.1369) -- (-0.9220,-0.1921) -- (-0.9289,-0.2492) -- (-0.9274,-0.3078) -- (-0.9172,-0.3676) -- (-0.8984,-0.4282) -- (-0.8711,-0.4891) -- (-0.8374,-0.5465);
  \draw[orbitback] (-0.2080,-0.9781) -- (-0.1655,-0.9855) -- (-0.1285,-0.9887) -- (-0.0952,-0.9881) -- (-0.0662,-0.9837) -- (-0.0397,-0.9746) -- (-0.0213,-0.9625) -- (-0.0085,-0.9470) -- (-0.0014,-0.9283) -- (-0.0003,-0.9066) -- (-0.0050,-0.8824) -- (-0.0156,-0.8559) -- (-0.0319,-0.8275) -- (-0.0535,-0.7977) -- (-0.0801,-0.7668) -- (-0.1113,-0.7352) -- (-0.1507,-0.6999) -- (-0.1898,-0.6683) -- (-0.2366,-0.6338) -- (-0.2860,-0.6005) -- (-0.3374,-0.5686) -- (-0.4480,-0.5077) -- (-0.5055,-0.4797) -- (-0.5665,-0.4525) -- (-0.6296,-0.4271) -- (-0.6927,-0.4045) -- (-0.7536,-0.3857) -- (-0.8064,-0.3721) -- (-0.8580,-0.3623) -- (-0.8976,-0.3587) -- (-0.9230,-0.3615) -- (-0.9286,-0.3654) -- (-0.9288,-0.3705);
  \draw[orbitback] (0.9969,0.0782) -- (0.9935,0.1044) -- (0.9870,0.1318) -- (0.9787,0.1563) -- (0.9667,0.1837) -- (0.9535,0.2081) -- (0.9359,0.2353) -- (0.8951,0.2864) -- (0.8447,0.3365) -- (0.7815,0.3883) -- (0.7089,0.4384) -- (0.6279,0.4865) -- (0.5549,0.5249) -- (0.4722,0.5640) -- (0.3800,0.6033) -- (0.2846,0.6402) -- (0.1646,0.6823) -- (0.0282,0.7256) -- (-0.1027,0.7637) -- (-0.3396,0.8291) -- (-0.4060,0.8492) -- (-0.4395,0.8615) -- (-0.4598,0.8723) -- (-0.4645,0.8768) -- (-0.4659,0.8810) -- (-0.4641,0.8849) -- (-0.4592,0.8883);
  \draw[orbit] (0.0000,0.9135) -- (0.1769,0.9118) -- (0.3216,0.9068) -- (0.4129,0.8991) -- (0.4422,0.8940) -- (0.4592,0.8883);
  \draw[orbit] (-0.9969,0.0782) -- (-0.9971,0.0408) -- (-0.9911,0.0019) -- (-0.9801,-0.0334) -- (-0.9622,-0.0710) -- (-0.9405,-0.1049) -- (-0.9112,-0.1406) -- (-0.8763,-0.1751) -- (-0.8360,-0.2083) -- (-0.7904,-0.2401) -- (-0.7399,-0.2702) -- (-0.6849,-0.2986) -- (-0.6256,-0.3252) -- (-0.5624,-0.3498) -- (-0.4957,-0.3725) -- (-0.4260,-0.3931) -- (-0.3480,-0.4128) -- (-0.2675,-0.4300) -- (-0.1792,-0.4457) -- (-0.0895,-0.4583) -- (0.0009,-0.4681) -- (0.0911,-0.4749) -- (0.1864,-0.4792) -- (0.2796,-0.4804) -- (0.3700,-0.4787) -- (0.4670,-0.4737) -- (0.5579,-0.4658) -- (0.6457,-0.4547) -- (0.7240,-0.4415) -- (0.8008,-0.4245) -- (0.8634,-0.4060) -- (0.9088,-0.3869) -- (0.9224,-0.3781) -- (0.9288,-0.3705);
  \draw[orbit] (0.2080,-0.9781) -- (0.2487,-0.9681) -- (0.2937,-0.9542) -- (0.3400,-0.9370) -- (0.3820,-0.9191) -- (0.4295,-0.8962) -- (0.4768,-0.8707) -- (0.5640,-0.8163) -- (0.6459,-0.7550) -- (0.6862,-0.7203) -- (0.7200,-0.6884) -- (0.7555,-0.6516) -- (0.7846,-0.6181) -- (0.8112,-0.5842) -- (0.8374,-0.5465);
  \draw[orbit] (-0.9975,0.0713) -- (-0.9966,0.0502) -- (-0.9906,0.0282) -- (-0.9793,0.0067) -- (-0.9628,-0.0138) -- (-0.9411,-0.0334) -- (-0.9143,-0.0516) -- (-0.8827,-0.0684) -- (-0.8495,-0.0823) -- (-0.8126,-0.0947) -- (-0.7720,-0.1054) -- (-0.7279,-0.1142) -- (-0.6806,-0.1211) -- (-0.6303,-0.1257) -- (-0.5822,-0.1280) -- (-0.5269,-0.1282) -- (-0.4748,-0.1262) -- (-0.3717,-0.1163) -- (-0.2704,-0.0990) -- (-0.1675,-0.0736) -- (-0.0705,-0.0418) -- (0.0191,-0.0049) -- (0.1053,0.0389) -- (0.1467,0.0634) -- (0.1865,0.0895) -- (0.2565,0.1430) -- (0.3150,0.1977) -- (0.3414,0.2268) -- (0.3657,0.2568) -- (0.3877,0.2877) -- (0.4073,0.3195) -- (0.4243,0.3520) -- (0.4367,0.3804) -- (0.4486,0.4140) -- (0.4563,0.4430) -- (0.4626,0.4771) -- (0.4654,0.5064) -- (0.4657,0.5405) -- (0.4633,0.5696) -- (0.4574,0.6030) -- (0.4496,0.6312) -- (0.4375,0.6634) -- (0.4244,0.6901) -- (0.4060,0.7202) -- (0.3877,0.7448) -- (0.3635,0.7720) -- (0.3403,0.7939) -- (0.3151,0.8143) -- (0.2879,0.8331) -- (0.2539,0.8528) -- (0.2230,0.8677) -- (0.1852,0.8826) -- (0.1515,0.8932) -- (0.1110,0.9028) -- (0.0754,0.9086) -- (0.0393,0.9122) -- (0.0030,0.9135) -- (-0.0333,0.9126) -- (-0.0694,0.9094) -- (-0.1051,0.9039) -- (-0.1458,0.8947) -- (-0.1797,0.8845) -- (-0.2178,0.8700) -- (-0.2489,0.8554) -- (-0.2832,0.8361) -- (-0.3107,0.8176) -- (-0.3362,0.7974) -- (-0.3598,0.7758) -- (-0.3844,0.7488) -- (-0.4031,0.7244) -- (-0.4219,0.6945) -- (-0.4354,0.6679) -- (-0.4481,0.6359) -- (-0.4563,0.6078) -- (-0.4626,0.5744) -- (-0.4654,0.5454) -- (-0.4656,0.5113) -- (-0.4632,0.4820) -- (-0.4574,0.4479) -- (-0.4500,0.4188) -- (-0.4386,0.3852) -- (-0.4265,0.3567) -- (-0.4099,0.3241) -- (-0.3907,0.2922) -- (-0.3690,0.2611) -- (-0.3450,0.2310) -- (-0.3189,0.2018) -- (-0.2908,0.1737) -- (-0.2565,0.1430) -- (-0.1865,0.0895) -- (-0.1467,0.0634) -- (-0.1053,0.0389) -- (-0.0191,-0.0049) -- (0.0705,-0.0418) -- (0.1675,-0.0736) -- (0.2704,-0.0990) -- (0.3717,-0.1163) -- (0.4748,-0.1262) -- (0.5269,-0.1282) -- (0.5822,-0.1280) -- (0.6303,-0.1257) -- (0.6806,-0.1211) -- (0.7279,-0.1142) -- (0.7720,-0.1054) -- (0.8126,-0.0947) -- (0.8495,-0.0823) -- (0.8827,-0.0684) -- (0.9143,-0.0516) -- (0.9411,-0.0334) -- (0.9628,-0.0138) -- (0.9793,0.0067) -- (0.9906,0.0282) -- (0.9966,0.0502) -- (0.9975,0.0713);
  \draw[orbit] (-0.8374,-0.5465) -- (-0.8112,-0.5842) -- (-0.7846,-0.6181) -- (-0.7555,-0.6516) -- (-0.7200,-0.6884) -- (-0.6862,-0.7203) -- (-0.6459,-0.7550) -- (-0.5640,-0.8163) -- (-0.4768,-0.8707) -- (-0.4295,-0.8962) -- (-0.3820,-0.9191) -- (-0.3400,-0.9370) -- (-0.2937,-0.9542) -- (-0.2487,-0.9681) -- (-0.2080,-0.9781);
  \draw[orbit] (-0.9288,-0.3705) -- (-0.9224,-0.3781) -- (-0.9088,-0.3869) -- (-0.8634,-0.4060) -- (-0.8008,-0.4245) -- (-0.7240,-0.4415) -- (-0.6457,-0.4547) -- (-0.5579,-0.4658) -- (-0.4670,-0.4737) -- (-0.3700,-0.4787) -- (-0.2796,-0.4804) -- (-0.1864,-0.4792) -- (-0.0911,-0.4749) -- (-0.0009,-0.4681) -- (0.0895,-0.4583) -- (0.1792,-0.4457) -- (0.2675,-0.4300) -- (0.3480,-0.4128) -- (0.4260,-0.3931) -- (0.4957,-0.3725) -- (0.5624,-0.3498) -- (0.6256,-0.3252) -- (0.6849,-0.2986) -- (0.7399,-0.2702) -- (0.7904,-0.2401) -- (0.8360,-0.2083) -- (0.8763,-0.1751) -- (0.9112,-0.1406) -- (0.9405,-0.1049) -- (0.9622,-0.0710) -- (0.9801,-0.0334) -- (0.9911,0.0019) -- (0.9971,0.0408) -- (0.9969,0.0782);
  \draw[orbit] (-0.4592,0.8883) -- (-0.4422,0.8940) -- (-0.4129,0.8991) -- (-0.3216,0.9068) -- (-0.1769,0.9118) -- (0.0000,0.9135);
\end{tikzpicture} &
\begin{tikzpicture}[x=1.32cm,y=1.32cm]
  \input{figures/tikz/frame_min.tikz}
  \draw[orbitback] (0.9962,0.0869) -- (0.9929,0.1057) -- (0.9872,0.1225) -- (0.9776,0.1416) -- (0.9664,0.1583) -- (0.9505,0.1772) -- (0.9340,0.1936) -- (0.8904,0.2279) -- (0.8363,0.2605) -- (0.7722,0.2909) -- (0.6988,0.3187) -- (0.6229,0.3418) -- (0.5344,0.3632) -- (0.4460,0.3797) -- (0.3530,0.3926) -- (0.2564,0.4017) -- (0.1571,0.4069) -- (0.0562,0.4079) -- (-0.0453,0.4048) -- (-0.1392,0.3982) -- (-0.2318,0.3879) -- (-0.3223,0.3740) -- (-0.4100,0.3567) -- (-0.4940,0.3359) -- (-0.5736,0.3117) -- (-0.6480,0.2845) -- (-0.7167,0.2543) -- (-0.7744,0.2239) -- (-0.8303,0.1885) -- (-0.8753,0.1538) -- (-0.9165,0.1141) -- (-0.9472,0.0759) -- (-0.9723,0.0329) -- (-0.9877,-0.0079) -- (-0.9956,-0.0495) -- (-0.9956,-0.0941);
  \draw[orbitback] (0.4584,-0.8887) -- (0.4409,-0.8945) -- (0.4142,-0.8991) -- (0.3226,-0.9068) -- (0.1799,-0.9118) -- (-0.0036,-0.9135) -- (-0.1799,-0.9118) -- (-0.3226,-0.9068) -- (-0.4142,-0.8991) -- (-0.4409,-0.8945) -- (-0.4584,-0.8887);
  \draw[orbitback] (0.9956,-0.0941) -- (0.9956,-0.0495) -- (0.9877,-0.0079) -- (0.9723,0.0329) -- (0.9472,0.0759) -- (0.9165,0.1141) -- (0.8753,0.1538) -- (0.8303,0.1885) -- (0.7744,0.2239) -- (0.7167,0.2543) -- (0.6480,0.2845) -- (0.5736,0.3117) -- (0.4940,0.3359) -- (0.4100,0.3567) -- (0.3223,0.3740) -- (0.2318,0.3879) -- (0.1392,0.3982) -- (0.0453,0.4048) -- (-0.0562,0.4079) -- (-0.1571,0.4069) -- (-0.2564,0.4017) -- (-0.3530,0.3926) -- (-0.4460,0.3797) -- (-0.5344,0.3632) -- (-0.6229,0.3418) -- (-0.6988,0.3187) -- (-0.7722,0.2909) -- (-0.8363,0.2605) -- (-0.8904,0.2279) -- (-0.9340,0.1936) -- (-0.9505,0.1772) -- (-0.9664,0.1583) -- (-0.9776,0.1416) -- (-0.9872,0.1225) -- (-0.9929,0.1057) -- (-0.9962,0.0869);
  \draw[orbit] (-0.0000,0.9135) -- (-0.0362,0.9124) -- (-0.0723,0.9090) -- (-0.1078,0.9034) -- (-0.1495,0.8937) -- (-0.1832,0.8833) -- (-0.2220,0.8681) -- (-0.2529,0.8533) -- (-0.2877,0.8331) -- (-0.3148,0.8143) -- (-0.3400,0.7940) -- (-0.3630,0.7721) -- (-0.3878,0.7442) -- (-0.4060,0.7196) -- (-0.4246,0.6887) -- (-0.4375,0.6619) -- (-0.4497,0.6289) -- (-0.4572,0.6007) -- (-0.4628,0.5664) -- (-0.4649,0.5374) -- (-0.4641,0.5024) -- (-0.4608,0.4732) -- (-0.4539,0.4382) -- (-0.4457,0.4093) -- (-0.4329,0.3750) -- (-0.4200,0.3468) -- (-0.4020,0.3137) -- (-0.3812,0.2813) -- (-0.3579,0.2498) -- (-0.3092,0.1948) -- (-0.2483,0.1392) -- (-0.2152,0.1133) -- (-0.1747,0.0847) -- (-0.0949,0.0365) -- (-0.0105,-0.0053) -- (0.0837,-0.0429) -- (0.1796,-0.0729) -- (0.2823,-0.0970) -- (0.3831,-0.1129) -- (0.4866,-0.1216) -- (0.5901,-0.1219) -- (0.6415,-0.1187) -- (0.6902,-0.1132) -- (0.7359,-0.1057) -- (0.7785,-0.0963) -- (0.8178,-0.0851) -- (0.8572,-0.0708) -- (0.8888,-0.0562) -- (0.9194,-0.0386) -- (0.9427,-0.0214) -- (0.9638,-0.0011) -- (0.9784,0.0182) -- (0.9897,0.0404) -- (0.9956,0.0633) -- (0.9962,0.0869);
  \draw[orbit] (-0.9956,-0.0941) -- (-0.9916,-0.1203) -- (-0.9853,-0.1454) -- (-0.9763,-0.1704) -- (-0.9648,-0.1955) -- (-0.9342,-0.2455) -- (-0.8937,-0.2950) -- (-0.8439,-0.3438) -- (-0.7808,-0.3947) -- (-0.7083,-0.4441) -- (-0.6275,-0.4916) -- (-0.5515,-0.5310) -- (-0.4708,-0.5687) -- (-0.3798,-0.6071) -- (-0.2857,-0.6431) -- (-0.1624,-0.6860) -- (-0.0262,-0.7287) -- (0.1044,-0.7663) -- (0.3389,-0.8304) -- (0.4067,-0.8507) -- (0.4386,-0.8624) -- (0.4584,-0.8726) -- (0.4634,-0.8772) -- (0.4650,-0.8815) -- (0.4631,-0.8854) -- (0.4584,-0.8887);
  \draw[orbit] (-0.4584,-0.8887) -- (-0.4631,-0.8854) -- (-0.4650,-0.8815) -- (-0.4634,-0.8772) -- (-0.4584,-0.8726) -- (-0.4386,-0.8624) -- (-0.4067,-0.8507) -- (-0.3389,-0.8304) -- (-0.1044,-0.7663) -- (0.0262,-0.7287) -- (0.1624,-0.6860) -- (0.2857,-0.6431) -- (0.3798,-0.6071) -- (0.4708,-0.5687) -- (0.5515,-0.5310) -- (0.6275,-0.4916) -- (0.7083,-0.4441) -- (0.7808,-0.3947) -- (0.8439,-0.3438) -- (0.8937,-0.2950) -- (0.9342,-0.2455) -- (0.9648,-0.1955) -- (0.9763,-0.1704) -- (0.9853,-0.1454) -- (0.9916,-0.1203) -- (0.9956,-0.0941);
  \draw[orbit] (-0.9962,0.0869) -- (-0.9956,0.0633) -- (-0.9897,0.0404) -- (-0.9784,0.0182) -- (-0.9638,-0.0011) -- (-0.9427,-0.0214) -- (-0.9194,-0.0386) -- (-0.8888,-0.0562) -- (-0.8572,-0.0708) -- (-0.8178,-0.0851) -- (-0.7785,-0.0963) -- (-0.7359,-0.1057) -- (-0.6902,-0.1132) -- (-0.6415,-0.1187) -- (-0.5901,-0.1219) -- (-0.4866,-0.1216) -- (-0.3831,-0.1129) -- (-0.2823,-0.0970) -- (-0.1796,-0.0729) -- (-0.0837,-0.0429) -- (0.0105,-0.0053) -- (0.0949,0.0365) -- (0.1747,0.0847) -- (0.2152,0.1133) -- (0.2483,0.1392) -- (0.3092,0.1948) -- (0.3579,0.2498) -- (0.3812,0.2813) -- (0.4020,0.3137) -- (0.4200,0.3468) -- (0.4329,0.3750) -- (0.4457,0.4093) -- (0.4539,0.4382) -- (0.4608,0.4732) -- (0.4641,0.5024) -- (0.4649,0.5374) -- (0.4628,0.5664) -- (0.4572,0.6007) -- (0.4497,0.6289) -- (0.4375,0.6619) -- (0.4246,0.6887) -- (0.4060,0.7196) -- (0.3878,0.7442) -- (0.3630,0.7721) -- (0.3400,0.7940) -- (0.3148,0.8143) -- (0.2877,0.8331) -- (0.2529,0.8533) -- (0.2220,0.8681) -- (0.1832,0.8833) -- (0.1495,0.8937) -- (0.1078,0.9034) -- (0.0723,0.9090) -- (0.0362,0.9124) -- (-0.0000,0.9135);
\end{tikzpicture} &
\begin{tikzpicture}[x=1.32cm,y=1.32cm]
  \input{figures/tikz/frame_min.tikz}\input{figures/tikz/f5_7_2.tikz}
\end{tikzpicture} \\[1pt]
\shortstack{{\footnotesize $\rho=2/1$}\\[1pt]{\footnotesize $\alpha=0.985821$}} &
\shortstack{{\footnotesize $\rho=5/2$}\\[1pt]{\footnotesize $\alpha=0.996936$}} &
\shortstack{{\footnotesize $\rho=3/1$}\\[1pt]{\footnotesize $\alpha=0.999357$}} &
\shortstack{{\footnotesize $\rho=7/2$}\\[1pt]{\footnotesize $\alpha=0.999866$}}
\end{tabular}
\caption{\label{fig.catalogue}A catalogue of closed Lissajous-type orbits, one for each of twelve rational windings, with the far side of each trajectory drawn faintly and the equator marked. Every panel is a complete orbit, returning to its starting point with its starting velocity after exactly $q$ oscillations in latitude and $p$ turns in longitude. The value of $\alpha$ under each is the unique one producing that winding. Note how the panels crowd towards the threshold, the last four needing an $\alpha$ within one and a half per cent of $v$.}
\end{figure*}

Table~\ref{tab.resonances} gives the first few values, obtained by solving $\rho(k) = p/q$ numerically; any root-finder will do, since $\rho$ is monotone and smooth. Figure~\ref{fig.catalogue} draws the orbits themselves, for those windings and for five more.

\begin{table}
\caption{\label{tab.resonances}What is needed to reproduce a closed orbit, at $v = 1$. The value of $\alpha$ is obtained by inverting the rotation number~\eqref{eq.rho}, and $q\,T_\theta$ is the time the trajectory takes to close, being $q$ polar periods~\eqref{eq.period}. The closed-form inversion and a direct integration of the equations of motion~\eqref{eq.eom} agree on the closure of each orbit to better than $3 \times 10^{-9}$. Note that the closure time is not monotone in the winding.}
\begin{tabular}{@{}lccccccc@{}}
\toprule
$\rho$ & $1/2$ & $2/3$ & $1$ & $3/2$ & $2$ & $5/2$ & $3$ \\
\midrule
$\alpha$ & 0.470184 & 0.598529 & 0.792726 & 0.939973 & 0.985821 & 0.996936 & 0.999357 \\
$q\,T_\theta$ & 13.363 & 20.995 & 7.926 & 20.053 & 12.747 & 31.512 & 18.862 \\
\bottomrule
\end{tabular}
\end{table}

It is worth noting how the entries of Table~\ref{tab.resonances} accumulate at $\alpha = v$. A winding of two demands an $\alpha$ within one and a half per cent of the threshold, and a winding of three one within a sixteenth of a per cent, because the divergence of $K$ is logarithmic and the rotation number must traverse all of $(0,\infty)$ before $k$ reaches one. The interesting orbits live where the pendulum barely goes over the top. At the other end, expanding the complete elliptic integral for a small modulus~\eqref{eq.modulus} gives
\beq \label{eq.weakfield}
\rho = k + \frac{k^3}{4} + \frac{9 k^5}{64} + O(k^7) ,
\eeq
so for a weak field the rotation number is the modulus itself, corrected at third order, while the Jacobi functions in~\eqref{eq.jacobi} degenerate to ${\rm sn}(v\tau,k) \to \sin(v\tau)$ and the orbit becomes a great circle precessing slowly at the rate $\alpha$. The familiar Larmor precession is thus the first term of a series whose higher terms we have in closed form.

The example has now delivered a complete answer to the question it was set. We know which orbits close, and we know that each rational winding belongs to one value of $\alpha$ and to no other. What it has not told us is what kind of object these orbits are, and to that we turn.

\section{No affine connection has these orbits}
\label{sec.noconnection}

\emph{Is there a connection whose geodesics these are?} The question can be settled without guessing at candidates. What we must compare is the family of curves traced on the sphere, so the first move is to remove the time and keep the paths alone. On any arc where the latitude is monotone we may use $\theta$ as the parameter and describe a path by its slope
\beq \label{eq.slope}
u = \frac{\d\varphi}{\d\theta} .
\eeq
Differentiating the slope~\eqref{eq.slope} along an orbit gives $u' = (\ddot\varphi \dot\theta - \dot\varphi \ddot\theta)/\dot\theta^{3}$, while the constancy of the speed~\eqref{eq.speed} gives $\dot\theta = v/\sqrt{1 + \sin^2\theta \, u^2}$. Feeding both into the equations of motion~\eqref{eq.eom}, every trace of the parametrisation cancels and there remains a single second-order equation for the paths,
\beq \label{eq.pathode}
u' = \Phi(\theta, u) ,
\qquad
\Phi = -\sin\theta\cos\theta \, u^3 - 2\cot\theta \, u
+ \frac{2\alpha}{v} \, \frac{\cos\theta}{\sin\theta}
\left( 1 + \sin^2\theta \, u^2 \right)^{3/2} .
\eeq
The path equation~\eqref{eq.pathode} carries no reference to $\ell$, so it governs the whole family at once; its first two terms are the geodesic equation of the round metric~\eqref{eq.metric} written in the same way, and the entire effect of the field sits in the last term. Rather than guess at candidate connections, we may ask what dependence on the slope an affine geodesic equation is able to produce at all.

\begin{figure}[tbp]
\centering
\plotunits{2.08}{2.04}%
\begin{tikzpicture}[x=\plotux,y=\plotuy]
  \draw[axis] (-1.04,0) -- (1.04,0);
  \draw[axis] (0,-1.02) -- (0,1.02);
  \draw[cubicpart] (-1.0000,0.9064) -- (-0.9428,0.7876) -- (-0.8856,0.6806) -- (-0.8270,0.5825) -- (-0.7970,0.5365) -- (-0.7655,0.4913) -- (-0.7341,0.4490) -- (-0.7026,0.4094) -- (-0.6712,0.3726) -- (-0.6383,0.3368) -- (-0.6054,0.3036) -- (-0.5711,0.2716) -- (-0.5368,0.2423) -- (-0.5025,0.2153) -- (-0.4453,0.1754) -- (-0.3853,0.1395) -- (-0.3224,0.1075) -- (-0.2538,0.0782) -- (-0.1951,0.0569) -- (-0.1294,0.0359) -- (0.0879,-0.0239) -- (0.1665,-0.0474) -- (0.2437,-0.0744) -- (0.3138,-0.1036) -- (0.3838,-0.1387) -- (0.4510,-0.1791) -- (0.5154,-0.2252) -- (0.5768,-0.2768) -- (0.6355,-0.3338) -- (0.6912,-0.3957) -- (0.7455,-0.4640) -- (0.7984,-0.5386) -- (0.8499,-0.6195) -- (0.9014,-0.7089) -- (0.9514,-0.8046) -- (1.0000,-0.9064);%
  \draw[curveB] (-1.0000,0.8580) -- (-0.9500,0.7563) -- (-0.8985,0.6613) -- (-0.8456,0.5733) -- (-0.7927,0.4946) -- (-0.7384,0.4227) -- (-0.6826,0.3579) -- (-0.6254,0.3001) -- (-0.5954,0.2730) -- (-0.5654,0.2481) -- (-0.5054,0.2042) -- (-0.4739,0.1842) -- (-0.4425,0.1662) -- (-0.3781,0.1348) -- (-0.3438,0.1208) -- (-0.3095,0.1086) -- (-0.2738,0.0976) -- (-0.2380,0.0883) -- (-0.1623,0.0737) -- (-0.0822,0.0647) -- (-0.0007,0.0616) -- (0.0808,0.0646) -- (0.1608,0.0734) -- (0.2366,0.0880) -- (0.3081,0.1081) -- (0.3424,0.1202) -- (0.3767,0.1341) -- (0.4096,0.1492) -- (0.4425,0.1662) -- (0.4739,0.1842) -- (0.5054,0.2042) -- (0.5654,0.2481) -- (0.5954,0.2730) -- (0.6254,0.3001) -- (0.6826,0.3579) -- (0.7384,0.4227) -- (0.7927,0.4946) -- (0.8456,0.5733) -- (0.8985,0.6613) -- (0.9500,0.7563) -- (1.0000,0.8580);
  \draw[orbit] (-1.0000,1.7644) -- (-0.9643,1.6151) -- (-0.9285,1.4753) -- (-0.8928,1.3447) -- (-0.8556,1.2182) -- (-0.8184,1.1008) -- (-0.7798,0.9882) -- (-0.7412,0.8847) -- (-0.7026,0.7896) -- (-0.6626,0.6997) -- (-0.6226,0.6180) -- (-0.5811,0.5416) -- (-0.5382,0.4707) -- (-0.4939,0.4057) -- (-0.4496,0.3483) -- (-0.4039,0.2965) -- (-0.3553,0.2488) -- (-0.3124,0.2124) -- (-0.2666,0.1790) -- (-0.2194,0.1495) -- (-0.1708,0.1238) -- (-0.1194,0.1010) -- (-0.0636,0.0806) -- (-0.0036,0.0626) -- (0.0622,0.0466) -- (0.1337,0.0325) -- (0.2152,0.0194) -- (0.3081,0.0071) -- (0.4153,-0.0047) -- (0.5411,-0.0164) -- (0.6841,-0.0279) -- (1.0000,-0.0484);
  \foreach \x/\lab in {-1/{-2.6}, 1/{2.6}}
    {\draw[axis] (\x,0) -- (\x,-0.05); \node[below,font=\footnotesize] at (\x,-0.05) {$\lab$};}
  \node[right,font=\footnotesize] at (1.06,0) {$u$};
\end{tikzpicture}
\caption{\label{fig.pathode}The obstruction, at $\theta = 1$ and $\alpha = 0.6$. The right-hand side of the path equation~\eqref{eq.pathode} is the heavy curve, and it splits exactly into the cubic which an affine connection would supply, dashed, and the term the field contributes. The cubic is antisymmetric; the field term is even, positive, and grows like $\lvert u\rvert^3$ without being a polynomial in $u$ at all. It is the second that the fourth derivative~\eqref{eq.obstruction} detects.}
\end{figure}
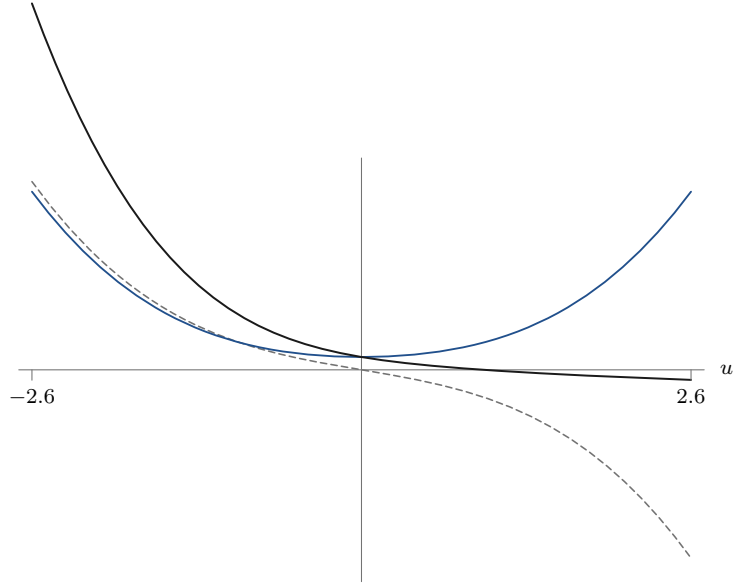

Let $\nabla$ be any affine connection on the sphere, with coefficients $\Gamma^a_{bc}$ in the coordinates $(\theta,\varphi)$, and describe a geodesic $\ddot x^a + \Gamma^a_{bc}\dot x^b \dot x^c = 0$ by its slope in the same manner. Substituting the two components into the same expression for $u'$ and cancelling $\dot\theta$ throughout gives
\beq \label{eq.cartan}
u' = -\Gamma^{\varphi}_{\theta\theta}
+ \left( \Gamma^{\theta}_{\theta\theta} - 2\Gamma^{\varphi}_{\theta\varphi} \right) u
+ \left( 2\Gamma^{\theta}_{\theta\varphi} - \Gamma^{\varphi}_{\varphi\varphi} \right) u^2
+ \Gamma^{\theta}_{\varphi\varphi} \, u^3 ,
\eeq
a cubic polynomial in the slope whose four coefficients are built from the connection and nothing else. Indeed, that is the whole content of the criterion. This is Cartan's criterion~\cite{cartan1924,ovsienko2005}, and it is necessary, so a single test settles our question. A fourth derivative in $u$ annihilates every cubic, and applying it to the path equation~\eqref{eq.pathode} leaves
\beq \label{eq.obstruction}
\frac{\partial^4 \Phi}{\partial u^4}
= \frac{18 \, \alpha \cos\theta \sin^3\theta}{v \left( 1 + \sin^2\theta \, u^2 \right)^{5/2}} ,
\eeq
a quantity which vanishes identically only when $\alpha = 0$ (Figure~\ref{fig.pathode}). For $\alpha \neq 0$, therefore, and away from the poles and the equator, no affine connection whatsoever --- metric or not, symmetric or not --- has these orbits among its geodesics. The obstruction is a statement about the family of paths itself, and not about any choice of parameter along them.

The statement is not peculiar to this system. The magnetic curves of any non-trivial field on any Riemannian manifold are the geodesics of no affine connection, as Barros, Cabrerizo, Fern\'andez and Romero prove~\cite{barros2005gauss}. What the computation above adds is the obstruction itself, in a form one can inspect.

The obstruction is worth examining at its zeros. It vanishes when the field is switched off, where the orbits become great circles and the Levi-Civita connection of the round metric~\eqref{eq.metric} returns, and on the equator, where the restricted field~\eqref{eq.field} itself vanishes. Elsewhere it does not, and the familiar chain of the introduction has broken at a named link.

Switch the field on, and the connection goes.

\section{A Finsler metric has all of them}
\label{sec.randers}

Nevertheless, the orbits of Section~\ref{sec.boundary} are as rigid as geodesics --- locally one through each point in each direction, no more and no fewer. Let us count degrees of homogeneity to see what could have them.

Write the equation of motion~\eqref{eq.lorentz} as $\ddot x^a + \Gamma^a_{bc}\dot x^b \dot x^c = (q/m) \, F^a{}_b \, \dot x^b$. The left-hand side is homogeneous of degree two in the velocity, as any geodesic equation is, whereas the Lorentz term on the right is homogeneous of degree one. That mismatch is what makes the system not a geodesic system, and it is repairable because our orbits all carry the same speed. On the set where $\lvert \dot x \rvert_g = v$ we may multiply the right-hand side by $\lvert \dot x\rvert_g / v$ without changing a single solution, and
\beq \label{eq.homogenised}
\ddot x^a + \Gamma^a_{bc}\dot x^b \dot x^c
= \frac{q}{mv} \, \lvert \dot x \rvert_g \, F^a{}_b \, \dot x^b
\eeq
is homogeneous of degree two throughout. It is, however, not \emph{polynomial} in the velocity, on account of the square root $\lvert\dot x\rvert_g$ carries --- and a non-polynomial right-hand side of degree two is exactly what Cartan's criterion~\eqref{eq.cartan} forbids to a connection, and exactly what a Finsler metric supplies. Equation~\eqref{eq.homogenised} thus tells us both why Section~\ref{sec.noconnection} had to come out negative and where to look next.

A Finsler metric on a surface assigns a length $F(\xi,\zeta) > 0$ to each non-zero tangent vector $\zeta$ at a point $\xi$, homogeneous of degree one in $\zeta$ and strongly convex in it; the length of a curve is $\int F(\gamma,\dot\gamma)\,\d\tau$, which the homogeneity makes independent of the parametrisation, and a Riemannian metric is the special case $F = \lvert \zeta \rvert_g$. The simplest departure from it adds a one-form,
\beq \label{eq.randers}
F_{\rm R}(\xi,\zeta) = \lvert \zeta \rvert_g + b(\zeta) ,
\eeq
and is called a Randers metric~\cite{randers1941,baochernshen2000}. The expression~\eqref{eq.randers} is homogeneous of degree one for any $b$, and is positive and strongly convex precisely when
\beq \label{eq.randerscondition}
\lvert b \rvert_g < 1
\eeq
at every point, which is the one condition a Randers metric must satisfy. It differs from a Riemannian length by a term of degree exactly one in the velocity, which is the homogeneity the Lorentz force carries.

For our system the one-form is fixed by the field, and this is where the vanishing flux noted in Section~\ref{sec.model} is needed. The restricted field~\eqref{eq.field} is exact, and a primitive of it is
\beq \label{eq.primitive}
a = \alpha \sin^2\theta \, \d\varphi ,
\qquad
\d a = 2\alpha \sin\theta\cos\theta \, \d\theta \wedge \d\varphi = \frac{q}{m} F ,
\eeq
as one checks by differentiating. Dividing it by the speed and inserting it into the Randers form~\eqref{eq.randers} with a minus sign gives the metric we want,
\beq \label{eq.FR}
F_{\rm R}(\zeta) = \lvert \zeta \rvert_g
 - \frac{\alpha}{v} \sin^2\theta \, \d\varphi(\zeta)
= \sqrt{ \dot\theta^2 + \sin^2\theta \, \dot\varphi^2 }
- \frac{\alpha}{v} \sin^2\theta \, \dot\varphi ,
\eeq
and the threshold of Section~\ref{sec.boundary} is waiting inside it. What we have written down is the explicit form, for this sphere and this field, of the classical correspondence between magnetic flows and Randers metrics~\cite{randers1941,baochernshen2000,baoroblesshen2004,gibbons2009}.

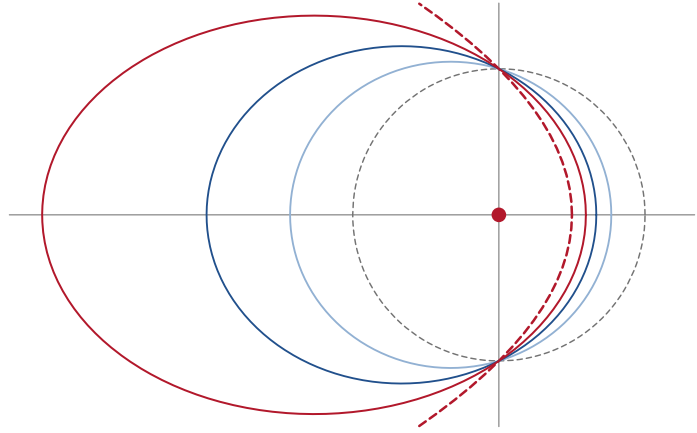
\begin{figure}[tbp]
\centering
\plotunits{4.69}{2.90}%
\begin{tikzpicture}[x=\plotux,y=\plotuy]
  \draw[axis] (-3.35,0) -- (1.34,0);
  \draw[axis] (0,-1.45) -- (0,1.45);
  \fill[dotmark] (0,0) circle (0.05);
  \draw[indcrit] (0.5000,0.0022) -- (0.4957,0.0923) -- (0.4835,0.1816) -- (0.4632,0.2714) -- (0.4342,0.3628) -- (0.3967,0.4546) -- (0.3518,0.5444) -- (0.2983,0.6352) -- (0.2381,0.7238) -- (0.1695,0.8131) -- (0.0926,0.9027) -- (0.0078,0.9922) -- (-0.0892,1.0855) -- (-0.1934,1.1777) -- (-0.3030,1.2672) -- (-0.4234,1.3590) -- (-0.5448,1.4455);
  \draw[indcrit] (-0.5448,-1.4455) -- (-0.4234,-1.3590) -- (-0.3030,-1.2672) -- (-0.1934,-1.1777) -- (-0.0892,-1.0855) -- (0.0078,-0.9922) -- (0.0926,-0.9027) -- (0.1695,-0.8131) -- (0.2381,-0.7238) -- (0.2983,-0.6352) -- (0.3518,-0.5444) -- (0.3967,-0.4546) -- (0.4342,-0.3628) -- (0.4632,-0.2714) -- (0.4835,-0.1816) -- (0.4957,-0.0923) -- (0.5000,-0.0022);%
  \draw[ind1] (1.0000,0.0000) -- (0.9988,0.0494) -- (0.9951,0.0986) -- (0.9890,0.1477) -- (0.9805,0.1963) -- (0.9696,0.2445) -- (0.9564,0.2921) -- (0.9246,0.3809) -- (0.9047,0.4261) -- (0.8825,0.4702) -- (0.8343,0.5512) -- (0.8061,0.5918) -- (0.7759,0.6309) -- (0.7438,0.6684) -- (0.7099,0.7043) -- (0.6404,0.7680) -- (0.6017,0.7987) -- (0.5615,0.8275) -- (0.5200,0.8542) -- (0.4771,0.8788) -- (0.4332,0.9013) -- (0.3881,0.9216) -- (0.3421,0.9396) -- (0.2953,0.9554) -- (0.2478,0.9688) -- (0.1996,0.9799) -- (0.1510,0.9885) -- (0.1020,0.9948) -- (0.0527,0.9986) -- (0.0034,1.0000) -- (-0.0460,0.9989) -- (-0.0953,0.9954) -- (-0.1443,0.9895) -- (-0.1930,0.9812) -- (-0.2412,0.9705) -- (-0.2889,0.9574) -- (-0.3358,0.9419) -- (-0.3819,0.9242) -- (-0.4271,0.9042) -- (-0.4712,0.8820) -- (-0.5142,0.8577) -- (-0.5559,0.8312) -- (-0.6317,0.7752) -- (-0.6692,0.7430) -- (-0.7051,0.7091) -- (-0.7393,0.6734) -- (-0.7716,0.6361) -- (-0.8021,0.5972) -- (-0.8306,0.5569) -- (-0.8571,0.5152) -- (-0.8815,0.4722) -- (-0.9037,0.4281) -- (-0.9238,0.3829) -- (-0.9416,0.3369) -- (-0.9570,0.2899) -- (-0.9702,0.2423) -- (-0.9810,0.1941) -- (-0.9894,0.1454) -- (-0.9953,0.0964) -- (-1.0000,0.0022) -- (-0.9989,-0.0471) -- (-0.9953,-0.0964) -- (-0.9894,-0.1454) -- (-0.9810,-0.1941) -- (-0.9702,-0.2423) -- (-0.9570,-0.2899) -- (-0.9416,-0.3369) -- (-0.9238,-0.3829) -- (-0.9037,-0.4281) -- (-0.8815,-0.4722) -- (-0.8571,-0.5152) -- (-0.8306,-0.5569) -- (-0.8021,-0.5972) -- (-0.7716,-0.6361) -- (-0.7083,-0.7059) -- (-0.6726,-0.7400) -- (-0.6352,-0.7723) -- (-0.5596,-0.8287) -- (-0.5180,-0.8554) -- (-0.4752,-0.8799) -- (-0.4311,-0.9023) -- (-0.3861,-0.9225) -- (-0.3400,-0.9404) -- (-0.2932,-0.9561) -- (-0.2456,-0.9694) -- (-0.1974,-0.9803) -- (-0.1488,-0.9889) -- (-0.0998,-0.9950) -- (-0.0505,-0.9987) -- (-0.0011,-1.0000) -- (0.0931,-0.9957) -- (0.1421,-0.9899) -- (0.1908,-0.9816) -- (0.2391,-0.9710) -- (0.2867,-0.9580) -- (0.3337,-0.9427) -- (0.3798,-0.9251) -- (0.4250,-0.9052) -- (0.4692,-0.8831) -- (0.5123,-0.8588) -- (0.5541,-0.8325) -- (0.5945,-0.8041) -- (0.6335,-0.7738) -- (0.6709,-0.7415) -- (0.7067,-0.7075) -- (0.7408,-0.6717) -- (0.7731,-0.6343) -- (0.8034,-0.5954) -- (0.8319,-0.5550) -- (0.8583,-0.5132) -- (0.8825,-0.4702) -- (0.9229,-0.3850) -- (0.9408,-0.3390) -- (0.9564,-0.2921) -- (0.9696,-0.2445) -- (0.9805,-0.1963) -- (0.9890,-0.1477) -- (0.9951,-0.0986) -- (0.9988,-0.0494) -- (1.0000,-0.0000);
  \draw[ind2] (0.7692,0.0000) -- (0.7679,0.0518) -- (0.7639,0.1035) -- (0.7571,0.1551) -- (0.7484,0.2031) -- (0.7364,0.2542) -- (0.7228,0.3015) -- (0.7055,0.3518) -- (0.6868,0.3983) -- (0.6641,0.4474) -- (0.6403,0.4926) -- (0.6141,0.5370) -- (0.5852,0.5807) -- (0.5539,0.6233) -- (0.5224,0.6620) -- (0.4860,0.7025) -- (0.4498,0.7389) -- (0.4113,0.7741) -- (0.3706,0.8079) -- (0.3275,0.8402) -- (0.2857,0.8685) -- (0.2381,0.8975) -- (0.1923,0.9225) -- (0.1445,0.9457) -- (0.0948,0.9669) -- (0.0432,0.9861) -- (-0.0056,1.0017) -- (-0.0606,1.0164) -- (-0.1124,1.0276) -- (-0.1655,1.0365) -- (-0.2198,1.0430) -- (-0.2753,1.0470) -- (-0.3266,1.0483) -- (-0.3839,1.0470) -- (-0.4366,1.0433) -- (-0.4898,1.0371) -- (-0.5434,1.0283) -- (-0.5972,1.0167) -- (-0.6457,1.0040) -- (-0.6994,0.9872) -- (-0.7476,0.9695) -- (-0.8006,0.9471) -- (-0.8478,0.9245) -- (-0.8943,0.8993) -- (-0.9399,0.8718) -- (-0.9846,0.8418) -- (-1.0280,0.8094) -- (-1.0701,0.7746) -- (-1.1063,0.7417) -- (-1.1453,0.7025) -- (-1.1785,0.6658) -- (-1.2139,0.6225) -- (-1.2435,0.5822) -- (-1.2746,0.5351) -- (-1.3002,0.4917) -- (-1.3237,0.4469) -- (-1.3450,0.4009) -- (-1.3641,0.3538) -- (-1.3808,0.3056) -- (-1.3951,0.2566) -- (-1.4070,0.2068) -- (-1.4163,0.1565) -- (-1.4230,0.1056) -- (-1.4271,0.0545) -- (-1.4286,0.0032) -- (-1.4274,-0.0481) -- (-1.4236,-0.0993) -- (-1.4172,-0.1501) -- (-1.4083,-0.2006) -- (-1.3968,-0.2504) -- (-1.3828,-0.2995) -- (-1.3663,-0.3478) -- (-1.3475,-0.3951) -- (-1.3265,-0.4412) -- (-1.3032,-0.4862) -- (-1.2779,-0.5298) -- (-1.2471,-0.5771) -- (-1.2177,-0.6175) -- (-1.1825,-0.6611) -- (-1.1496,-0.6980) -- (-1.1107,-0.7374) -- (-1.0747,-0.7706) -- (-1.0328,-0.8056) -- (-0.9895,-0.8383) -- (-0.9450,-0.8686) -- (-0.8994,-0.8964) -- (-0.8530,-0.9218) -- (-0.8059,-0.9447) -- (-0.7529,-0.9674) -- (-0.7048,-0.9853) -- (-0.6511,-1.0025) -- (-0.6026,-1.0154) -- (-0.5488,-1.0272) -- (-0.4952,-1.0363) -- (-0.4419,-1.0428) -- (-0.3891,-1.0467) -- (-0.3318,-1.0483) -- (-0.2804,-1.0472) -- (-0.2248,-1.0435) -- (-0.1704,-1.0372) -- (-0.1171,-1.0285) -- (-0.0652,-1.0175) -- (-0.0101,-1.0030) -- (0.0388,-0.9876) -- (0.0905,-0.9686) -- (0.1404,-0.9475) -- (0.1884,-0.9245) -- (0.2344,-0.8996) -- (0.2821,-0.8708) -- (0.3241,-0.8426) -- (0.3673,-0.8104) -- (0.4083,-0.7768) -- (0.4469,-0.7417) -- (0.4833,-0.7053) -- (0.5199,-0.6649) -- (0.5515,-0.6263) -- (0.5831,-0.5837) -- (0.6121,-0.5402) -- (0.6386,-0.4958) -- (0.6625,-0.4507) -- (0.6854,-0.4016) -- (0.7042,-0.3552) -- (0.7217,-0.3049) -- (0.7364,-0.2542) -- (0.7484,-0.2031) -- (0.7571,-0.1551) -- (0.7639,-0.1035) -- (0.7679,-0.0518) -- (0.7692,-0.0000);%
  \draw[ind3] (0.6667,0.0000) -- (0.6650,0.0569) -- (0.6602,0.1138) -- (0.6520,0.1707) -- (0.6412,0.2245) -- (0.6265,0.2814) -- (0.6092,0.3352) -- (0.5887,0.3890) -- (0.5648,0.4426) -- (0.5390,0.4931) -- (0.5098,0.5435) -- (0.4771,0.5935) -- (0.4429,0.6403) -- (0.4054,0.6866) -- (0.3642,0.7324) -- (0.3221,0.7746) -- (0.2766,0.8161) -- (0.2307,0.8540) -- (0.1815,0.8909) -- (0.1288,0.9267) -- (0.0766,0.9587) -- (0.0211,0.9892) -- (-0.0377,1.0182) -- (-0.0951,1.0432) -- (-0.1556,1.0665) -- (-0.2191,1.0877) -- (-0.2800,1.1051) -- (-0.3435,1.1203) -- (-0.4036,1.1320) -- (-0.4720,1.1423) -- (-0.5365,1.1492) -- (-0.6029,1.1534) -- (-0.6644,1.1547) -- (-0.7273,1.1535) -- (-0.7915,1.1496) -- (-0.8569,1.1429) -- (-0.9234,1.1331) -- (-0.9907,1.1201) -- (-1.0512,1.1057) -- (-1.1119,1.0884) -- (-1.1728,1.0683) -- (-1.2336,1.0451) -- (-1.2941,1.0189) -- (-1.3541,0.9894) -- (-1.4059,0.9610) -- (-1.4641,0.9254) -- (-1.5140,0.8916) -- (-1.5626,0.8552) -- (-1.6097,0.8163) -- (-1.6551,0.7749) -- (-1.6987,0.7311) -- (-1.7401,0.6850) -- (-1.7737,0.6436) -- (-1.8106,0.5932) -- (-1.8400,0.5484) -- (-1.8716,0.4944) -- (-1.8962,0.4466) -- (-1.9219,0.3893) -- (-1.9412,0.3390) -- (-1.9604,0.2792) -- (-1.9740,0.2271) -- (-1.9862,0.1654) -- (-1.9937,0.1120) -- (-1.9983,0.0584) -- (-2.0000,0.0045) -- (-1.9988,-0.0494) -- (-1.9947,-0.1031) -- (-1.9877,-0.1565) -- (-1.9760,-0.2183) -- (-1.9629,-0.2706) -- (-1.9442,-0.3306) -- (-1.9253,-0.3810) -- (-1.9001,-0.4385) -- (-1.8759,-0.4865) -- (-1.8447,-0.5408) -- (-1.8156,-0.5859) -- (-1.7791,-0.6365) -- (-1.7401,-0.6850) -- (-1.6987,-0.7311) -- (-1.6551,-0.7749) -- (-1.6097,-0.8163) -- (-1.5626,-0.8552) -- (-1.5140,-0.8916) -- (-1.4641,-0.9254) -- (-1.4059,-0.9610) -- (-1.3541,-0.9894) -- (-1.2941,-1.0189) -- (-1.2412,-1.0420) -- (-1.1804,-1.0656) -- (-1.1195,-1.0861) -- (-1.0587,-1.1037) -- (-0.9983,-1.1184) -- (-0.9308,-1.1318) -- (-0.8643,-1.1420) -- (-0.7987,-1.1490) -- (-0.7343,-1.1532) -- (-0.6713,-1.1547) -- (-0.6097,-1.1536) -- (-0.5430,-1.1497) -- (-0.4784,-1.1431) -- (-0.4097,-1.1330) -- (-0.3494,-1.1215) -- (-0.2857,-1.1066) -- (-0.2245,-1.0894) -- (-0.1607,-1.0683) -- (-0.1001,-1.0453) -- (-0.0424,-1.0203) -- (0.0167,-0.9915) -- (0.0724,-0.9611) -- (0.1249,-0.9292) -- (0.1779,-0.8935) -- (0.2273,-0.8567) -- (0.2734,-0.8189) -- (0.3192,-0.7774) -- (0.3615,-0.7352) -- (0.4029,-0.6895) -- (0.4407,-0.6432) -- (0.4750,-0.5964) -- (0.5079,-0.5464) -- (0.5373,-0.4961) -- (0.5648,-0.4426) -- (0.5887,-0.3890) -- (0.6092,-0.3352) -- (0.6265,-0.2814) -- (0.6412,-0.2245) -- (0.6520,-0.1707) -- (0.6602,-0.1138) -- (0.6650,-0.0569) -- (0.6667,-0.0000);
  \draw[ind4] (0.5952,0.0000) -- (0.5930,0.0669) -- (0.5863,0.1339) -- (0.5749,0.2013) -- (0.5594,0.2665) -- (0.5392,0.3324) -- (0.5150,0.3962) -- (0.4872,0.4580) -- (0.4543,0.5207) -- (0.4178,0.5813) -- (0.3757,0.6428) -- (0.3322,0.6992) -- (0.2831,0.7563) -- (0.2308,0.8108) -- (0.1723,0.8658) -- (0.1146,0.9149) -- (0.0509,0.9640) -- (-0.0147,1.0099) -- (-0.0867,1.0554) -- (-0.1600,1.0972) -- (-0.2339,1.1352) -- (-0.3138,1.1721) -- (-0.3933,1.2049) -- (-0.4712,1.2335) -- (-0.5543,1.2604) -- (-0.6428,1.2853) -- (-0.7281,1.3058) -- (-0.8182,1.3240) -- (-0.9035,1.3379) -- (-0.9928,1.3492) -- (-1.0862,1.3576) -- (-1.1726,1.3622) -- (-1.2621,1.3639) -- (-1.3546,1.3623) -- (-1.4380,1.3579) -- (-1.5358,1.3493) -- (-1.6235,1.3383) -- (-1.7128,1.3237) -- (-1.8036,1.3054) -- (-1.8954,1.2831) -- (-1.9748,1.2606) -- (-2.0545,1.2349) -- (-2.1343,1.2057) -- (-2.2137,1.1731) -- (-2.2925,1.1368) -- (-2.3704,1.0969) -- (-2.4342,1.0607) -- (-2.5091,1.0138) -- (-2.5699,0.9719) -- (-2.6405,0.9180) -- (-2.6971,0.8703) -- (-2.7512,0.8200) -- (-2.8027,0.7673) -- (-2.8513,0.7122) -- (-2.8966,0.6548) -- (-2.9385,0.5952) -- (-2.9767,0.5337) -- (-3.0109,0.4703) -- (-3.0410,0.4053) -- (-3.0667,0.3388) -- (-3.0879,0.2711) -- (-3.1044,0.2025) -- (-3.1161,0.1330) -- (-3.1230,0.0631) -- (-3.1250,0.0070) -- (-3.1238,-0.0491) -- (-3.1179,-0.1191) -- (-3.1071,-0.1886) -- (-3.0916,-0.2575) -- (-3.0713,-0.3254) -- (-3.0465,-0.3921) -- (-3.0173,-0.4574) -- (-2.9838,-0.5211) -- (-2.9464,-0.5831) -- (-2.9053,-0.6430) -- (-2.8606,-0.7009) -- (-2.8127,-0.7565) -- (-2.7617,-0.8097) -- (-2.7081,-0.8605) -- (-2.6520,-0.9087) -- (-2.5819,-0.9631) -- (-2.5214,-1.0056) -- (-2.4468,-1.0532) -- (-2.3832,-1.0898) -- (-2.3056,-1.1304) -- (-2.2269,-1.1673) -- (-2.1475,-1.2005) -- (-2.0678,-1.2303) -- (-1.9881,-1.2566) -- (-1.9086,-1.2796) -- (-1.8166,-1.3025) -- (-1.7257,-1.3213) -- (-1.6361,-1.3364) -- (-1.5483,-1.3479) -- (-1.4501,-1.3571) -- (-1.3664,-1.3618) -- (-1.2735,-1.3638) -- (-1.1836,-1.3626) -- (-1.0968,-1.3583) -- (-1.0030,-1.3503) -- (-0.9132,-1.3393) -- (-0.8275,-1.3256) -- (-0.7369,-1.3078) -- (-0.6511,-1.2875) -- (-0.5621,-1.2628) -- (-0.4785,-1.2360) -- (-0.4002,-1.2075) -- (-0.3202,-1.1749) -- (-0.2399,-1.1381) -- (-0.1655,-1.1002) -- (-0.0917,-1.0584) -- (-0.0193,-1.0130) -- (0.0467,-0.9671) -- (0.1108,-0.9180) -- (0.1689,-0.8689) -- (0.2277,-0.8139) -- (0.2803,-0.7593) -- (0.3297,-0.7022) -- (0.3735,-0.6458) -- (0.4159,-0.5842) -- (0.4527,-0.5236) -- (0.4858,-0.4608) -- (0.5150,-0.3962) -- (0.5392,-0.3324) -- (0.5594,-0.2665) -- (0.5749,-0.2013) -- (0.5863,-0.1339) -- (0.5930,-0.0669) -- (0.5952,-0.0000);
\end{tikzpicture}
\caption{\label{fig.indicatrix}The unit circle of the metric~\eqref{eq.FR}, drawn in a tangent plane for four values of $\lvert b\rvert_g$ and, dashed, for the critical value $1$. The set $F_{\rm R} = 1$ is the polar curve $r = 1/(1 + \lvert b\rvert_g\cos\chi)$, a conic with the origin --- the marked point --- at a focus and eccentricity exactly $\lvert b\rvert_g$; at $\lvert b\rvert_g = 0$ it is the round circle of the metric~\eqref{eq.metric}. As the field strengthens it grows eccentric and slides off the origin, and at $\lvert b\rvert_g = 1$, which by the norm~\eqref{eq.bnorm} is $\alpha = v$ on the equator, the ellipse opens into a parabola and closes no longer. That is the threshold~\eqref{eq.threshold1again} seen in one tangent plane.}
\end{figure}

The threshold is the existence of the metric. The one-form of the metric~\eqref{eq.FR} has $\lvert \d\varphi \rvert_g = 1/\sin\theta$, so its norm is
\beq \label{eq.bnorm}
\lvert b \rvert_g = \frac{\alpha}{v} \, \sin^2\theta \cdot \frac{1}{\sin\theta}
= \frac{\alpha}{v} \, \sin\theta ,
\eeq
whose largest value over the sphere is $\alpha/v$, on the equator. Comparing it against the Randers condition~\eqref{eq.randerscondition},
\beq \label{eq.threshold1again}
F_{\rm R} \ \text{is a Finsler metric}
\qquad \Longleftrightarrow \qquad
\alpha < v ,
\eeq
which is the threshold~\eqref{eq.threshold1} of Section~\ref{sec.boundary}, arrived at from an entirely different direction. Figure~\ref{fig.indicatrix} shows what fails at it: the unit circle of $F_{\rm R}$ is a conic with the origin at a focus, and its eccentricity is exactly $\lvert b\rvert_g$.

The two are more closely related than a shared number. Since the one-form is the primitive~\eqref{eq.primitive} divided by the speed, the norm~\eqref{eq.bnorm} is $\lvert b \rvert_g = \lvert a \rvert_g / v$ with
\beq \label{eq.anorm}
\lvert a \rvert_g = \alpha \sin\theta ,
\eeq
and substituting the norm~\eqref{eq.anorm} into the polar quadrature~\eqref{eq.polar} of Section~\ref{sec.boundary} turns it into
\beq \label{eq.quadnorm}
\dot\theta^2 = v^2 - \lvert a \rvert_g^2
\eeq
identically. The effective potential which confined the latitude was the squared length of the primitive all along. Both readings are the one comparison $\sup_\theta \lvert a \rvert_g < v$ --- a single quantity measured twice, in the base as an energy the particle must overcome, on the tangent spaces as the size of the one-form which deforms the length. There it was the separatrix of a pendulum; here it is the condition for an expression to define a length. \emph{The pendulum goes over the top exactly while the geometry exists.}

The orbits are its geodesics. Since the Randers metric~\eqref{eq.FR} is homogeneous of degree one, its length functional is parametrisation-free, and we may compute with $\theta$ as the parameter exactly as in Section~\ref{sec.noconnection}. Writing the slope~\eqref{eq.slope} again, the length of a path is $\int f(\theta,u) \, \d\theta$ with
\beq \label{eq.f}
f(\theta, u) = \sqrt{ 1 + \sin^2\theta \, u^2 } - \frac{\alpha}{v} \sin^2\theta \, u .
\eeq
The Lagrangian~\eqref{eq.f} does not contain $\varphi$, so its Euler--Lagrange equation delivers a conserved momentum at once,
\beq \label{eq.finslermomentum}
\frac{\partial f}{\partial u}
= \frac{\sin^2\theta \, u}{\sqrt{1 + \sin^2\theta \, u^2}} - \frac{\alpha}{v}\sin^2\theta
= \text{constant} ,
\eeq
and it is strictly convex in the slope, because $\partial^2 f/\partial u^2 = \sin^2\theta \, (1 + \sin^2\theta u^2)^{-3/2} > 0$. Now evaluate the momentum~\eqref{eq.finslermomentum} along a magnetic orbit. There $\sqrt{1+\sin^2\theta u^2} = v/\dot\theta$ and $u = \dot\varphi/\dot\theta$, so the first term is $\sin^2\theta \, \dot\varphi / v$ and the whole expression collapses to
\beq \label{eq.identification}
\frac{\partial f}{\partial u}
= \frac{\sin^2\theta \left( \dot\varphi - \alpha \right)}{v} = \frac{\ell}{v} ,
\eeq
which is constant by the first integral~\eqref{eq.ell}. Thus every magnetic orbit satisfies the conservation law of the Randers length functional. The converse costs nothing further. Strict convexity makes the momentum~\eqref{eq.finslermomentum} strictly increasing in the slope, so $\partial f/\partial u = c$ determines $u$ uniquely from $\theta$ and $c$, and the two families --- magnetic orbits at level $\ell$, $F_{\rm R}$-geodesics at level $c$ --- coincide with $c = \ell/v$.

That is the whole proof, and it uses no machinery beyond a first course in the calculus of variations. The identification~\eqref{eq.identification} also explains what the conserved quantity $\ell$ of Section~\ref{sec.model} has been all along: it is the momentum conjugate to the longitude in a geometry we had not yet named.

It settles a debt from Section~\ref{sec.closure} as well. There we determined which orbits close without saying what closure was a statement about. The closed trajectories of Table~\ref{tab.resonances} are closed geodesics of the metric~\eqref{eq.FR}, and each entry of that table is a value of $\alpha$ at which this geometry possesses one. The round sphere is the extreme case. At $\alpha = 0$ the metric is the round one, every geodesic is a great circle, and every one of them closes after a single turn with no arithmetic involved anywhere. Switching the field on destroys that at once, and closure survives only where the rotation number~\eqref{eq.rho} is rational. The criterion of Section~\ref{sec.closure} is what remains of ``all geodesics closed'' once the sphere carries a field.

\medskip
\begin{remark}[Two properties of the metric]
\label{rem.gauge}
Replacing the primitive~\eqref{eq.primitive} by $a + \d\nu$ leaves the field, and with it the orbits, untouched, but it does change the metric~\eqref{eq.FR}. Since $\d\nu$ contributes to the length integral a term depending only on the endpoints, the two metrics have the same geodesics while assigning different lengths, so the gauge freedom of the potential appears here as a freedom to alter the Finsler length without altering its geodesics. Among the primitives with the symmetry of the problem, ours makes the norm~\eqref{eq.bnorm} smallest. And unlike a Riemannian length, $F_{\rm R}$ charges $v-\alpha$ per unit time to a particle circulating one way about the equator and $v+\alpha$ to one circulating the other. The field distinguishes the two senses, and a geometry encoding it must do the same.
\end{remark}

\section{Closing remarks}
\label{sec.closing}

Let us gather what the example has produced. A charged particle on a round sphere in a uniform ambient field reduces to the quadrature~\eqref{eq.quadrature}, and on the family of orbits which reaches the poles it becomes a pendulum~\eqref{eq.pendulum} whose separatrix~\eqref{eq.threshold1} divides orbits crossing the poles from orbits trapped in a cap. On that family the winding per oscillation is the rotation number~\eqref{eq.rho}, a strictly increasing bijection onto the positive numbers, so a rational winding belongs to exactly one value of $\alpha$.

The geometry then explained what the mechanics had found. No affine connection has these orbits, by the fourth derivative~\eqref{eq.obstruction}; the Randers metric~\eqref{eq.FR} has all of them, by the convexity argument around the identification~\eqref{eq.identification}. The threshold is the condition~\eqref{eq.threshold1again} for that metric to exist, and the pendulum's effective potential is the squared length~\eqref{eq.quadnorm} of the one-form which builds it --- one quantity, measured once in the base and once on the tangent spaces. Closure, which Section~\ref{sec.closure} settled by arithmetic alone, is what survives of a property the round sphere has outright. The sphere carries a geometry that is not Riemannian as soon as the field is switched on.

Two directions lie outside what we have done. The orbits with $\ell \neq 0$ are integrable as well, but their azimuthal rate varies along the trajectory, so the winding is a complete elliptic integral of the third kind and deciding which windings occur requires ordering the roots of a quartic~\cite{lm2026closure}. The regime $\alpha > v$, where the pendulum librates, is exactly where the metric~\eqref{eq.FR} stops being one, so the geometric description ends precisely where the polar motion becomes trapped; what replaces it there we do not know.

It is worth noting what the example does \emph{not} require. No part of the argument used the sphere's constant curvature or the uniformity of the field in the ambient space, and the same three steps --- reduce to a quadrature, locate the threshold by hand, then name the Randers metric --- apply on any surface of revolution carrying an exact field. What the sphere supplies is that every step can be finished in closed form.

The value of the example in a course is not that any one ingredient is unfamiliar, but that they arrive in a single continuous calculation, each forced by the one before it, so that a reader who begins with the Lorentz force ends at a non-Riemannian geometry without having been asked to accept anything in advance.

\section*{Exercises}

\begin{enumerate}
\item Obtain the equations of motion~\eqref{eq.eom} from the covariant Lorentz
force~\eqref{eq.lorentz}, and verify in one line that~\eqref{eq.ell} is conserved.
\item Show that the substitution~\eqref{eq.psi} carries the polar
equation~\eqref{eq.polar} into the pendulum~\eqref{eq.pendulum}, and locate the separatrix. Which configuration of the particle is the pendulum hanging at rest?
\item Verify that the Jacobi functions~\eqref{eq.jacobi} solve the polar
equation~\eqref{eq.polar}, derive the period~\eqref{eq.period}, and obtain the first two terms of~\eqref{eq.weakfield} from the series for $K$.
\item Reproduce the second row of Table~\ref{tab.resonances} by bisection on the rotation
number~\eqref{eq.rho}. For each entry, integrate the equations of motion~\eqref{eq.eom} from the north pole over the time given in the third row and confirm that the trajectory closes, using the embedding~\eqref{eq.embedding} to compare positions so that the polar passages are handled automatically.
\item Derive the cubic form~\eqref{eq.cartan}, confirm that its coefficients reproduce the
first two terms of~\eqref{eq.pathode} when the connection is Levi-Civita for the round metric, and compute the obstruction~\eqref{eq.obstruction}.
\item Verify the identification~\eqref{eq.identification} directly: take an orbit at some
$\ell \neq 0$, compute the momentum~\eqref{eq.finslermomentum} numerically along it, and confirm that the value is $\ell/v$. Show also, from the norm~\eqref{eq.bnorm}, that the metric fails to be positive on some direction as soon as $\alpha \geq v$, and identify it.
\end{enumerate}

\section*{Acknowledgements}

In memory of Jos\'e Antonio Eduardo Roa Neri.

S.I.-R. acknowledges a master's scholarship from the Secretar\'ia de Ciencia, Humanidades, Tecnolog\'ia e Innovaci\'on (SECIHTI), Mexico. The authors declare no conflict of interest.


\end{document}